\documentclass[8pt]{elsarticle}
\usepackage{booktabs}
\usepackage{makecell}
\usepackage{multirow}
\usepackage{longtable}
\usepackage{caption}
\usepackage{footnote}
\usepackage{graphicx}
\usepackage{float}
\usepackage{graphicx}
\usepackage{geometry}
\usepackage{url}
\usepackage{textcomp}
\usepackage{comment}
\usepackage{xurl}
\usepackage{xcolor}
\usepackage{soul}
\usepackage{algorithm}
\usepackage{algorithmic}
\usepackage{amssymb}
\usepackage{geometry}
\usepackage{amsmath}

\journal{arXiv}
\begin{document}
\begin{frontmatter}
\title{Zero Day Malware Detection with Alpha: Fast DBI with Transformer Models for Real World Application}

\author[inst1]{Matthew Gaber}

\affiliation[inst1]{organization={Edith Cowan University School of Science},
            addressline={270 Joondalup Dr}, 
            city={Joondalup },
            postcode={6027}, 
            state={WA},
            country={Australia}}

\author[inst1]{Mohiuddin Ahmed}
\author[inst1]{Helge Janicke}


\begin{abstract}
The effectiveness of an AI model in accurately classifying novel malware hinges on the quality of the features it is trained on, which in turn depends on the effectiveness of the analysis tool used. Peekaboo, a Dynamic Binary Instrumentation (DBI) tool, defeats malware evasion techniques to capture authentic behavior at the Assembly (ASM) instruction level. This behavior exhibits patterns consistent with Zipf's law, a distribution commonly seen in natural languages, making Transformer models particularly effective for binary classification tasks.

We introduce Alpha, a framework for zero-day malware detection that leverages Transformer models and ASM language. Alpha is trained on malware and benign software data collected through Peekaboo, enabling it to identify entirely new samples with exceptional accuracy. Alpha eliminates any common functions from the test samples that are in the training dataset. This forces the model to rely on contextual patterns and novel ASM instruction combinations to detect malicious behavior, rather than memorizing familiar features. By combining the strengths of DBI, ASM analysis, and Transformer architectures, Alpha offers a powerful approach to proactively addressing the evolving threat of malware. Alpha demonstrates perfect accuracy for Ransomware, Worms and APTs with flawless classification for both malicious and benign samples. The results highlight the model’s exceptional performance in detecting truly new malware samples. 
\end{abstract}


\end{frontmatter}

\section{Introduction}
The increasing reliance on digital devices and widespread internet access has led to a rise in cyber attacks targeting information systems. Cyber criminals primarily use malicious software (malware) to carry out these attacks. Malware comes in various forms, including Ransomware, Worms, Trojans, Botnets, Spyware, Advanced Persistent Threats (APT), and post exploitation Tools, and continues to evolve with advanced obfuscation and anti-analysis techniques \cite{gabercsur2024}. Each type of malware serves distinct purposes and exhibits different behaviors \cite{gaber2024nature}. This study focuses on the development of Alpha, a framework that leverages Transformer models and Peekaboo Dynamic Binary Instrumentation (DBI) data to detect zero-day malware \cite{gaberro2024}.

Cyber attacks are becoming more frequent and sophisticated. IBM Security's 2024 report highlights that the average cost of a data breach reached USD 4.88 million, an increase of 10\% compared to 2023, while the average time time to identify and contain breaches involving stolen credentials was 292 days. The Blackberry Cyber Threat Intelligence Report that covers April to June 2024 states that Blackberry cybersecurity solutions prevented 800,000 attacks against critical infrastructure \cite{blackberry2024}. Almost half of the attacks were against businesses in the financial sector. Further, 49\% of the total unique malware hashes, captured in the same period, were used in attacks on critical infrastructure \cite{blackberry2024}. Between 2019 and 2024, the total number of malware targeting platforms like Windows, Android, macOS, and Linux has more than doubled, growing from 540 million to 1.2 billion \cite{avtest2024}. The rise of advanced malware is driven by the availability of feature-rich programming languages, along with powerful open-source encryption and AI libraries \cite{Sharmeen2020}

Software and malware are composed of data, instructions, conditions and functions built for specific tasks and are normally distributed in an executable binary format \cite{ahn2022}. Signature and heuristic based malware detection are typically used by anti-malware vendors \cite{aurangzeb2021,HIRANO2022,khan2020}. A signature is a unique sequence of bytes extracted from information in the binary \cite{khan2020, CARLIN2019, gibert2021}. When new malware is discovered, vendors analyze the sample and create a signature based on distinctive patterns, URLs, strings, and code segments, within the binary \cite{khan2020, CARLIN2019, gibert2021}. This process is time-intensive, requiring not only the creation and testing of the signature but also its distribution through updates \cite{khan2020}. Further, malware that has been recycled and slightly altered can bypass signature based detection \cite{gibert2021}. The process of defining signatures for new or recycled malware is time consuming due to the extensive analysis required and the delays in client updates, which can take several months \cite{ye2017, khan2020, CARLIN2019}. The need for automated AI-driven methods to defend against zero day malware has become urgent. However, the effectiveness of AI in accurately classifying zero day malware depends on the quality of the features it is trained on. Moreover, the authenticity and reliability of these features rely heavily on the analysis tools and datasets used \cite{GABER2025COSE, gabercsur2024, Kajiwara2021}.

Sophisticated and evasive malware can easily bypass commonly used dynamic and static analysis tools due to their various limitations. In contrast, Dynamic Binary Instrumentation (DBI) allows for the capture of the genuine behavior of sophisticated malware \cite{gaber2024nature, galloro2022, kim2022}. Further, if behavioral data is collected from malware while it is concealing it's true behavior in an analysis environment, any AI trained on that data would fail to accurately identify the malware when it runs in a real-world environment, where its malicious actions are fully revealed \cite{galloro2022, nunes2022}. Consequently the Peekaboo DBI data was used in this research \cite{gaber2024nature}. Peekaboo is a DBI tool that comprehensively instruments evasive techniques used by malware. It extracts the malware's true malicious behavior by executing the sample and logging every Assembly (ASM) instruction processed by the CPU. As the malware is forced to expose its malicious functionality, Peekaboo surpasses both static and dynamic analysis tools in extracting its genuine behavior \cite{gaber2024nature}. However, comprehending the contextual information embedded in a binary is a complex task and faces several challenges. First, the content and structure of semantically equivalent binaries can vary based on factors such as obfuscation, architectures, compilers and optimizations. Second, key semantic details, including class hierarchies, function names, variable names and data structures and types, are typically removed by the compiler \cite{ahn2022}. Fundamentally, the challenge is leveraging the contextual meaning derived from binary code, allowing for the inference of code semantics even across syntactically distinct samples.

In previous work, we demonstrated that the Peekaboo DBI data with Transformer models can accurately detect zero-day ransomware \cite{GABER2025COSE}. In this study, we expand this approach to include other types of malware. Additionally, the Peekaboo DBI data was collected by running the samples for 10 to 15 minutes, which has limited real-world application. In Experiment A, we use the complete Peekaboo dataset to assess whether DistilBERT can accurately classify the various types of zero day malware. In Experiments B and C, we examine the per-minute information density within the samples to determine how early during execution in Peekaboo a sample can be accurately classified. Specifically, we aim to identify the minimum amount of time required for effective classification. 

At its core, the challenge lies in identifying previously unseen malicious behaviors by interpreting contextual information and code semantics. In other words, the key question is whether a Transformer model can accurately classify functions it has never encountered before as malicious or benign. Given the limitations of current zero day malware detection techniques, the central aim of this research is to explore how authentic behavioral data can be combined with Transformer models to detect truly new malicious samples.

\subsection{Key Contributions}
Our main contributions are summarized as follows:
\begin{itemize}
    \item We developed Alpha, a framework that uses the Peekaboo DBI data and integrates a stacked architecture consisting of a input Support Vector Machine (SVM), a DistilBERT Transformer model, and an output SVM. 
    \item Alpha employs feature engineering to extract code semantics and contextual meaning from ASM language data across the various malware types.
    \item The experimental results demonstrate the effectiveness of Alpha, which is robust to detecting truly new malicious functions and samples. 
    \item Alpha outperforms the previous state-of-the-art approaches for zero day malware detection.
    \item In real-world applications, Peekaboo's 10-15 minute analysis time is a limitation, we demonstrate that only a 1 minute data slice is required for highly accurate zero day malware detection.
    \item The complete Alpha algorithm is presented.
    \item To the best of our knowledge, the use of ASM language with Transformer models for detecting zero-day malware has not been explored in previous research. In line with the principles of open science, we are releasing the fine tuned DistilBERT models and scripts. \cite{gabertpro2024}.
\end{itemize}

\subsection{Paper Roadmap}
The structure of this paper is as follows: Section 2 provides an overview of the fundamental concepts underlying this research. Section 3 reviews related work in the field. Section 4 details the proposed malware detection pipeline. Section 5 outlines the experimental analysis and implementation specifics. Section 6 discusses the results of the study, and Section 7 concludes the paper with a summary and directions for future research.

\section{Background}
\label{sec:background}
In this section, we provide background information on the different types of malware availabel in the Peekaboo dataset, DistilBERT Transformer model, the DBI data and Zipfs law. Each type of malware serves a different purpose, but the overarching goal of most is to gain some level of control over a system, either for financial gain, espionage, disruption, or continued exploitation. 

Ransomware is designed to block access to a computer system or its data by encrypting files or locking the system. The malware then demands a ransom, typically in cryptocurrency, in exchange for restoring access. The primary objective of ransomware is financial gain through extortion \cite{madani2022ransomwareclassification}. Attackers aim to leverage victims' fear of losing critical data or access to systems to pressure them into paying the ransom \cite{alenezi2020malwareevolution}. Ransomware attacks can target individuals, businesses, and critical infrastructure.

Worms are self-replicating malware programs that spread autonomously over a network. The main purpose of worms is to infect as many systems as possible, often for disruptive purposes. They may overload networks, consume bandwidth, or create back doors for further exploitation. Some worms also serve as delivery mechanisms for other types of malware, such as Ransomware \cite{kara2019maltypes}.

Trojans are a type of malware that trick users into installing it by disguising itself as legitimate software. Once executed, Trojans can perform a variety of malicious actions, such as granting remote access to an attacker, stealing data, and installing additional malicious payloads. The main goal of Trojans is to create a foothold in a victim’s system for further exploitation. Trojans can be used to steal sensitive information, and provide attackers with persistent access to networks \cite{kara2019maltypes}.

A botnet is a network of infected devices called bots, they are remotely controlled by the threat actor. Botnets are normally used to perform large scale malicious attacks including sending spam emails, distributing other malware, and Distributed Denial of Service (DDoS) attacks. The objective of a botnet is to use the collective computing power of compromised devices for illicit purposes. Botnets are offered as a service for attacks, used to harvest personal data, or leveraged for large-scale cyber crime operations \cite{kara2019maltypes}.

Spyware is designed to secretly monitor and collect information about a user’s activities. It can collect, login credentials, browsing habits, keystrokes, or sensitive data \cite{kara2019maltypes}. Spyware is often bundled with other software or installed through phishing or malicious downloads. The objective of spyware is to covertly gather personal or financial information from users for malicious purposes, such as identity theft, fraud, or espionage \cite{alenezi2020malwareevolution}. 

 APTs are highly sophisticated, targeted attacks usually carried out by nation-states or organized cyber criminal groups. These attacks involve prolonged, stealthy intrusions into an organization’s networks, with the aim of stealing sensitive data or intellectual property over time \cite{ahmed2024cyespionage}. APTs often employ a variety of techniques, including social engineering, zero day exploits, and custom malware. The primary objective of an APT is typically espionage, either for corporate or state-sponsored motives. Attackers aim to infiltrate and maintain persistent access to a target’s systems for an extended period, extracting valuable information without detection \cite{ahmed2024cyespionage}.

Post-exploitation tools are used by attackers after they have successfully compromised a system. These tools are designed to maintain access, escalate privileges, move laterally within a network, and gather additional intelligence \cite{benito2023postexpl}. Common post-exploitation techniques include credential dumping, keylogging, and establishing backdoors. The objective of post-exploitation tools is to maintain control of the compromised system, elevate the attacker’s privileges, and expand the scope of the attack within the target network \cite{benito2023postexpl}. These tools help attackers maximize the value of their initial breach, often by further compromising other systems or exfiltrating more sensitive data.

Each malware type exhibits unique functionality and purpose, yet some overlap exists in certain behaviors. The Peekaboo DBI data captures these behaviors across all types and was generated by running the malware and benign samples in a DBI environment for durations ranging between 10 and 15 minutes per sample. While this extended runtime allows for comprehensive behavior analysis of the samples, it lacks practical applicability in real world scenarios where time sensitive detection is crucial. Long analysis times are often impractical for deployment in environments such as endpoint security systems, where decisions need to be made quickly.

To address this limitation, firstly experiment A was conducted to evaluate whether DistilBERT Transformer models, could classify the various types malware and benign samples effectively using the full 10-15 minutes of data. This provided a baseline for the model's classification accuracy. Subsequently, Experiments B and C focused on determining the minimum runtime required to maintain comparable classification performance. These experiments involved truncating the DBI data to shorter durations, that is down to 1 minute data slices, and assessing how well the DistilBERT models performed with the reduced dataset. This approach aimed to identify the optimal balance between analysis time and classification accuracy, thereby making Alpha more applicable to real world scenarios where quick and efficient malware detection is essential.

\begin{table}[htbp]
\centering
\footnotesize
\captionsetup{justification=centering}
  \caption{Binary corpus of the training malware and benign samples from Peekaboo.}
  \label{tab:binarycorpus}
  \begin{tabular}{ccccc}
    \toprule
     \makecell{Family} & \makecell{Samples} & \makecell{Total\\Instructions} & \makecell{Unique\\Instructions} & \makecell{Unique\\Instructions\\Freq. \textgreater{}10}\\
    \midrule
        \makecell{Ransomware} & \makecell{416} & \makecell{182,778,191} & \makecell{3,778,938} & \makecell{256,150} \\
        \makecell{Worm} & \makecell{242} & \makecell{146,895,568} & \makecell{1,951,239} & \makecell{302,759} \\
        \makecell{Trojan} & \makecell{1829} & \makecell{1,455,993,153} & \makecell{3,901,713} & \makecell{469,891} \\
        \makecell{Spyware} & \makecell{900} & \makecell{491,529,663} & \makecell{9,686,779} & \makecell{1,053,880} \\
        \makecell{Botnet} & \makecell{448} & \makecell{130,793,446} & \makecell{1,378,721} & \makecell{209,438} \\
        \makecell{Tool} & \makecell{328} & \makecell{152,110,725} & \makecell{1,596,088} & \makecell{276,001} \\
        \makecell{APT} & \makecell{65} & \makecell{21,958,221} & \makecell{562,284} & \makecell{117,366} \\
        \makecell{Benign} & \makecell{277} & \makecell{70,128,596} & \makecell{1,574,835} & \makecell{177,512} \\ 
    \bottomrule
\end{tabular}
\end{table}

\subsection{Peekaboo DBI}
The Peekaboo dataset includes authentic behavioral data from malware samples, covering Worms, Advanced Persistent Threats (APT), Trojans, Ransomware, post-exploitation Tools, Spyware, and Botnets. Each sample is labeled with its type, family, and variant (e.g., Spyware-SnakeKeylogger-SHA256). In addition, the dataset contains benign software samples \cite{gaber2024nature}. For this study, we utilized every type in the Peekaboo dataset as shown in Table \ref{tab:binarycorpus}. 

\subsection{Zipfs law}
\begin{figure}[htbp]
  \centering
  \includegraphics[width=0.6\textwidth]{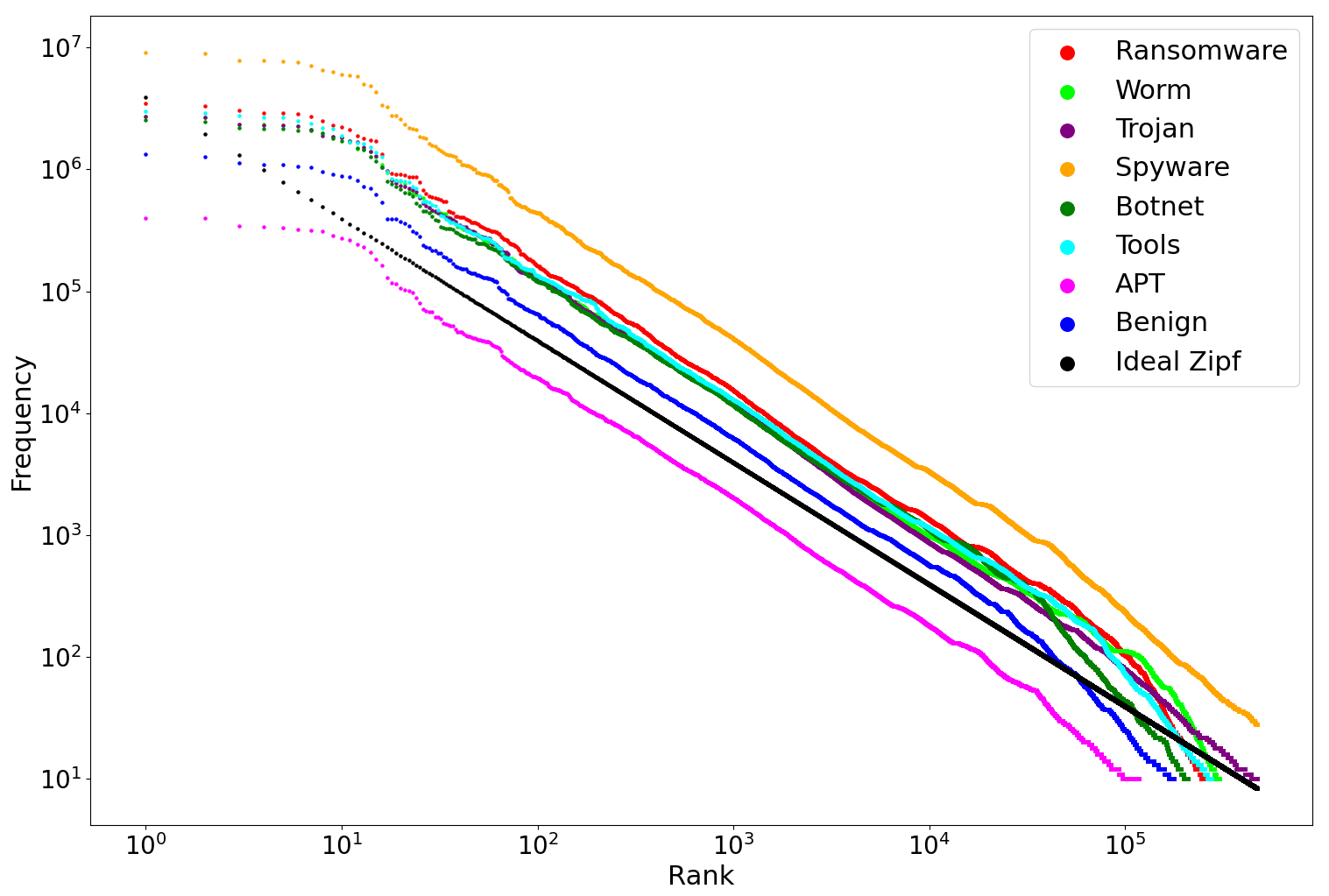}
  \caption{Log-log plot of ASM instruction frequency versus rank across Peekaboo samples.}
  \label{fig:Zip Law}
\end{figure}

Assembly (ASM) language of malware and software exhibits patterns that reveal its structure and usage, with instruction frequencies following Zipf’s law, as seen in natural language \cite{thurner2015}. Zipf’s law describes an inverse relationship between word frequency and rank, resulting in a power law distribution where frequency drops with rank, as shown for all Peekaboo malware and benign ASM instructions in Figure \ref{fig:Zip Law}. Transformer models, which excel at capturing contextual relationships, are well-suited for classifying ASM instructions due to their ability to interpret high and low frequency terms within context, as they do with natural language \cite{koo2021}. By fine-tuning DistilBERT Transformer models on Peekaboo ASM data, instructions and functions can be analyzed similarly to words and sentences, enabling effective binary classification \cite{GABER2025COSE}.

\subsection{Transformer Models}

Transformer models have revolutionized Natural Language Processing (NLP) by leveraging self-attention mechanisms to effectively capture dependencies and contextual information across long sequences without using recurrence or convolutions \cite{vaswani2017, devlin2019}. These models consist of several key components:
\begin{enumerate}
    \item Attention Mechanism: This allows the model to focus on the most relevant parts of an input sequence by calculating attention scores between all token pairs, generating context-aware representations \cite{vaswani2017}. 
    \item Positional Encodings: To account for sequence order, positional information is added to token embeddings, enabling the model to consider sequence structure \cite{vaswani2017}.  
    \item Encoder: A stack of self-attention and feed-forward layers processes the input sequence, progressively refining token representations by capturing dependencies across positions \cite{vaswani2017}.   
    \item Decoder: Also consisting of self-attention and feed-forward layers, it generates output sequences based on the encoder’s final representations, typically for tasks like text generation \cite{vaswani2017}.
    \item Classification Head: For classification tasks, a simple feed-forward network is added atop the encoder, transforming the encoded input into class predictions \cite{devlin2019}. 
\end{enumerate}

BERT (Bidirectional Encoder Representations from Transformers) advanced the Transformer framework by capturing bidirectional context, crucial for understanding semantics in tasks like language modeling, text classification, and question answering \cite{devlin2019}. BERT is pre-trained on large text corpora using two objectives:  
\begin{itemize}
    \item Masked Language Modeling (MLM): Random tokens are masked, and the model predicts them based on surrounding context, learning deeper semantic relationships. 
    \item Next Sentence Prediction (NSP): The model predicts if two sentences follow logically, improving its understanding of inter-sentence relationships.
\end{itemize}

DistilBERT, a smaller, more efficient version of BERT, reduces model size and computational demands while retaining much of its performance. It achieves this through parameter sharing and a combination of supervised learning and knowledge distillation from the larger BERT model \cite{sanh2020}. In previous work, we demonstrated that DistilBERT achieves high accuracy in ransomware classification, making it a suitable choice for this research \cite{GABER2025COSE}.

\section{Related Works}
\begin{table}[htbp]
\centering
\captionsetup{justification=centering}
  \caption{Summary and comparison of recent Transformer model research with our paper}
  \label{tab:litreviews}
  \begin{tabular}{cccccc}
    \toprule
     \makecell{Paper}&\makecell{Dynamic\\Analysis}& \makecell{Assembly\\Language}&\makecell{Function\\Level}&\makecell{Malware\\Classification}&\makecell{Diverse\\Malware}\\
    \midrule
    \cite{li2021}&\texttimes&\checkmark&\checkmark&\texttimes&n.a\\
    \cite{ahn2022}&\texttimes&\checkmark&\checkmark&\texttimes&n.a\\
    \cite{koo2021}&\texttimes&\checkmark&\checkmark&\texttimes&n.a\\
    \cite{saracino2023}&\checkmark&\texttimes&\texttimes&\checkmark&\texttimes\\
    \cite{maniriho2024}&\texttimes&\texttimes&\texttimes&\checkmark&\checkmark\\
    \cite{LIU2024SeMalBERT}&\texttimes&\texttimes&\texttimes&\checkmark&\texttimes\\
    \cite{lu2024malsightexploringmalicioussource}&\texttimes&\texttimes&\checkmark&\checkmark&\texttimes\\
    \cite{GABER2025COSE}&\checkmark&\checkmark&\checkmark&\checkmark&\texttimes\\
    This paper&\checkmark&\checkmark&\checkmark&\checkmark&\checkmark\\
    \bottomrule
\end{tabular}
\end{table}
Research in this field using ASM instructions with Transformer models has primarily focused on BCSD tasks, such as vulnerability identification and software plagiarism detection. Additionally, Transformer models have been applied to malware detection using API call sequences and features derived from static analysis. This section highlights recent advancements in this area, with Table \ref{tab:litreviews} summarizing their key contributions.

Pulse, a framework that employs Transformer models trained with multiple feature engineering techniques using Peekaboo DBI data was introduced in \cite{GABER2025COSE}. The test dataset included ransomware samples from diverse families and benign samples. A key focus of this research was detecting novel ransomware. To achieve this, any functions present in both the training and test samples were removed from the test data. This ensured that the model relied solely on inferring malicious behavior from novel ASM instruction combinations and contextual patterns to classify samples correctly. The results demonstrated the robustness of Pulse in detecting previously unseen malicious functions. It outperformed prior state-of-the-art methods for novel ransomware detection. Unlike other models that may rely on overlapping functions between training and test datasets, Pulse uniquely removes familiar functions, requiring the model to identify malicious behavior based on truly new ASM instruction patterns and context \cite{GABER2025COSE}.

A novel method for analyzing and categorizing Android malware using BERT (Bidirectional Encoder Representations from Transformers) to classify Android API call sequences extracted from a call graph was introduced in \cite{saracino2023}. By leveraging the call graph, this approach captured the dependencies and complex relationships between API calls, offering an in depth understanding of Android malware behavior. The results demonstrate high accuracy in classifying API call sequences as benign or malicious, highlighting the potential for effectively categorizing Android malware and mitigating the risks associated with malicious Android applications\cite{saracino2023}.

EarlyMalDetect utilizes a fine-tuned GPT-2 model to predict future API calls for the early detection of malware attacks \cite{maniriho2024}. By combining Transformer models, bidirectional GRUs, and fully connected neural networks, it effectively detects unknown Windows malware. However, the small dataset and occasional irrelevant sequence generation are noted limitations \cite{maniriho2024}.

SeMalBERT employs a BERT-based architecture with CNN and LSTM layers for API call sequence analysis \cite{LIU2024SeMalBERT}. Using static analysis, API function sequences are extracted from binaries, pre-processed by BERT for semantic relationships, and used to train the CNN-LSTM model. SeMalBERT demonstrates strong resilience to malware transformations, achieving an accuracy of 98.81\%, surpassing previous API-based methods \cite{LIU2024SeMalBERT}.

MALSIGHT is a framework for summarizing benign and malicious source code and pseudocode \cite{lu2024malsightexploringmalicioussource}. A Transformer-based code model called MalT5 was trained on curated datasets. MalT5 processes the code to produce summaries of code structure and interactions. Despite its smaller size (0.77 billion parameters), MalT5 achieved comparable performance to larger models such as ChatGPT-3.5, highlighting its efficiency and accuracy \cite{lu2024malsightexploringmalicioussource}.

DeepSemantic employs an instruction normalization process that uses fine-tuned BERT models for BCSD \cite{koo2021}. Normalizing ASM instructions is crucial Transformer models and prior methods often oversimplified or used extremely granular instruction decomposition. These extremes resulted in issues like loss of semantic meaning, embedding challenges, and Out-of-Vocabulary (OOV) problems. DeepSemantic strikes a balance by preserving critical details, such as memory access patterns, jump targets or call destinations, and register sizes, while managing the token count to reflect accurate semantics \cite{koo2021}.

PalmTree introduced an encoding technique using a BERT model for ASM language and benchmarked it against methods like Asm2Vec and Instruction2Vec \cite{li2021}. Evaluations on basic block search and outlier detection demonstrated PalmTree's superior ability to identify semantic differences in ASM instructions, demonstrating its effectiveness in modeling ASM language \cite{li2021}. PalmTree consistently excelled, even across diverse datasets and compilers, outperforming baseline methods. Function Type Signature Analysis accurately categorized function types, leveraging information like function names, return types, and parameters. In VSA, PalmTree surpassed DeepVSA and other baselines, particularly in understanding global and heap memory usage, aiding tasks such as memory bug detection, code optimization, and program behavior analysis \cite{li2021}. This robust performance underscores PalmTree's enhanced generalization and practical applicability in binary analysis.

BinShot, introduced by \cite{ahn2022}, is a BERT model that uses a Siamese architecture for BCSD. Using IDAPro, non-obfuscated binary files were statically disassembled, and functions were normalized following DeepSemantic's rules \cite{koo2021}. However, this approach is unsuitable for heavily obfuscated malware \cite{gibert2021, khan2020, xiao2020}. BinShot embeds normalized functions with a BERT model and implements a Siamese neural network to determine similarity between function pairs. A binary classifier is then trained on the similarity. Unlike other approaches, BinShot excludes Next Sentence Prediction (NSP), as relationships in binary code are defined by function calls rather than sequential order. Despite potential limitations, such as mapping distinct functions to the same normalized representation, BinShot achieves state-of-the-art performance, outperforming models like DeepSemantic, Gemini and PalmTree in function similarity prediction \cite{ahn2022}. 

\section{Proposed malware detection pipeline}
\label{sec:Proposed malware detection pipeline}
\begin{figure}[htbp]
  \centering
  \includegraphics[width=0.7\textwidth]{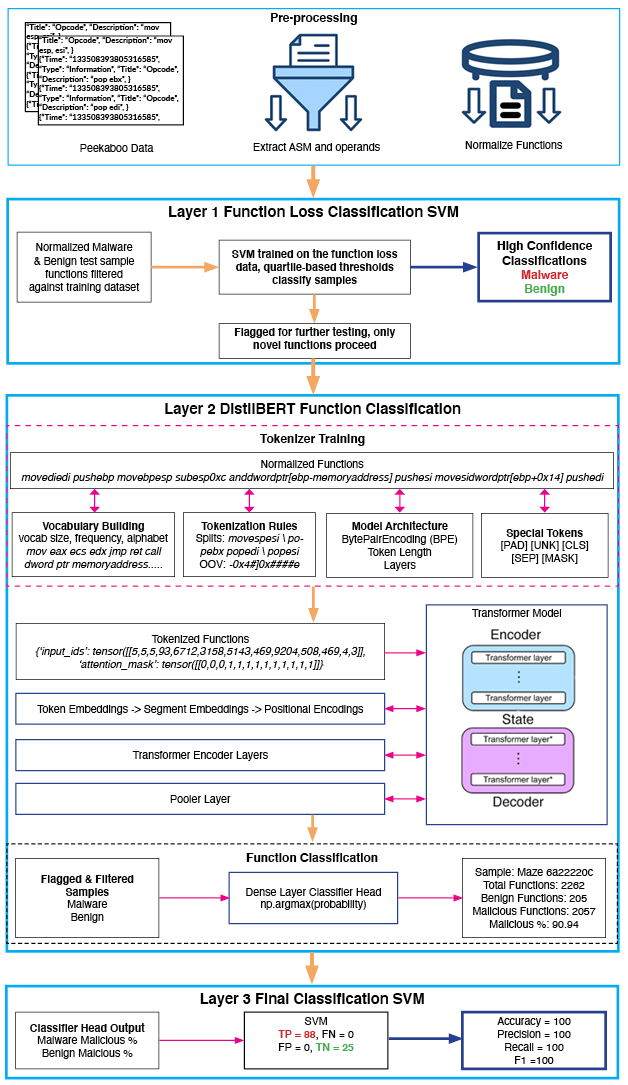}
  \caption{Alpha architecture}
  \label{fig:Alpha Classifier}
\end{figure}
DistilBERT transformer models were fine-tuned with the Peekaboo malware and benign data detailed in Table \ref{tab:binarycorpus}. The models were then evaluated on their ability to classify truly new and never before seen functions, as benign or malicious. The pre-trained DistilBERT models are fine tuned on specific tasks to perform more effectively \cite{devlin2019}. The architecture of Alpha is shown in Figure \ref{fig:Alpha Classifier}. There are three layers:

\textbf{Layer 1: Function Loss SVM}
Each function in every benign and malware test sample is checked against the training dataset. Functions labeled as malicious or benign in the training set are removed from the test sample. This ensures that the remaining functions represent entirely new, unseen behaviors. However, the filtering process can lead to significant information loss for some samples, especially those with a high proportion of overlapping functions. The functions that were removed from a sample because they were in the training dataset, and their labels, that is benign or malicious are evaluated by a SVM, which classifies the sample based on the ratio of malicious functions to benign functions removed during filtering. This layer serves as a first pass filter and determines whether further testing is required. If the SVM cannot classify a sample with high confidence, specifically ensuring no chance of false negatives (FN) or false positives (FP) the filtered sample, containing only truly new functions that the Trasnformer model has never seen before, is passed to Layer 2 for further testing.

\textbf{Layer 2: DistilBERT Function Classification}
This layer leverages the power of DistilBERT, a pre-trained Transformer model, which is fine tuned on the Peekaboo DBI data for function classification specific to each type of malware, that is each type of malware has it's own model. It analyzes the functions in the test sample and classifies them as either benign or malicious based on the semantic relationships and patterns in the code. A simple feed forward neural network and a classification layer were added to the DistilBERT model, creating an end-to-end trainable system using the labeled Peekaboo dataset. The classification layer employs a softmax activation function, which converts the model's outputs into probabilities for each class. Each output node represents the likelihood that the input belongs to a specific class. During prediction, the model processes new input data through its layers and outputs a probability distribution over two classes: benign or malicious. The class with the highest probability is selected as the final prediction.

\textbf{Layer 3: Final Classification SVM}
The final classification is crucial for ensuring accuracy by utilizing the outputs from Layer 2. Specifically, it examines the labeling of functions within each sample, which were categorized as either benign or malicious. This SVM operates by constructing a linear hyperplane that serves as a decision boundary. This boundary is determined based on the proportion of functions labeled as malicious in each malware type sample versus the benign test samples. If the malicious percentage exceeds a certain threshold, the SVM classifies the sample as malicious; otherwise, it classifies the sample as benign. The hyperplane effectively separates the samples into these two categories with the highest possible margin to ensure robust classification.

Overall, this multi layered approach of the DistilBERT model sandwiched between SVMs, enables the system to deliver reliable and accurate predictions for zero day malware. It achieves this by integrating detailed function level analysis with consideration to whether a sample, after significant loss of information during pre-processing, shares substantial similarities with the training dataset. In such cases, the sample can be quickly identified as a variant of a training dataset sample and efficiently classified in Layer 1.

\subsection{Feature Engineering}
In previous work, we demonstrated that pre-processing ASM language instructions by concatenating the instruction itself with its corresponding operands significantly improved the effectiveness of the Transformer models \cite{GABER2025COSE}. Additionally, we replaced hexadecimal memory address values with the placeholder text \texttt{memoryaddress}, as shown in Table \ref{tab:ASM Instruction transformations}. This transformation helped standardize the data by reducing variability caused by unique memory addresses, which are often irrelevant to the functional behavior of the code. Together, these techniques enhanced the ability of the model to focus on meaningful patterns in the instructions and their relationships, leading to improved classification performance \cite{GABER2025COSE}.

\begin{table}[htbp]
\centering
\footnotesize
\captionsetup{justification=centering}
  \caption{ASM instruction transformation}
  \label{tab:ASM Instruction transformations}
  \begin{tabular}{ccc}
    \toprule
     \makecell{Original Instruction} & \makecell{Transformed}\\
    \midrule
        \makecell{mov esp, esi} & \makecell{movespesi} \\
        \makecell{pop esi} & \makecell{popesi} \\
        \makecell{mov byte ptr [ebp-0x2], al} & \makecell{movbyteptr[ebp-0x2]al} \\
        \makecell{call 0x229eda3b} & \makecell{callmemoryaddress} \\
        \makecell{mov eax, dword ptr fs:[0x1]} & \makecell{moveaxdwordptrfs:[0x1]} \\
    \bottomrule
\end{tabular}
\end{table}

\begin{table}[htbp]
\centering
\footnotesize
\captionsetup{justification=centering}
  \caption{Function normalisation}
  \label{tab:Function normalisation}
  \begin{tabular}{cc}
    \toprule
     \makecell{Raw Peekaboo data} & \makecell{Normalised functions}\\
    \midrule
        \makecell{mov esp, esi} & \makecell{}\\
        \makecell{pop ebx} & \makecell{}\\
        \makecell{pop edi} & \makecell{}\\
        \makecell{pop esi} & \makecell{}\\
        \makecell{pop ebp} & \makecell{}\\
        \makecell{\textbf{ret 0x20}} & \makecell{movespesi popebx popedi popesi popebp \textbf{ret0x20}}\\
       \noalign{\vskip 5pt}
        \makecell{mov byte ptr [ebp-0x21], al} & \makecell{}\\
        \makecell{mov dword ptr [ebp-0x1], 0xff12bc11} & \makecell{}\\
        \makecell{mov dword ptr [ebp-0x21], 0x0} & \makecell{movbyteptr[ebp-0x21]al movdwordptr[ebp-0x1]memoryaddress}\\
        \makecell{\textbf{call 0x5577deba}} & \makecell{movdwordptr[ebp-0x21]0x0 \textbf{callmemoryaddress}}\\
       \noalign{\vskip 5pt}
        \makecell{mov eax, dword ptr fs:[0x10]} & \makecell{}\\
        \makecell{mov eax, dword ptr [eax+0x70]} & \makecell{}\\
        \makecell{test eax, eax} & \makecell{}\\
        \makecell{jnz 0x115a3b1e} & \makecell{moveaxdwordptrfs:[0x10] moveaxdwordptr[eax+0x70]}\\
        \makecell{\textbf{ret}} & \makecell{testeaxeax jnz memoryaddress \textbf{ret}}\\

    \bottomrule
\end{tabular}
\end{table}

The ASM instructions captured by Peekaboo for both malware and benign samples adhere to Zipf’s law, which describes the frequency distribution of words in natural languages. Consequently, the transformed ASM instructions and operands shown in Table \ref{tab:ASM Instruction transformations} were engineered into structures analogous to words and sentences as shown in Table \ref{tab:Function normalisation}. 

\subsection{Tokenizer Training}
\begin{table}[htbp]
\centering
\footnotesize
\captionsetup{justification=centering}
  \caption{Tokenizer details}
  \label{tab:tokenizer details}
  \begin{tabular}{ccc}
    \toprule
     \makecell{Tokenizer} & \makecell{Vocab} & \makecell{Tokens}\\
    \midrule
         \makecell{WordPiece distilbert-base-uncased} & \makecell{30,000} & \makecell{mo \#\#v ea \#\#x ,d \#\#word pt \#\#r f \#\#s : [ 0 \#\#x \#\#70 ]\\ mo \#\#v ea \#\#x ,d \#\#word  pt \#\#r [ ea \#\#x + 0 \#\#x \#\#10 ] test\\ ea \#\#x , ea \#\#x j \#\#nz memory \#\#ad \#\#dre \#\#ss re \#\#t} \\
        \noalign{\vskip 5pt}

        \makecell{WordPiece distilbert-custom} & \makecell{30,522} & \makecell{moveaxdwordptrfs : [ 0x70 ] moveaxdwordptr [ eax + 0x10 ] \\testeaxeax jn \#\#z memoryaddress ret}\\
        \noalign{\vskip 5pt}

        \makecell{WordPiece distilbert-trojan} & \makecell{30,522} & \makecell{moveaxdwordptrfs : [ 0x70 ] moveaxdwordptr [ eax + 0x10 ] \\testeaxeax jnzmem ret} \\
        \noalign{\vskip 5pt}
        
    \bottomrule
\end{tabular}
\end{table}
Custom tokenizers were trained to process this data and are compared to the standard distilbert-base-uncased in Table \ref{tab:tokenizer details}. In this framework, each ASM instruction corresponds to a word, and functions are treated as sentences. The training dataset was then processed at the function level, with each function labeled as benign or malicious. At this granular level of ASM instruction analysis, malware samples were found to contain functions also present in benign samples. To ensure accuracy, any functions in the malware samples that also appeared in the benign dataset were excluded from the malicious dataset.

\section{Experimental Analysis}
\begin{table}[htbp]
\centering
\footnotesize
\caption{Training testing dataset details}
\label{tab:Training testing dataset details}
\begin{tabular}{cccccc}
\hline
\makecell{Type} & \multicolumn{2}{c}{\makecell{Training}} & \multicolumn{2}{c}{\makecell{Testing}} \\
\hline
\multirow{13}{*}{\makecell{Ransomware}} & BlackCat & 62 & LockBit & 26 \\
& Chaos & 2 & Maze & 9 \\
& Clop & 2 & Petya & 5 \\
& Conti & 51 & WannaCry & 48 \\
& DarkSide & 47 & & \\
& Dharma & 10 & & \\
& Hive & 15 & & \\
& Locky & 17 & & \\
& NetWalker & 33 & & \\
& RagnarLocker & 2 & & \\
& Ryuk & 26 & & \\
& Sodinokibi & 70 & & \\
& Stop & 80 & & \\
\hline
\multirow{3}{*}{\makecell{Worm}} & \makecell{N-W0rm} & \makecell{139} & \makecell{Worm.Sfone} & \makecell{4} \\
& Worm.Shodi & 5 & Worm.Virut & 22 \\
& Worm.Ramnit & 100 & & \\
\hline
\multirow{14}{*}{\makecell{Trojan}} & AgentTesla & 35-150 & Dyre & 1 \\
& AveMariaRAT & 35-150 & Emotet & 63 \\
& DarkComet & 35-150 & Quakbot & 2 \\
& Dridex & 15 & SchoolBoy & 9 \\
& Glupteba & 35-150 & YoungLotus & 10 \\
& Heodo & 35-150 & & \\
& IcedID & 14 & & \\
& Loki & 35-150 & & \\
& NanoCore & 35-150 & & \\
& njrat & 35-150 & & \\
& PrivateLoader & 35-150 & & \\
& SmokeLoader & 35-150 & & \\
& Trickbot & 35-150 & & \\
& XWorm & 35-150 & & \\
\hline
\multirow{6}{*}{\makecell{Spyware}} & \makecell{RedLineStealer} & \makecell{150} & \makecell{MysticStealer} & \makecell{50} \\
& Formbook & 150 & Vidar & 49 \\
& Stealc & 150 & SnakeKeylogger & 34 \\
& RecordBreaker & 150 & & \\
& MarsStealer & 150 & & \\
& RacoonStealer & 150 & & \\
\hline
\multirow{3}{*}{\makecell{Botnet}} & \makecell{Amadey} & \makecell{150} & \makecell{Lu0Bot} & \makecell{62} \\
& Socks5Systemz & 150 \\
& ZeuS & 148 & & \\
\hline
\multirow{2}{*}{\makecell{Tool}} & \makecell{CobaltStrike} & \makecell{288} & \makecell{Backdoor.TeamViewer} & \makecell{50} \\
& Mimikatz & 40  \\
\hline
\multirow{3}{*}{\makecell{APT}} & \makecell{APT29} & \makecell{4} & \makecell{APT28} & \makecell{6} \\
& Lazarus & 15 & Turla & 8 \\
& RustyStealer & 46 & & \\
\hline
\multirow{1}{*}{\makecell{Benign}} & \makecell{Benign} & \makecell{277} & \makecell{Benign} & \makecell{25} \\
\hline
\end{tabular}
\end{table}

To determine the effectiveness of the DistilBERT models with the Peekaboo DBI data 3 experiments were performed. For all experiments the test data, including the benign test samples, were filtered and any function that was in the train data was removed from the test data. For more detail please see the source code at \cite{gaberro2024}.  The train and test samples are listed in Table \ref{tab:Training testing dataset details} and the hyperparameters for the DistilBERT models are shown in Table \ref{tab:Hyperparameters}. 
\begin{table}[htbp]
\centering
\footnotesize
\captionsetup{justification=centering}
  \caption{Transformer models hyperparameters}
  \label{tab:Hyperparameters}
  \begin{tabular}{ccccc}
    \toprule
     \makecell{Model} & \makecell{Layers} & \makecell{Hidden Layers} & \makecell{Attention Heads} & \makecell{Parameters}\\
    \midrule
        \makecell{DistilBERT} & \makecell{6 Encoder} & \makecell{768} & \makecell{12} & \makecell{66M}\\  
    \bottomrule
\end{tabular}
\end{table}

Experiment A used DistilBERT 256, custom tokenizers, and the normalized functions. Experiment B focused on analyzing the distribution and density of ASM instructions within the test samples by dividing the data into one-minute time slices. A key objective of this research was to evaluate whether DBI data and Transformer models, such as DistilBERT, have practical real-world applications. To achieve this, the study assessed the ability of the DistilBERT models trained in Experiment A to classify samples accurately, using only one minute of data from these time slices, instead of relying on the complete dataset. Experiment C introduced the Layer 1 Function Loss Classification SVM, as shown in Figure \ref{fig:Alpha Classifier}. A primary objective of this research was to evaluate whether the DistilBERT model could accurately classify entirely new, never-before seen functions and samples. Consequently, each test sample was filtered against the training data to remove any functions that had been previously encountered. This filtering process resulted in significant information loss for certain samples, depending on the extent of overlap with the training data, where different malware families can use similar functionality. Layer 1 utilized a SVM to determine whether a sample was benign or malicious, based on the proportion of functions removed during filtering. Specifically, the SVM analyzed how many of these functions had been labeled as malicious versus benign in the training dataset. For this final experiment, Layer 1, the DistilBERT models trained in Experiment A and the one-minute slices of data analyzed in Experiment B were used. 

\subsection{Experiment setup and evaluation}
This section outlines the computational environment and performance metrics employed in the study. The experiments were conducted using Google Colaboratory to train and test the Transformer models, with Python 3 as the programming environment \cite{googlecolab}. Depending on availability, various GPUs were utilized for model processing and optimization, enabling efficient training and fine-tuning in a cloud-based setting.

To evaluate the performance of the Transformer models in accurately identifying and classifying both malware and benign software, several metrics were computed: precision, recall, and F1 score. While accuracy indicates the overall proportion of correct predictions, it can be misleading in cases of class imbalance. Precision focuses on the correctness of the model’s ransomware predictions by measuring the proportion of true ransomware predictions among all ransomware predictions made. Recall, also known as sensitivity, evaluates the model’s ability to detect ransomware by measuring the proportion of actual ransomware samples correctly identified. The F1 score, a harmonic mean of precision and recall, provides a comprehensive measure of the model’s performance, especially in scenarios where there is a trade-off between precision and recall.

\subsection{Experiment A}
\begin{table}[htbp]
\centering
\footnotesize
\captionsetup{justification=centering}
  \caption{Experiment A Normalised functions training corpus}
  \label{tab:Experiment A training Corpus}
  \begin{tabular}{ccccccc}
    \toprule
     \makecell{Type} & \makecell{Initial} & \makecell{Deduplicated} & \makecell{Filtered} & \makecell{Avg Function\\Length} & \makecell{\%\textgreater256} & \makecell{Unique Words}\\
    \midrule
    \makecell{Benign} & \makecell{1,599,629} & \makecell{267,937} & \makecell{267,937}  & \makecell{28.5} & \makecell{257}& \makecell{731,997}\\
    \makecell{Ransomware} & \makecell{4,189,859} & \makecell{518,829} & \makecell{489,204}  & \makecell{37.26} & \makecell{1047}& \makecell{1,174,268}\\
    \makecell{Worm} & \makecell{3,209,618} & \makecell{315,923} & \makecell{286,951}  & \makecell{34.22} & \makecell{1265}& \makecell{604,083}\\
    \makecell{Trojan 150} & \makecell{44,855,742} & \makecell{5,113,889} & \makecell{4,935,699}  & \makecell{32.15} & \makecell{14,287}& \makecell{10,553,658}\\
    \makecell{Trojan 75} & \makecell{16,675,577} & \makecell{2,454,363} & \makecell{2,291,770}  & \makecell{31.51} & \makecell{6,537}& \makecell{5,629,423}\\
    \makecell{Trojan 35} & \makecell{8,298,014} & \makecell{1,484,448} & \makecell{1,337,475}  & \makecell{30.92} & \makecell{3375}& \makecell{3,373,513}\\
    \makecell{Trojan 35 JMPNZ} & \makecell{8,298,014} & \makecell{1,226,776} & \makecell{1,096,636}  & \makecell{32.61} & \makecell{3304}& \makecell{1,132,150}\\
    \makecell{Spyware} & \makecell{10,926,205} & \makecell{1,604,553} & \makecell{1,465,581}  & \makecell{33.00} & \makecell{4810}& \makecell{3,924,394}\\
    \makecell{APT} & \makecell{484,371} & \makecell{76,347} & \makecell{53,918}  & \makecell{41.98} & \makecell{585}& \makecell{228,458}\\
    \makecell{Botnet} & \makecell{2,984,318} & \makecell{222,013} & \makecell{184,117}  & \makecell{31.6} & \makecell{600}& \makecell{568,748}\\
    \makecell{Tools} & \makecell{3,589,840} & \makecell{268,025} & \makecell{214,487}  & \makecell{33.43} & \makecell{745}& \makecell{645,724}\\
    \bottomrule
\end{tabular}
\end{table}
\begin{table}[htbp]
\centering
\footnotesize
\captionsetup{justification=centering}
  \caption{Experiment A DistilBERT 256 and fine tuning details}
  \label{tab:finetuningA}
  \begin{tabular}{cccccc}
    \toprule
    \makecell{Type} & \makecell{Epochs} & \makecell{Batch Size} & \makecell{Tokenizer} & \makecell{Validation\\Loss}& \makecell{Validation\\Accuracy \%}\\
    \midrule      
        \makecell{Ransomware} & \makecell{3} & \makecell{16} & \makecell{distilbert-custom} &  \makecell{0.1004} & \makecell{95.46} \\
        \makecell{Worm} & \makecell{3} & \makecell{16} & \makecell{distilbert-custom} &  \makecell{0.1120} & \makecell{94.75} \\
        \makecell{Trojan 150} & \makecell{3} & \makecell{16} & \makecell{distilbert-custom} & \makecell{0.3014} & \makecell{84.73} \\
        \makecell{Trojan 75} & \makecell{3} & \makecell{16} & \makecell{distilbert-custom} & \makecell{0.2640} & \makecell{87.18} \\
        \makecell{Trojan 35} & \makecell{3} & \makecell{16} & \makecell{distilbert-custom} & \makecell{0.2506} & \makecell{89.16} \\
        \makecell{Trojan 35 JMPNZ} & \makecell{3} & \makecell{16} & \makecell{distilbert-custom} & \makecell{0.2522} & \makecell{88.80} \\
        \makecell{Spyware} & \makecell{3} & \makecell{16} & \makecell{distilbert-custom} & \makecell{0.2217} & \makecell{90.31} \\
        \makecell{APT} & \makecell{3} & \makecell{16} & \makecell{distilbert-custom} &  \makecell{0.1030} & \makecell{96.02} \\
        \makecell{Botnet} & \makecell{3} & \makecell{16} & \makecell{distilbert-custom} &  \makecell{0.2758} & \makecell{88.33} \\
        \makecell{Tools} & \makecell{3} & \makecell{16} & \makecell{distilbert-custom} &  \makecell{0.2916} & \makecell{87.01} \\
        
    \bottomrule
\end{tabular}
\end{table}

\begin{figure}[htbp]
    \centering
    \begin{minipage}{0.4\textwidth}
        \centering
        \includegraphics[width=\textwidth]{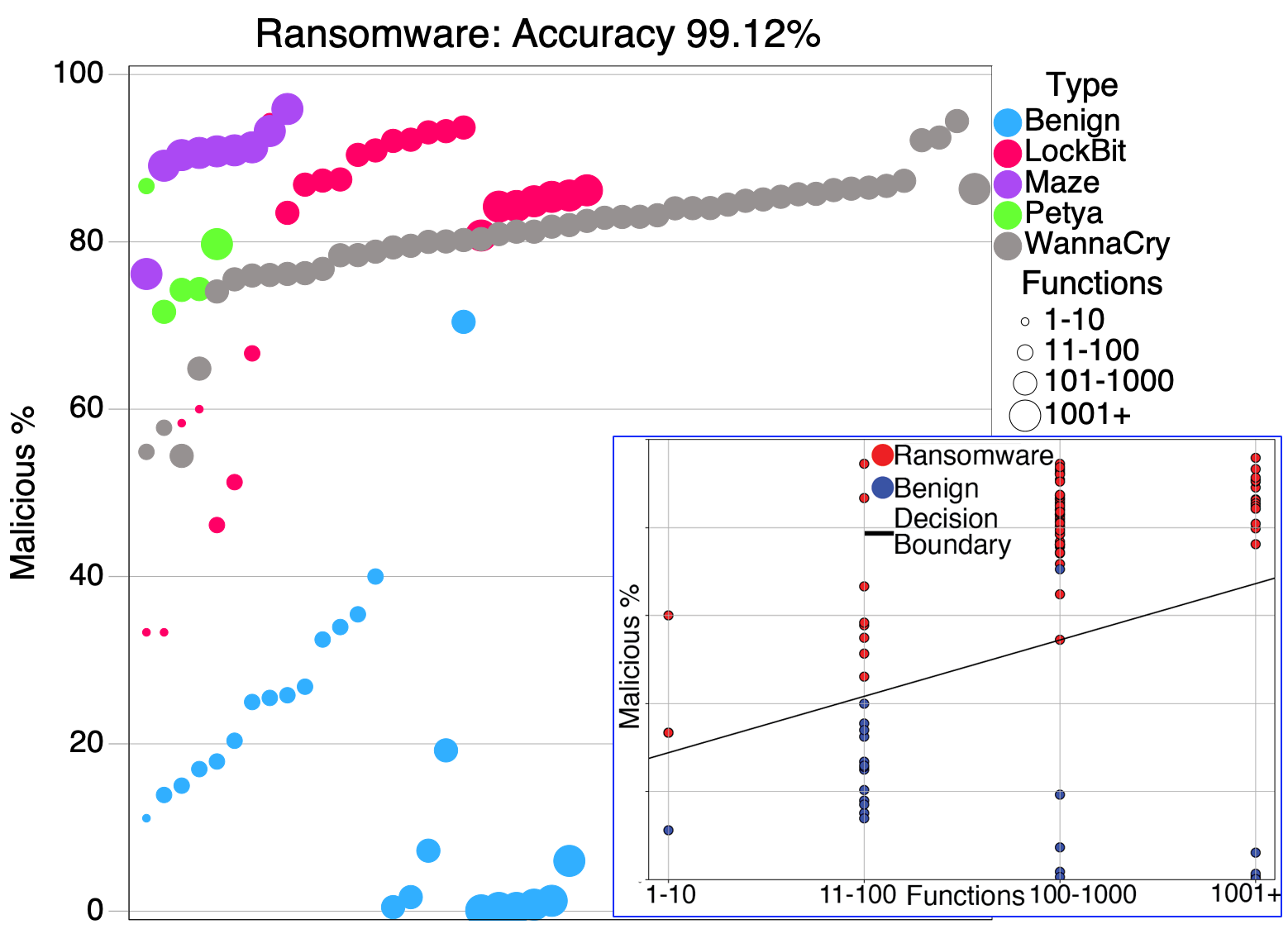}
    \end{minipage}
    \begin{minipage}{0.4\textwidth}
        \centering
        \includegraphics[width=\textwidth]{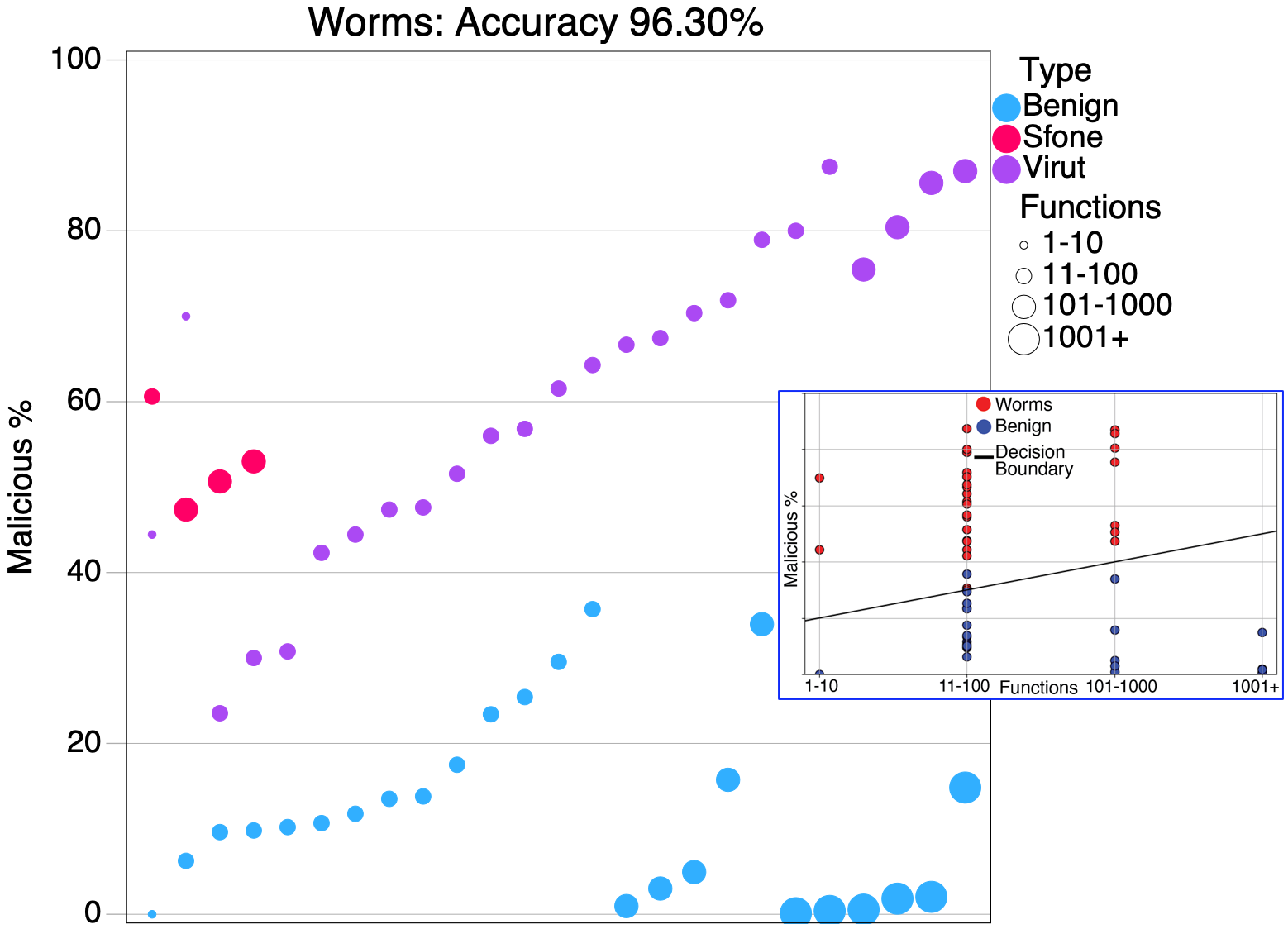}
    \end{minipage}


    \begin{minipage}{0.4\textwidth}
        \centering
        \includegraphics[width=\textwidth]{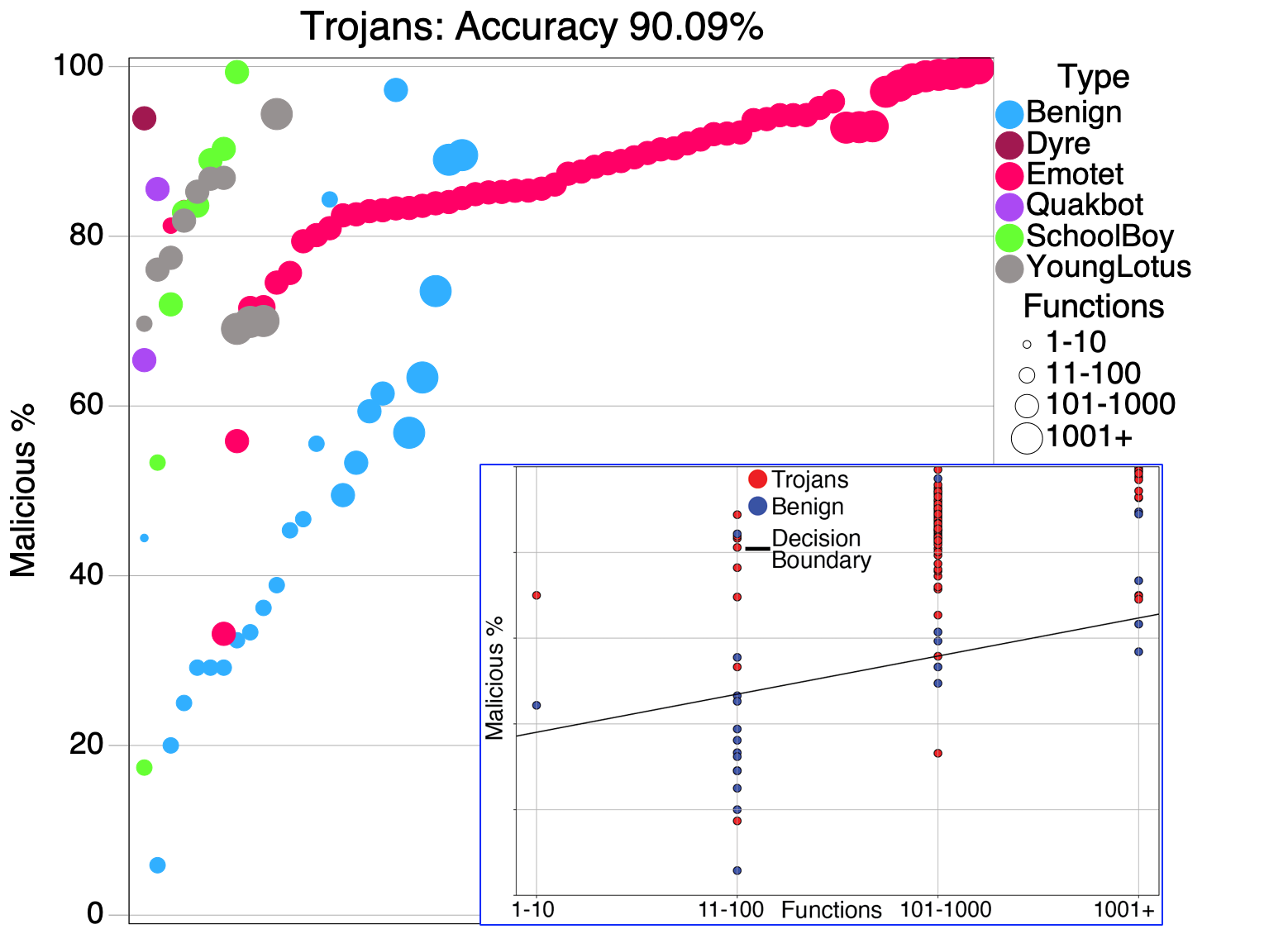}
    \end{minipage}
    \begin{minipage}{0.4\textwidth}
        \centering
        \includegraphics[width=\textwidth]{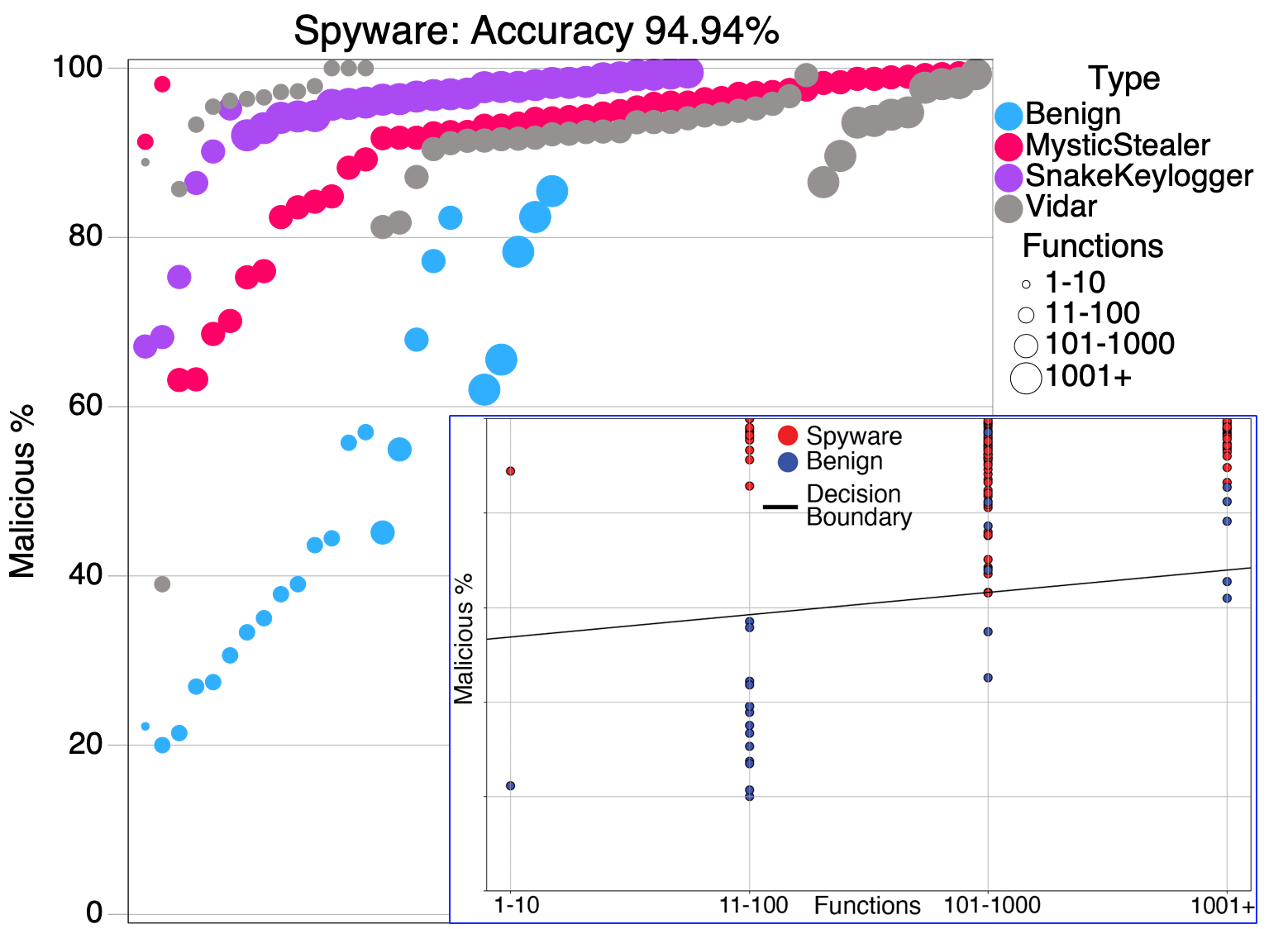}
    
    \end{minipage}


    \begin{minipage}{0.4\textwidth}
        \centering
        \includegraphics[width=\textwidth]{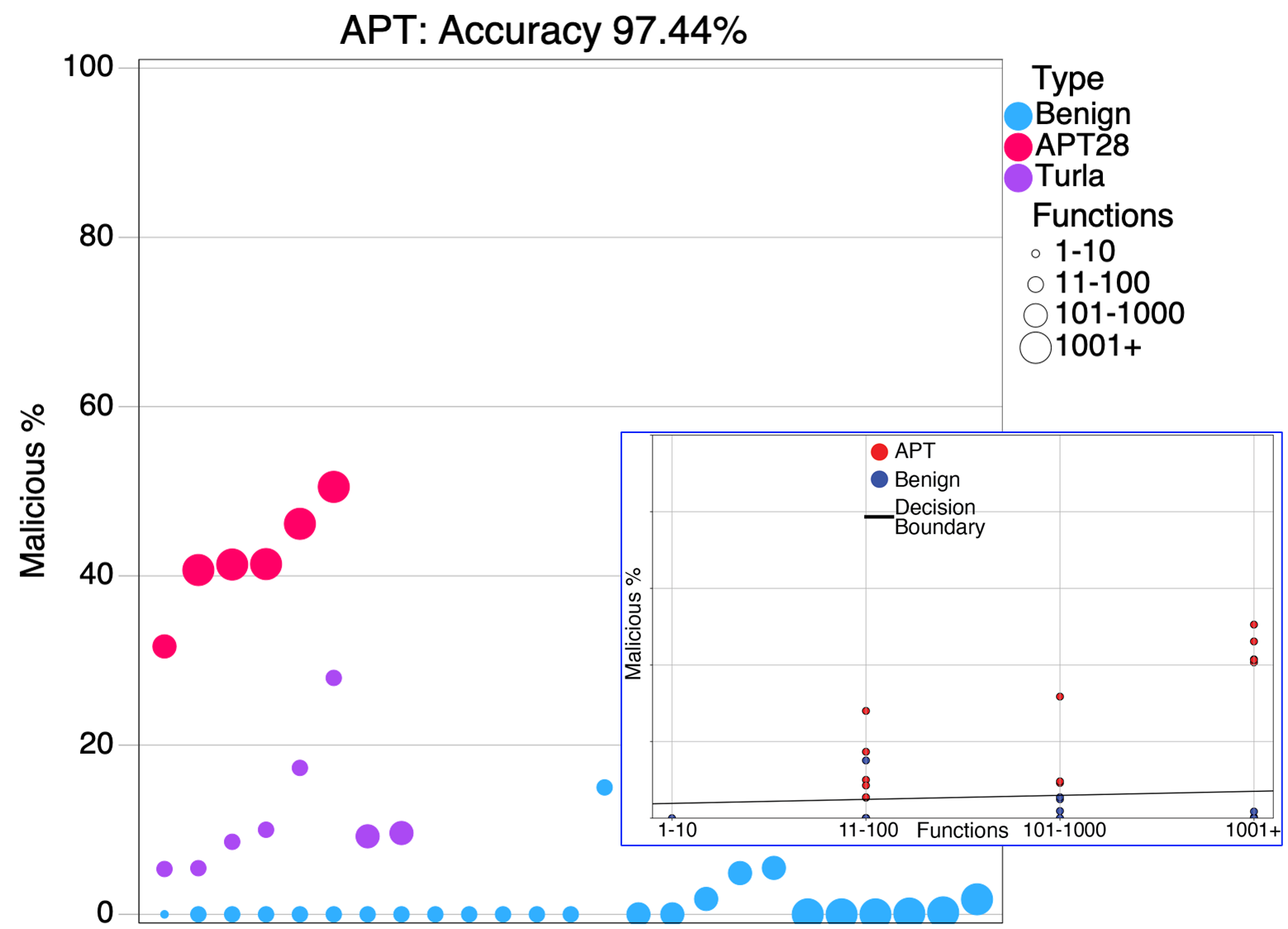}
    
    \end{minipage}
    \begin{minipage}{0.4\textwidth}
        \centering
        \includegraphics[width=\textwidth]{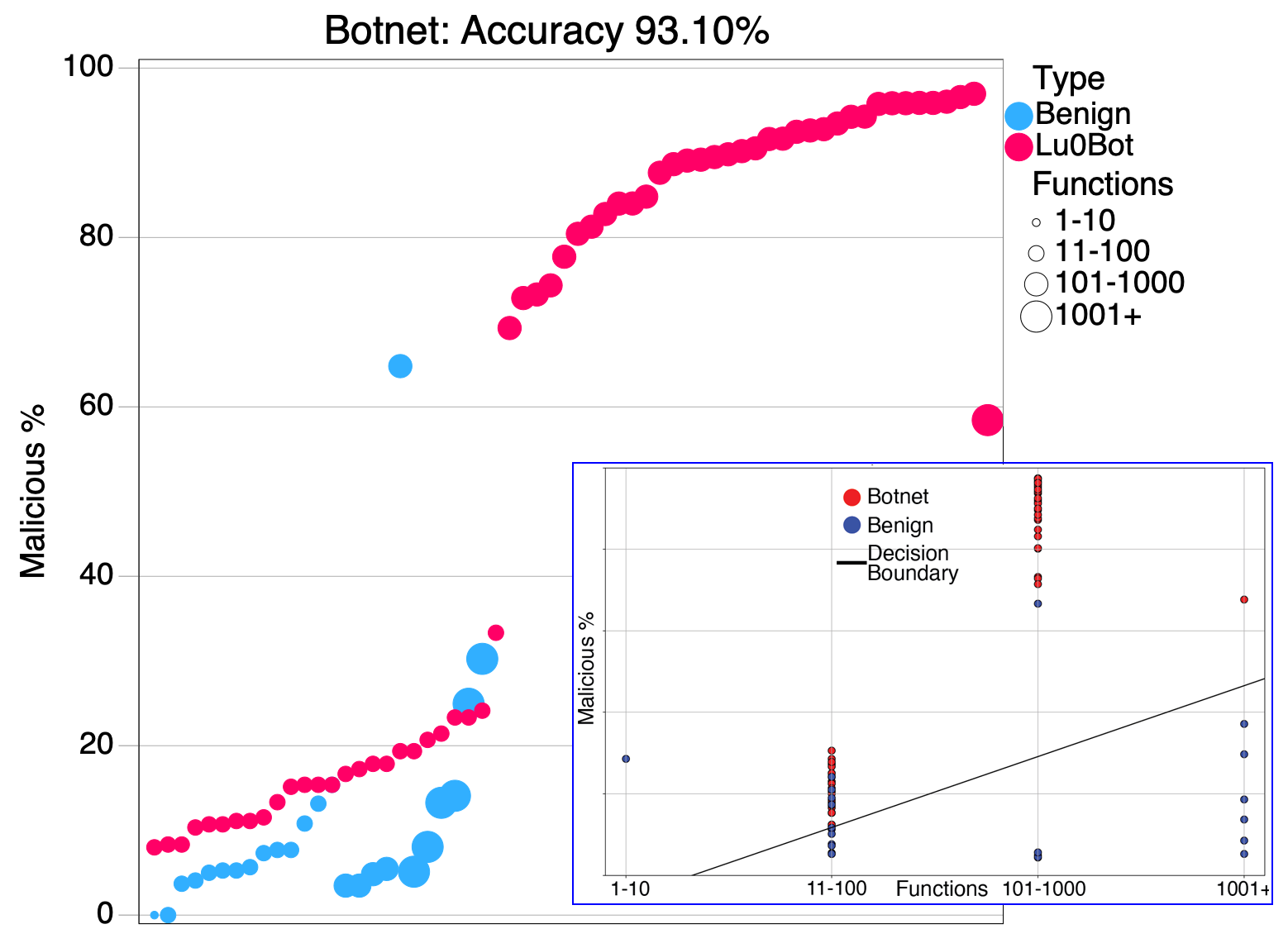}
        
    \end{minipage}

    \begin{minipage}{0.4\textwidth}
        \centering
        \includegraphics[width=\textwidth]{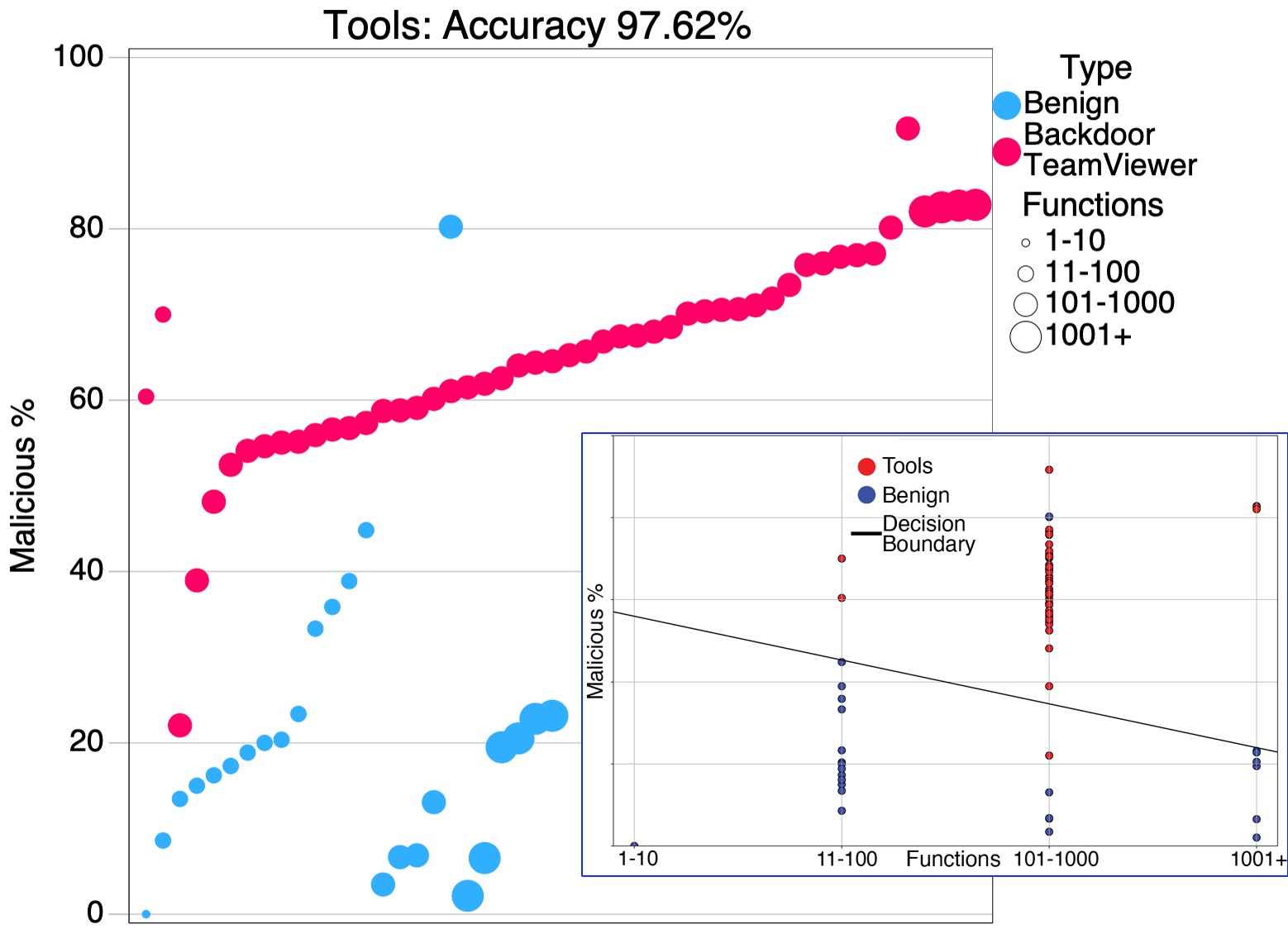}
        
    \end{minipage}
    \caption{Experiment A results}
    \label{fig:exparesults}
\end{figure}

\begin{table}[htbp]
\centering
\footnotesize
\captionsetup{justification=centering}
\caption{Experiment A final classification performance metrics}
\label{tab:expamodel performance}
\begin{tabular}{lcccccccc}
    \toprule
    \makecell{Type} & \makecell{TP} & \makecell{FN} & \makecell{FP} & \makecell{TN} & \makecell{Accuracy} & \makecell{Precision} & \makecell{Recall} & \makecell{F1 Score} \\
    \midrule
    \makecell{Ransomware} & \makecell{88} & \makecell{0} & \makecell{1} & \makecell{24} & \makecell{99.12} & \makecell{98.88} & \makecell{100} & \makecell{99.34} \\
    \makecell{Worm} & \makecell{28} & \makecell{1} & \makecell{1} & \makecell{24} & \makecell{96.30} & \makecell{96.55} & \makecell{96.55} & \makecell{96.55} \\
    \makecell{Trojan 35} & \makecell{84} & \makecell{2} & \makecell{9} & \makecell{16} & \makecell{90.09} & \makecell{90.32} & \makecell{97.67} & \makecell{93.85} \\
    \makecell{Spyware} & \makecell{132} & \makecell{1} & \makecell{7} & \makecell{18} & \makecell{94.94} & \makecell{94.96} & \makecell{99.25} & \makecell{97.07} \\
    \makecell{APT} & \makecell{14} & \makecell{0} & \makecell{1} & \makecell{24} & \makecell{97.44} & \makecell{93.33} & \makecell{100} & \makecell{96.55} \\
    \makecell{Botnet} & \makecell{62} & \makecell{0} & \makecell{6} & \makecell{19} & \makecell{93.10} & \makecell{91.18} & \makecell{100} & \makecell{95.53} \\
    \makecell{Tools} & \makecell{49} & \makecell{1} & \makecell{1} & \makecell{24} & \makecell{97.62} & \makecell{98.00} & \makecell{98.00} & \makecell{98.00} \\
    \bottomrule
\end{tabular}
\end{table}

The detailed breakdown of the number of functions, for the Peekaboo malware and benign samples, as detailed in Table \ref{tab:Training testing dataset details} is shown in Table \ref{tab:Experiment A training Corpus}. A DistilBERT model was fine tuned on each type of malware and in this experiment only layer 2 and 3 shown in Figure \ref{fig:Alpha Classifier} were used. As described in Section 4 duplicate functions were removed and any function present in the malware training samples that also appeared in the benign training samples was removed from the malicious dataset. Of these filtered functions a maximum of 0.2\% were longer than 256 tokens, that is for Trojan 150. The fine tuning and validation results for the models are shown in Table \ref{tab:finetuningA}. Trojans proved to be the most challenging to classify, exhibiting the lowest accuracy. As a result, multiple approaches were implemented during the training and testing phases to improve Trojan classification.

In Table \ref{tab:Experiment A training Corpus}, \texttt{Trojan 150} refers to a configuration where each Trojan family was limited to 150 training samples. This approach aimed to address the imbalance between Trojans and Benign samples, as there were significantly more Trojans overall. In the \texttt{Trojan 35} configuration, each Trojan family was further limited to 35 training samples. The \texttt{Trojan 35 JMPNZ} variant included additional pre-processing steps beyond the \texttt{memoryaddress} replacement described in Section 4. Specifically, any \texttt{jmp}, \texttt{jnz}, \texttt{jz} and the associated memory address were replaced with \texttt{jmpmem}, \texttt{jnzmem} \texttt{jzmem}. These replacements were introduced to address the imbalance in the number of unique words between the Trojan and Benign training sets. As shown in Table \ref{tab:finetuningA}, the \texttt{Trojan 35} configuration achieved the highest validation accuracy and the lowest loss among the tested configurations. Consequently, it was selected for the Final Classification SVM layer.

As described in Section 4, the DistilBERT Function Classification head predicted the number of malicious and benign functions within each sample. Layer 3 Final Classification SVM used a hyperplane serving as the decision boundary to distinguish between malicious and benign samples, as illustrated in Figure \ref{fig:Alpha Classifier}. The process involved the DistilBERT models classifying individual functions, while the SVM hyperplane classified the samples based on the percentage of malicious functions.

The test samples listed in Table \ref{tab:Training testing dataset details} were used to evaluate the DistilBERT models’ ability to identify novel malicious behaviors. To ensure rigorous evaluation on truly new samples, any function present in the test samples that also appeared in the training data was removed from the test samples. The function classification head predictions are the percentage of malicious and benign functions in each sample, as shown in Figure \ref{fig:exparesults}. The bubble size in the figure indicates the number of functions in each sample, ranging from 3 to over 1000, categorized into four distinct classes. Insets in Figure \ref{fig:exparesults} display the SVM decision boundary used to separate benign and malware samples. This SVM classifier distinguishes between two classes: Malware and Benign samples. The classifier uses a linear kernel and equal class weights to train on a the test samples with two features: \texttt{Functions} and \texttt{Malicious \%}. After training, it calculates the decision boundary, which is represented as a hyperplane that separates the two classes. Malware samples are plotted as red dots and Benign as blue dots, with the decision boundary visualized as a black line. The SVM decision function computes the distance of each sample from the hyperplane, where positive distances indicate malicious classification and negative distances indicate benign classification. All models achieved impressive accuracy levels exceeding 90\%, demonstrating their effectiveness. While each model produced at least one False Positive (FP), and the Worm, Trojan, Spyware, and Tool models each generated at least one False Negative (FN), these outcomes are still highly promising. Notably, the predictions were made on entirely novel functions, that is unique combinations of ASM instructions that the DistilBERT model had never encountered before, highlighting the model's strong generalization capabilities in classifying unseen behaviors.

\subsection{Experiment B}
\begin{figure}[htbp]
    \centering
    \begin{minipage}{0.3\textwidth}
        \centering
        \includegraphics[width=\textwidth]{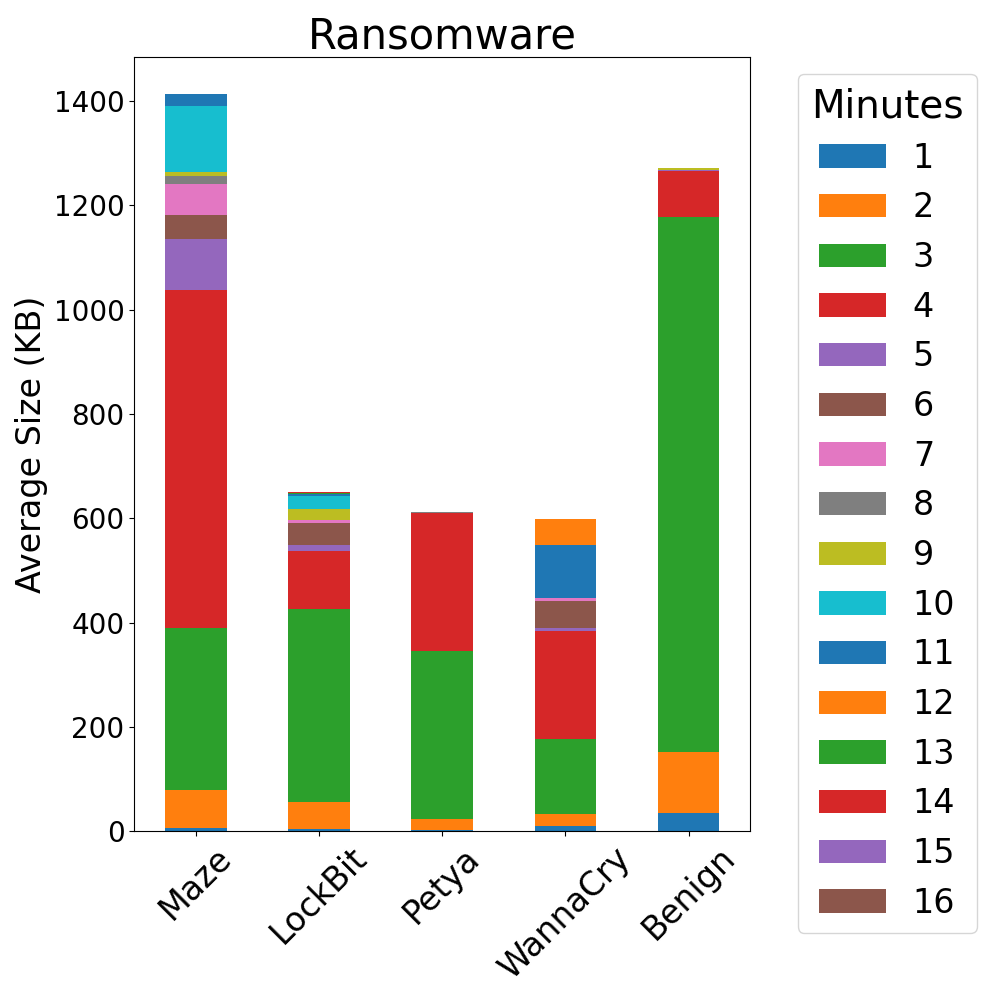}
    \end{minipage}
    \begin{minipage}{0.3\textwidth}
        \centering
        \includegraphics[width=\textwidth]{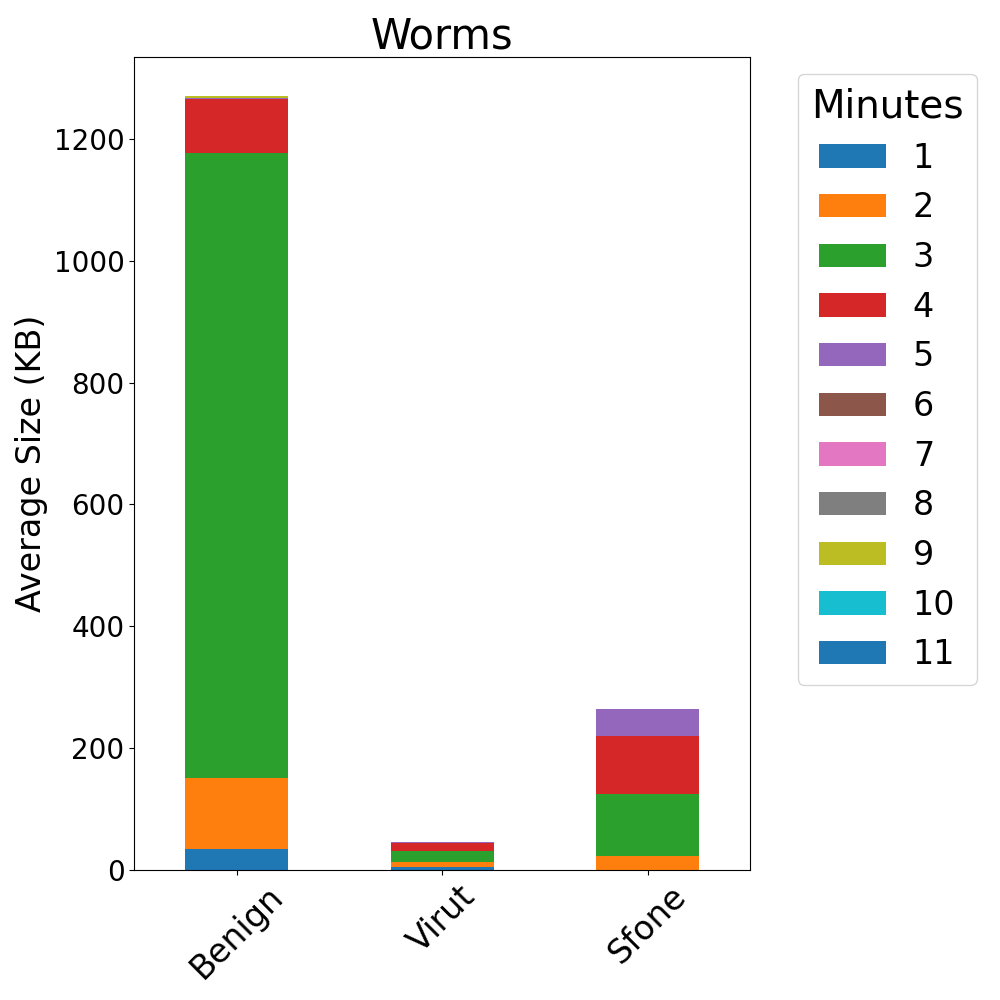}
    \end{minipage}


    \begin{minipage}{0.3\textwidth}
        \centering
        \includegraphics[width=\textwidth]{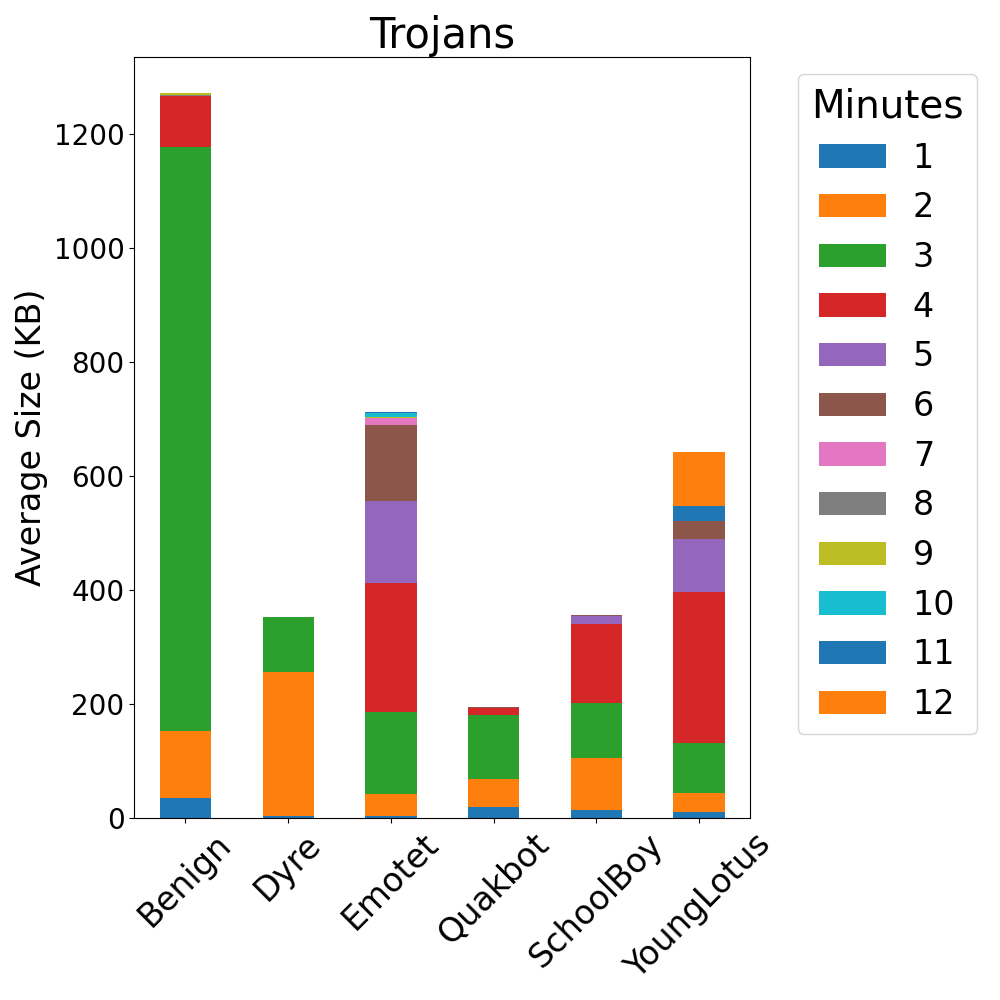}
    \end{minipage}
    \begin{minipage}{0.3\textwidth}
        \centering
        \includegraphics[width=\textwidth]{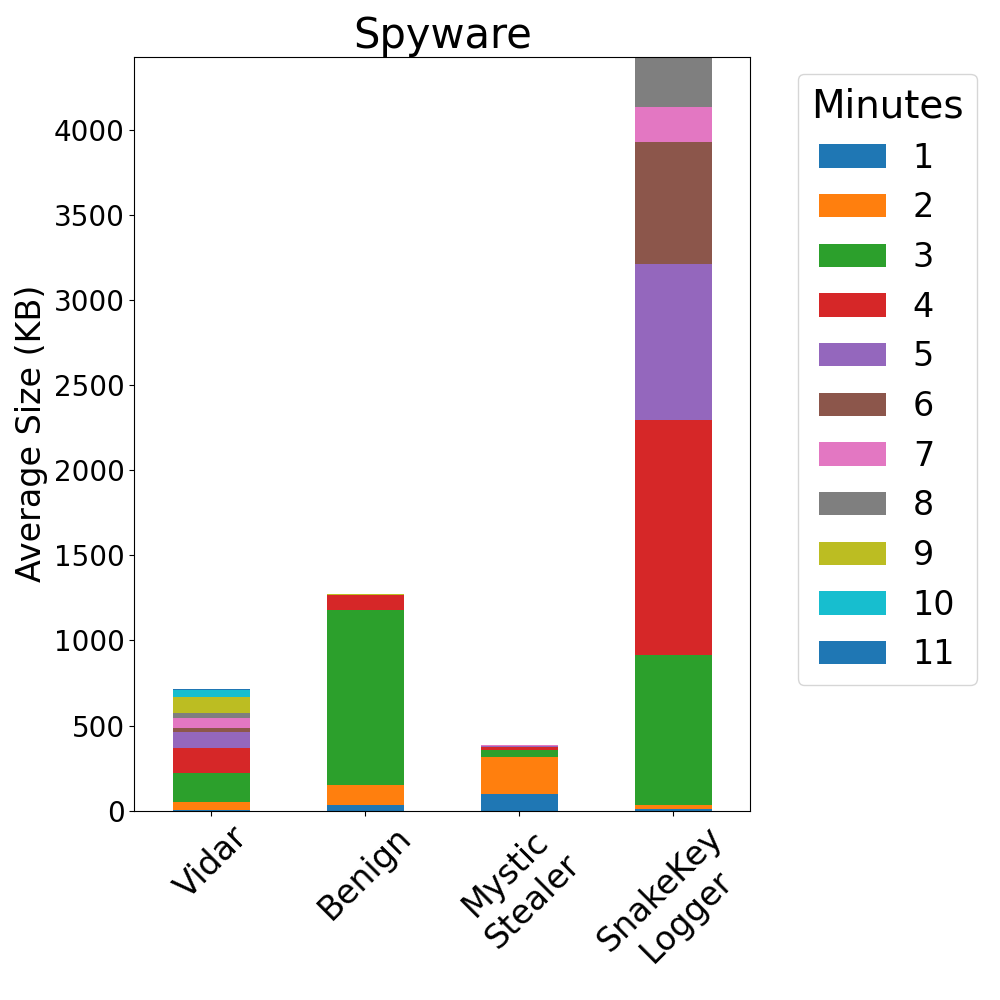}
    
    \end{minipage}


    \begin{minipage}{0.3\textwidth}
        \centering
        \includegraphics[width=\textwidth]{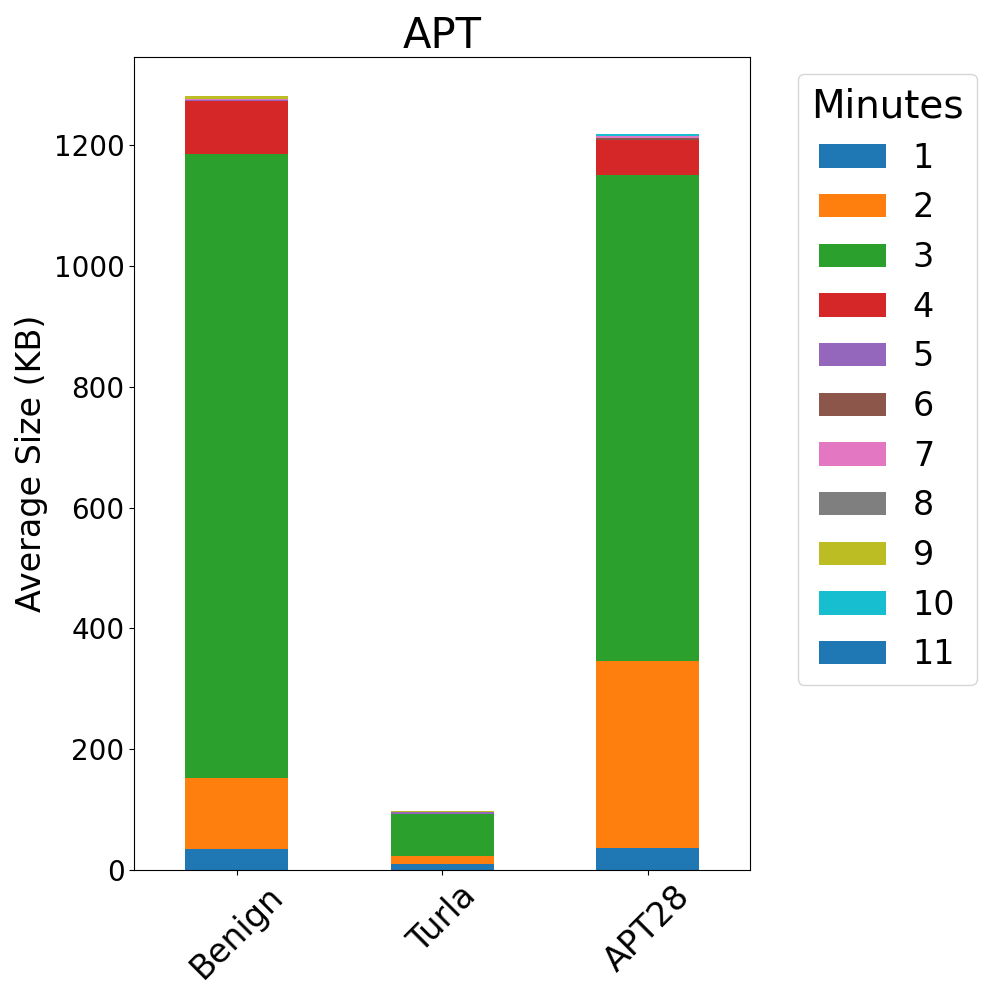}
    
    \end{minipage}
    \begin{minipage}{0.3\textwidth}
        \centering
        \includegraphics[width=\textwidth]{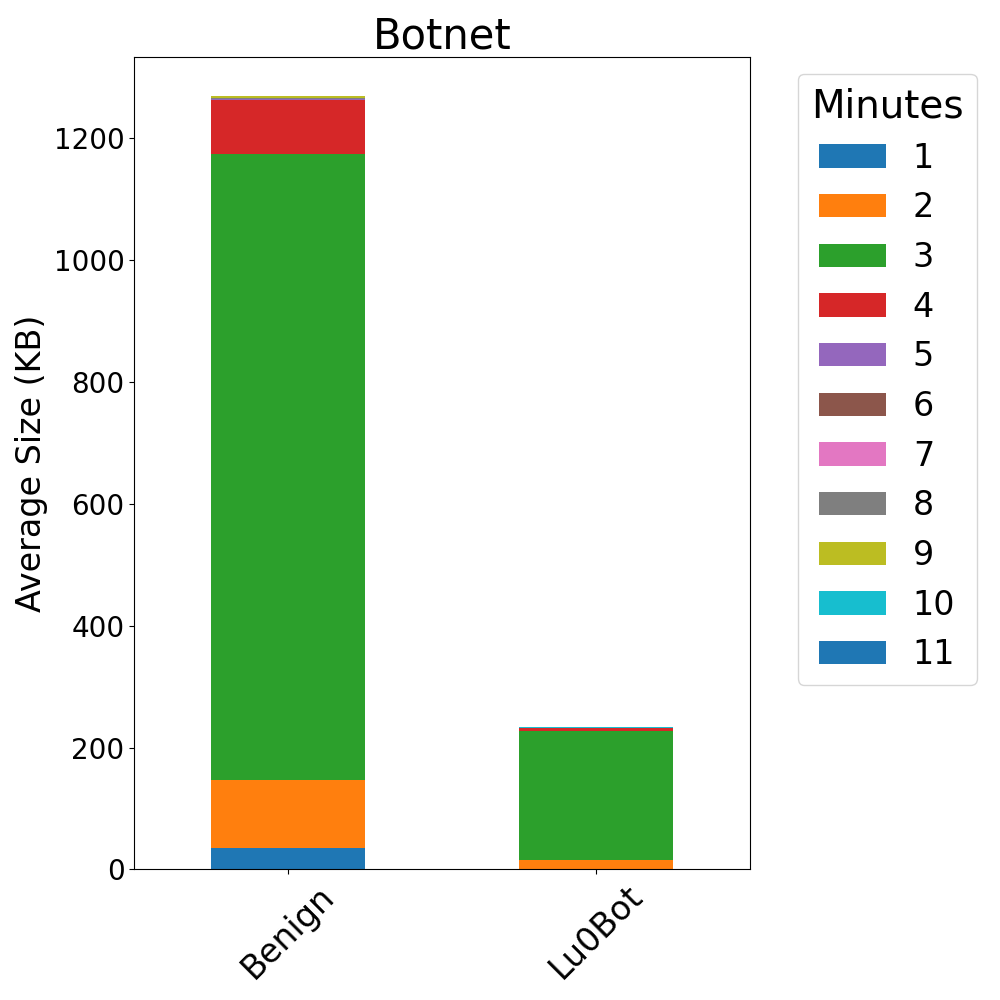}
        
    \end{minipage}

    \begin{minipage}{0.3\textwidth}
        \centering
        \includegraphics[width=\textwidth]{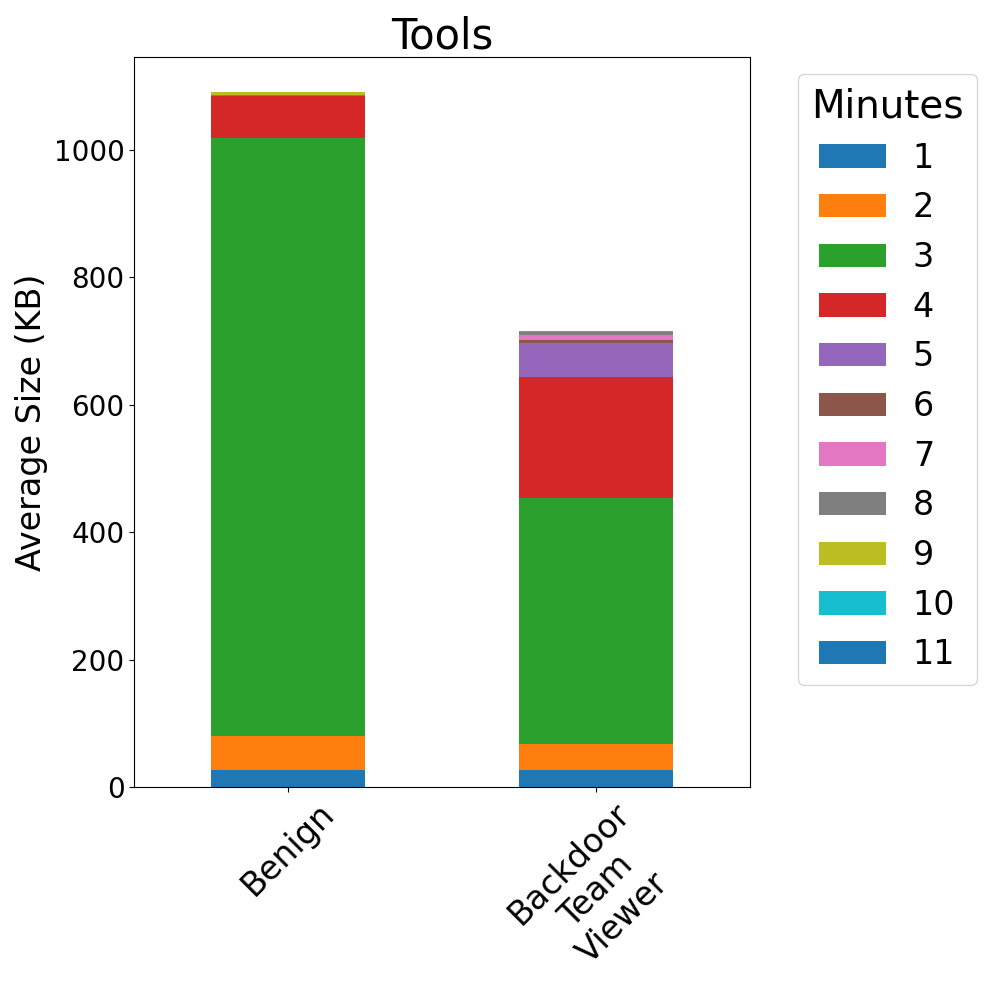}
        
    \end{minipage}
    \caption{ASM instruction density per minute}
    \label{fig:asm instruction density}
\end{figure}
\begin{figure}[htbp]
    \centering
    \begin{minipage}{0.3\textwidth}
        \centering
        \includegraphics[width=\textwidth]{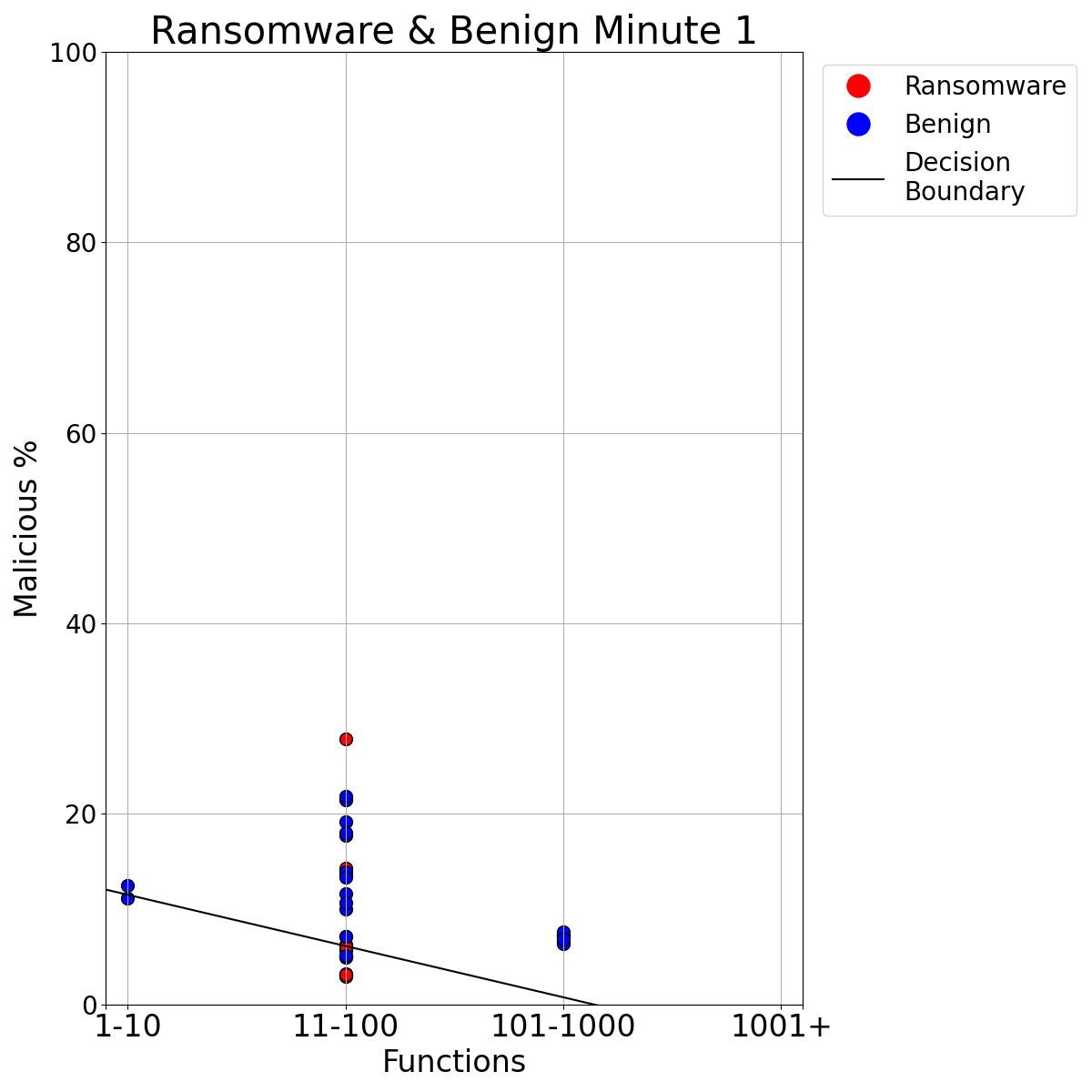}
    \end{minipage}
    \begin{minipage}{0.3\textwidth}
        \centering
        \includegraphics[width=\textwidth]{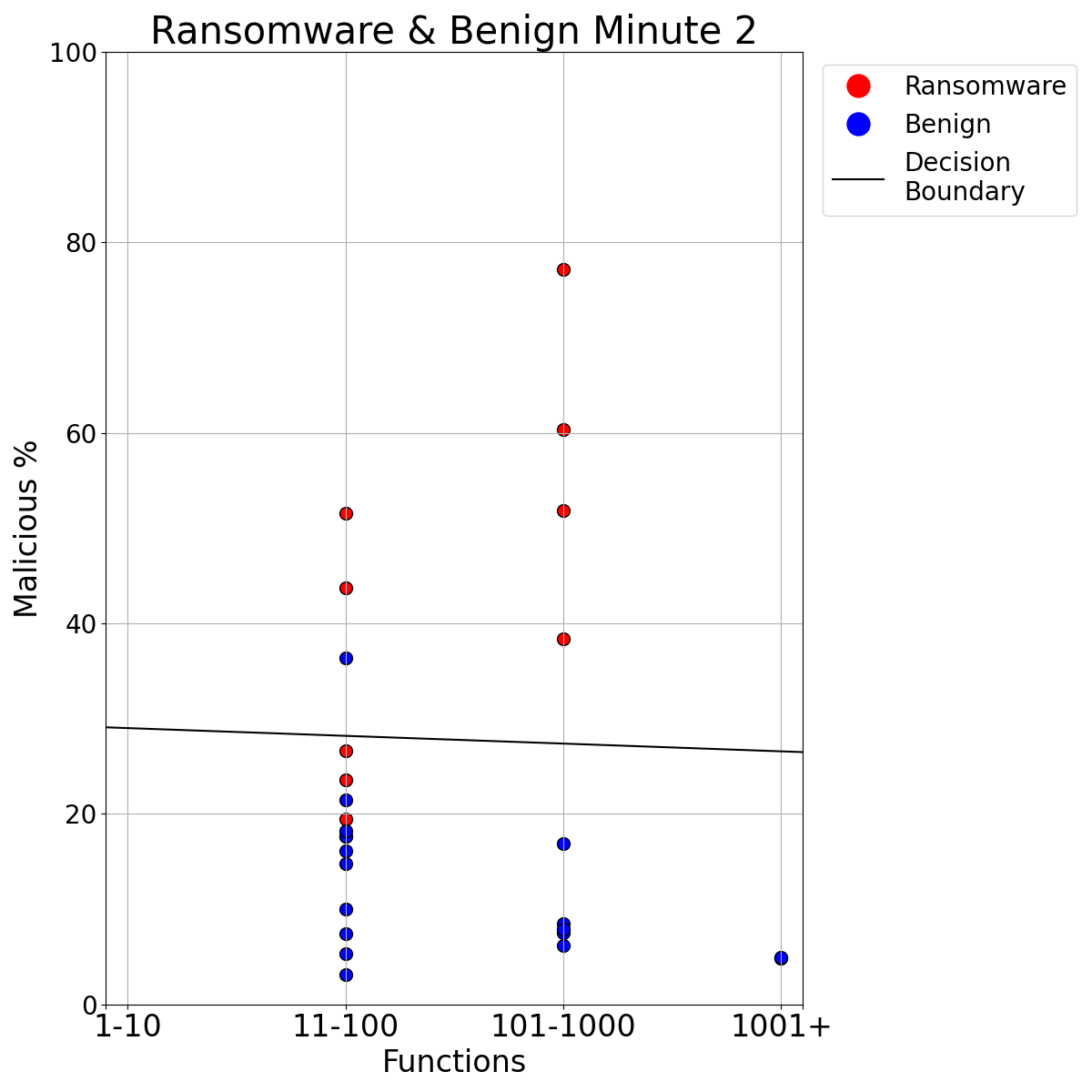}
    \end{minipage}

    \begin{minipage}{0.3\textwidth}
        \centering
        \includegraphics[width=\textwidth]{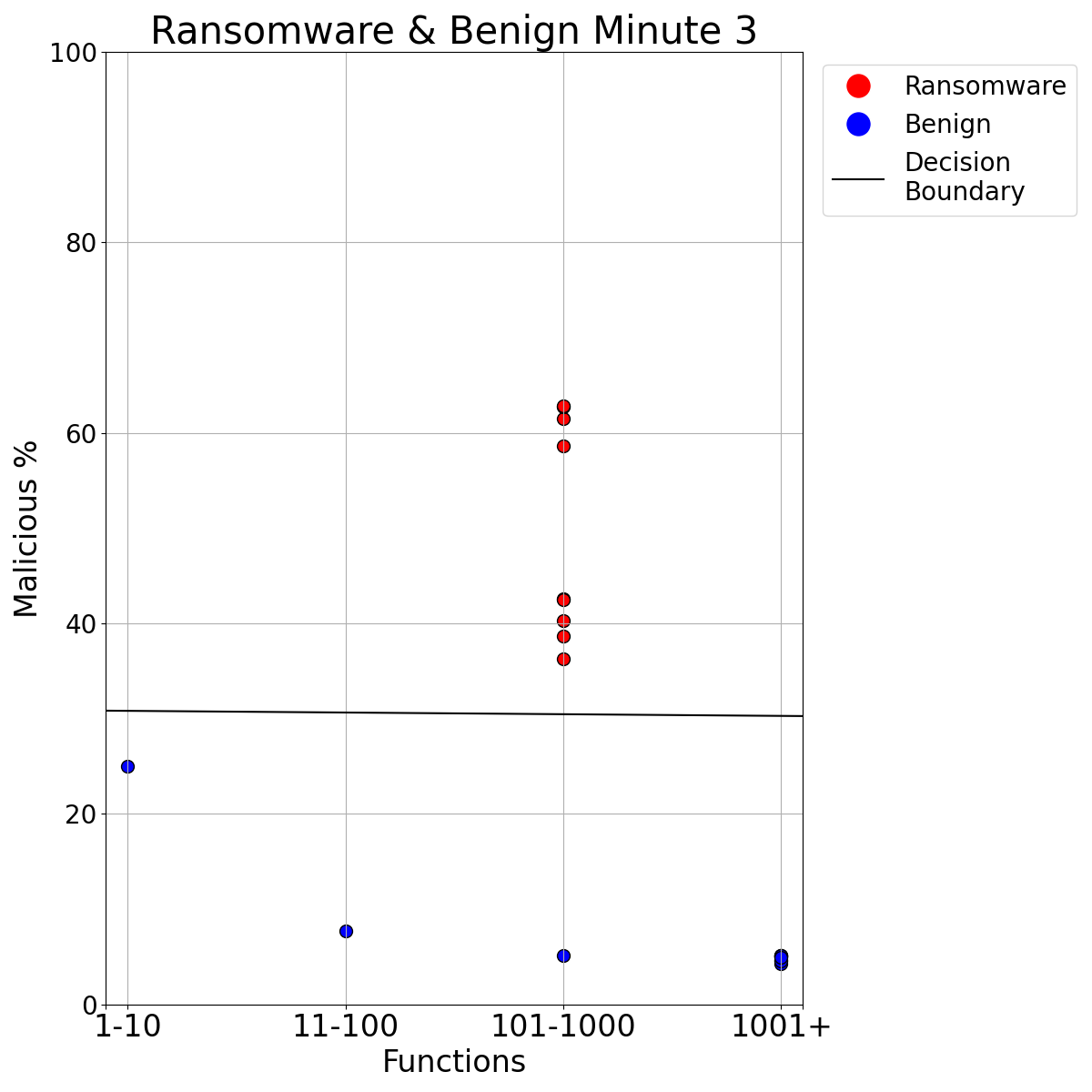}
    \end{minipage}
    \begin{minipage}{0.3\textwidth}
        \centering
        \includegraphics[width=\textwidth]{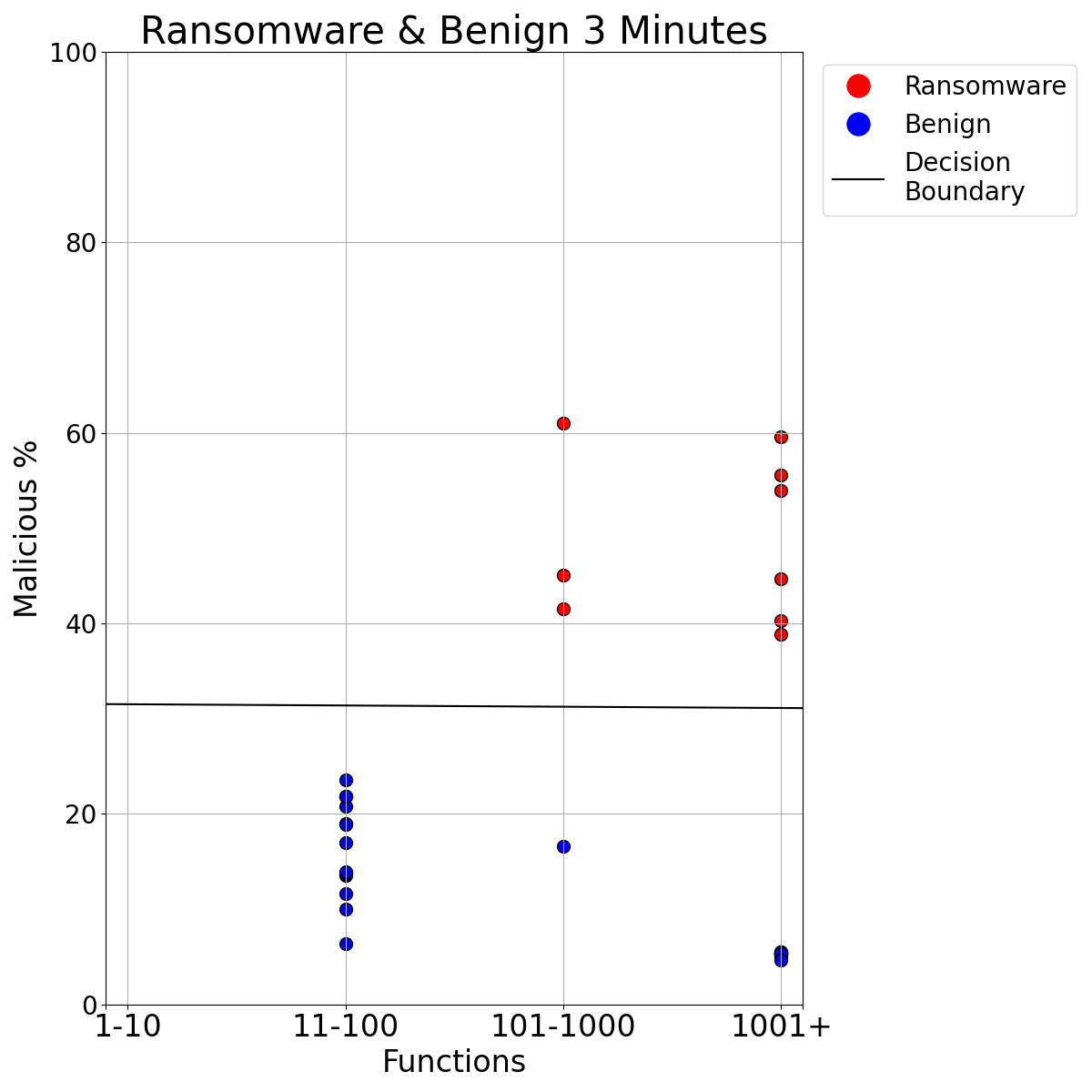}
    
    \end{minipage}
    \caption{Maze Final Classification SVM per minute results}
    \label{fig:Maze ransomware per minute tests}
\end{figure}

\begin{table}[htbp]
\centering
\footnotesize
\captionsetup{justification=centering}
\caption{Maze Final Classification SVM per minute performance metrics}
\label{tab:maze 1 minute output}
\begin{tabular}{lccccccc}
    \toprule
    \makecell{Time} & \makecell{TP} & \makecell{FN} & \makecell{FP} & \makecell{TN} & \makecell{Accuracy} & \makecell{Precision} & \makecell{Recall} \\
    \midrule
    \makecell{Minute 1} & \makecell{6} & \makecell{3} & \makecell{4} & \makecell{21} & \makecell{79.41} & \makecell{60.00} & \makecell{66.67} \\
    \makecell{Minute 2} & \makecell{6} & \makecell{3} & \makecell{1} & \makecell{17} & \makecell{85.19} & \makecell{86.71} & \makecell{66.67} \\
    \makecell{Minute 3} & \makecell{9} & \makecell{0} & \makecell{0} & \makecell{18} & \makecell{100} & \makecell{100} & \makecell{100} \\
    \makecell{Full 3 Minutes} & \makecell{9} & \makecell{0} & \makecell{0} & \makecell{25} & \makecell{100} & \makecell{100} & \makecell{100} \\
    \bottomrule
\end{tabular}
\end{table}

\begin{figure}[htbp]
    \centering
    \begin{minipage}{0.3\textwidth}
        \centering
        \includegraphics[width=\textwidth]{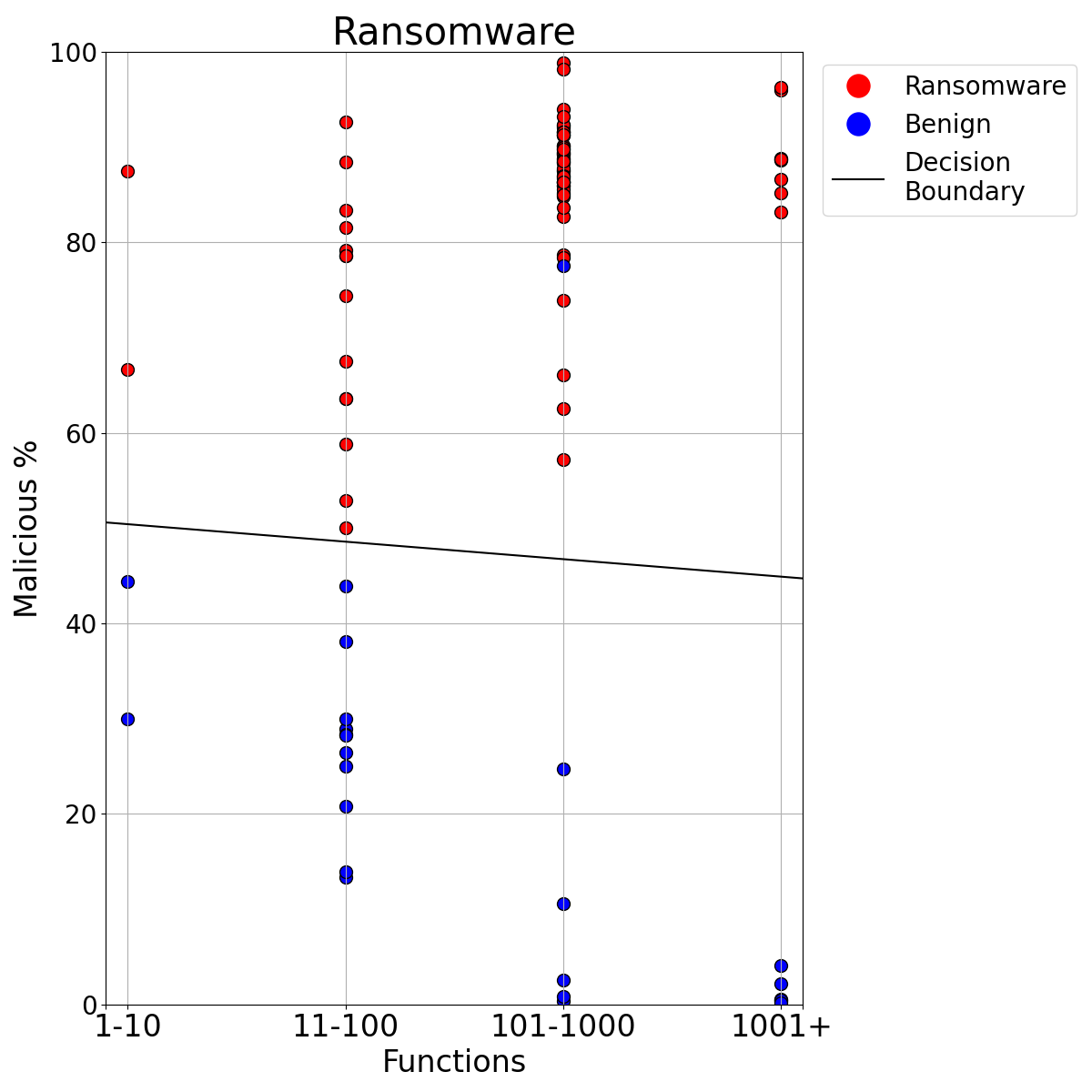}
    \end{minipage}
    \begin{minipage}{0.3\textwidth}
        \centering
        \includegraphics[width=\textwidth]{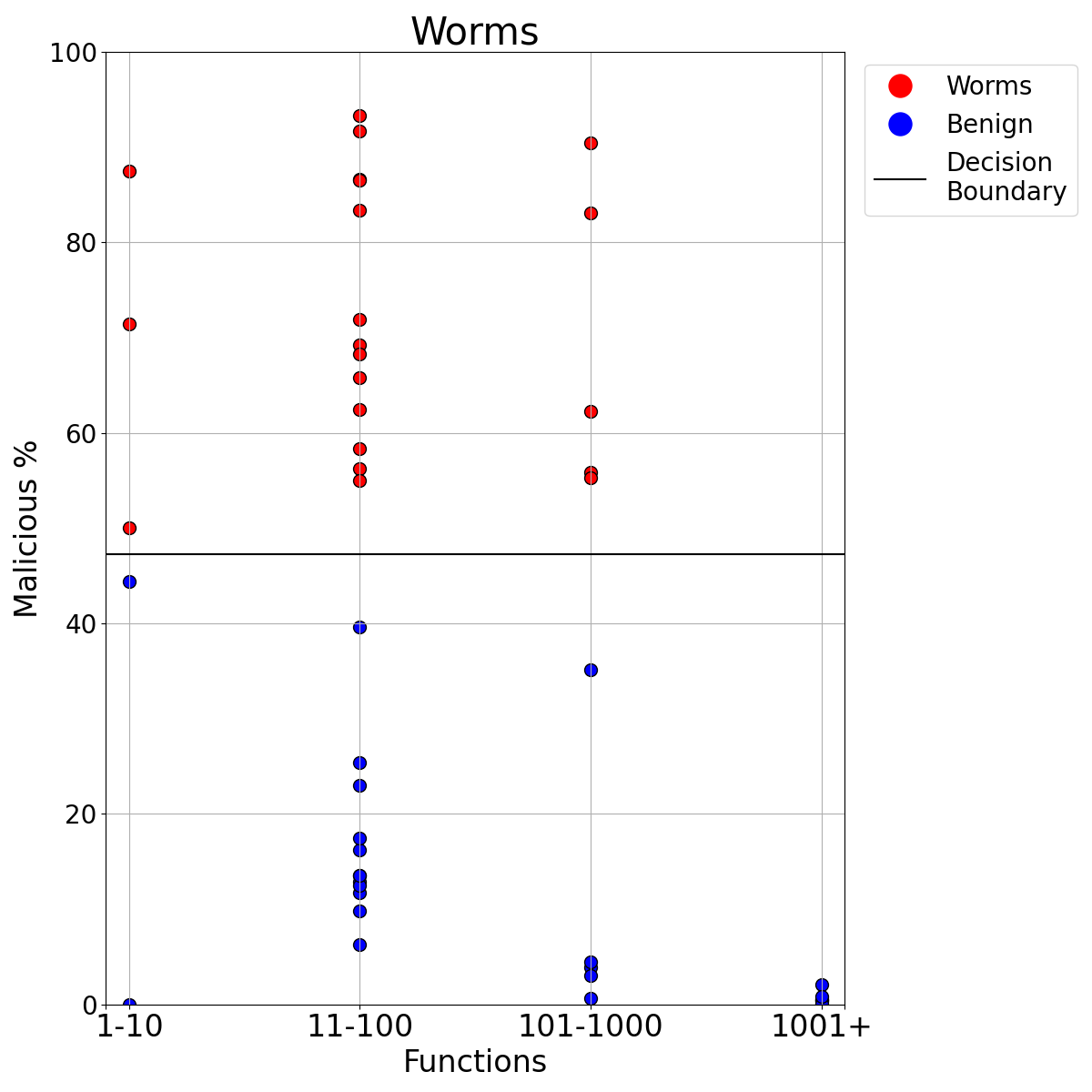}
    \end{minipage}


    \begin{minipage}{0.3\textwidth}
        \centering
        \includegraphics[width=\textwidth]{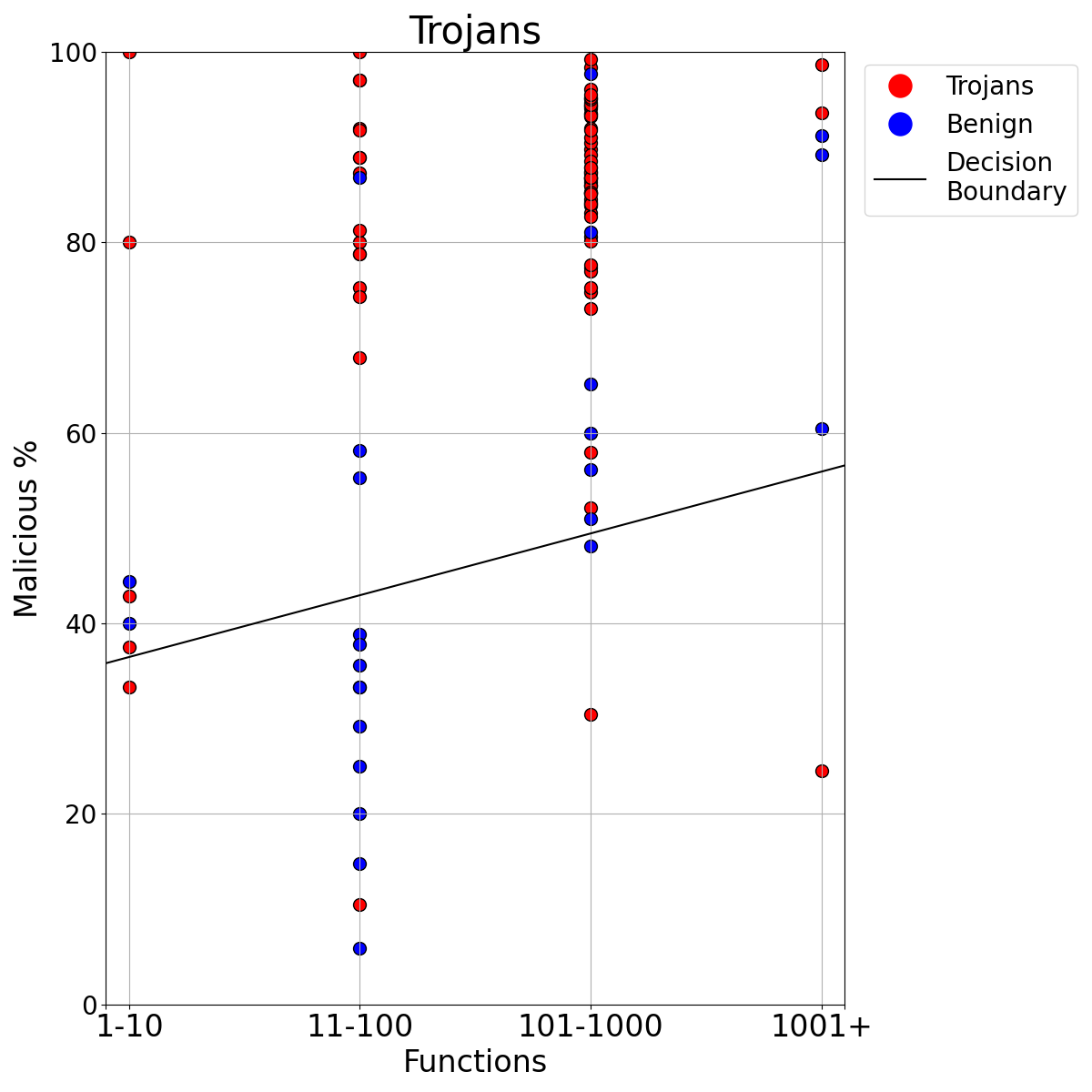}
    \end{minipage}
    \begin{minipage}{0.3\textwidth}
        \centering
        \includegraphics[width=\textwidth]{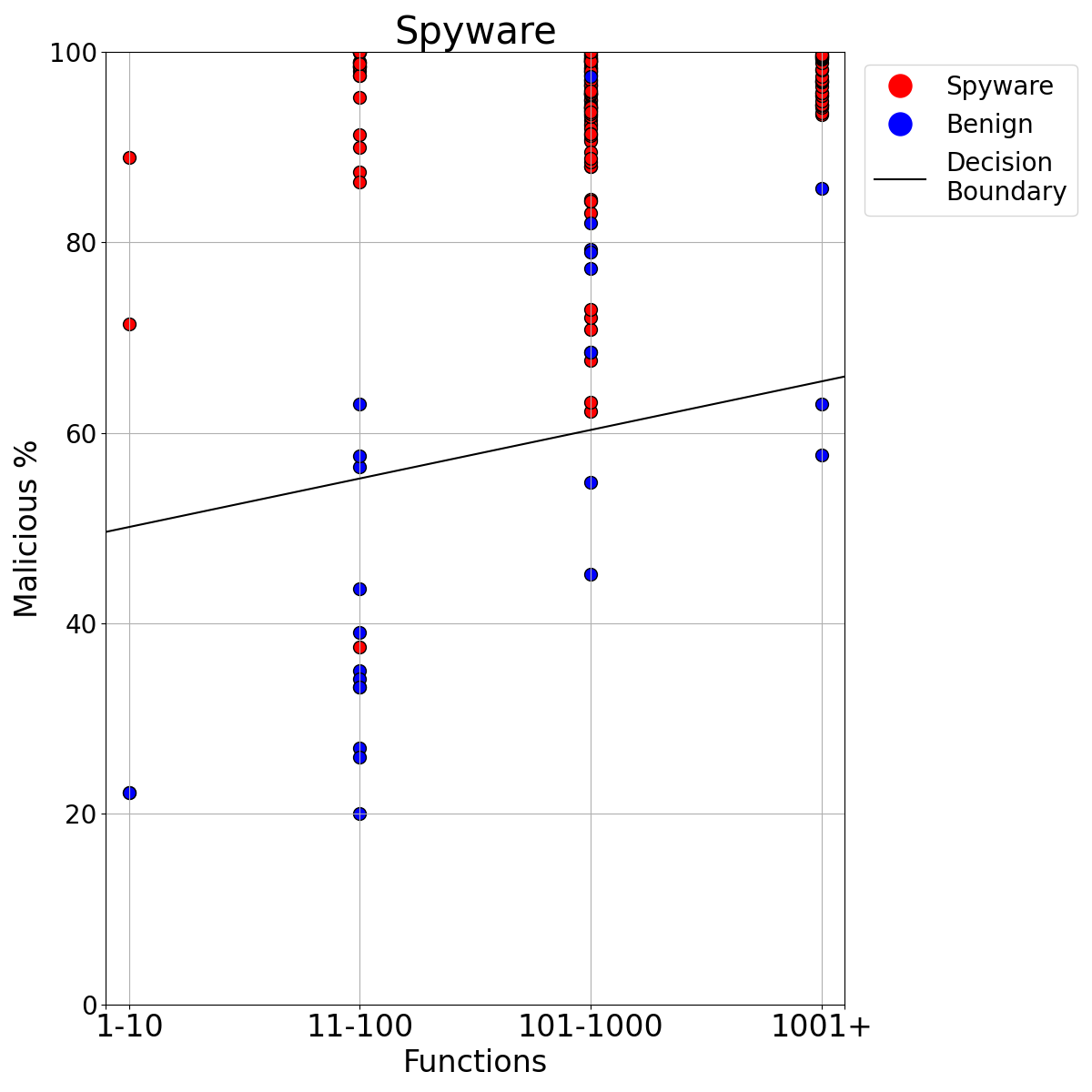}
    
    \end{minipage}


    \begin{minipage}{0.3\textwidth}
        \centering
        \includegraphics[width=\textwidth]{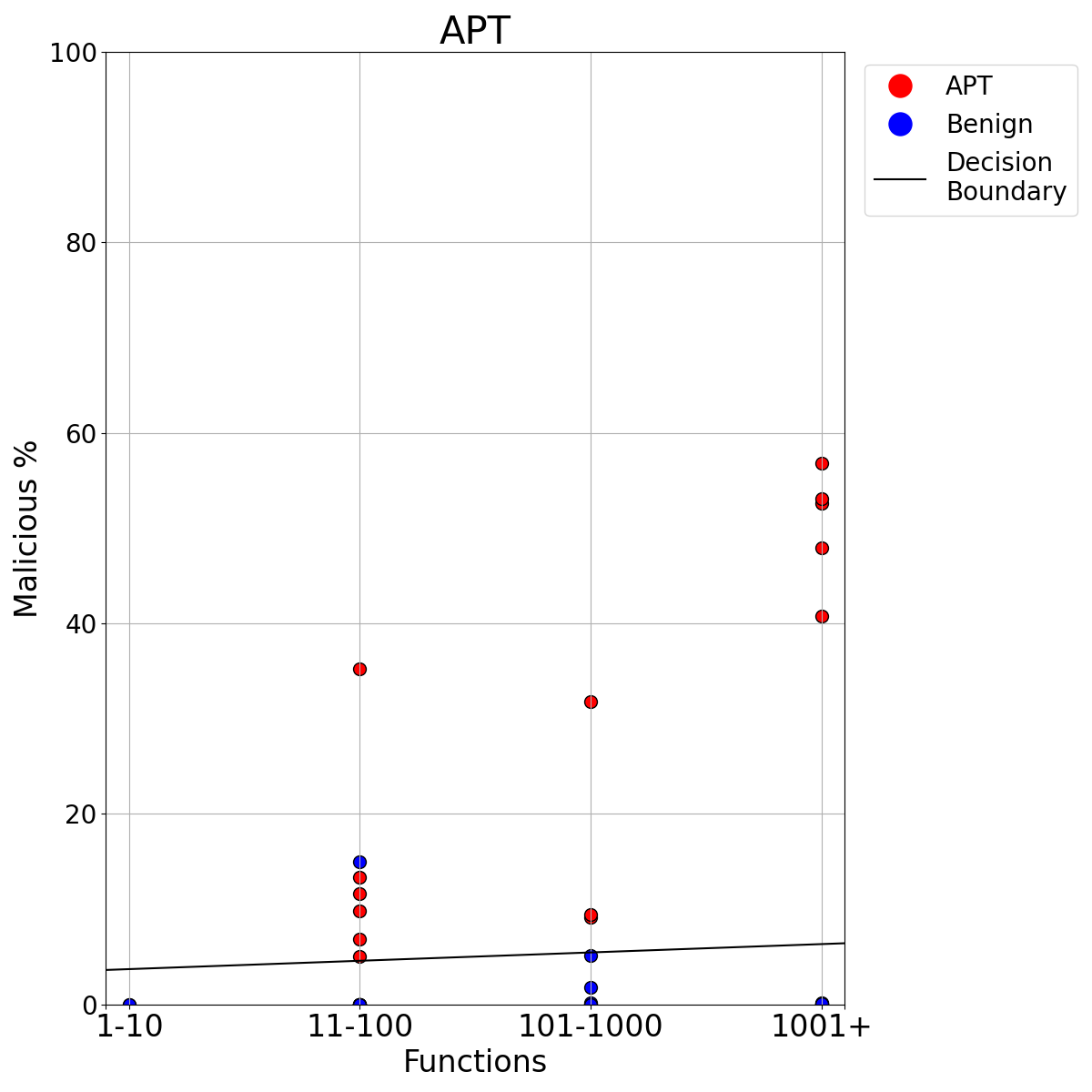}
    
    \end{minipage}
    \begin{minipage}{0.3\textwidth}
        \centering
        \includegraphics[width=\textwidth]{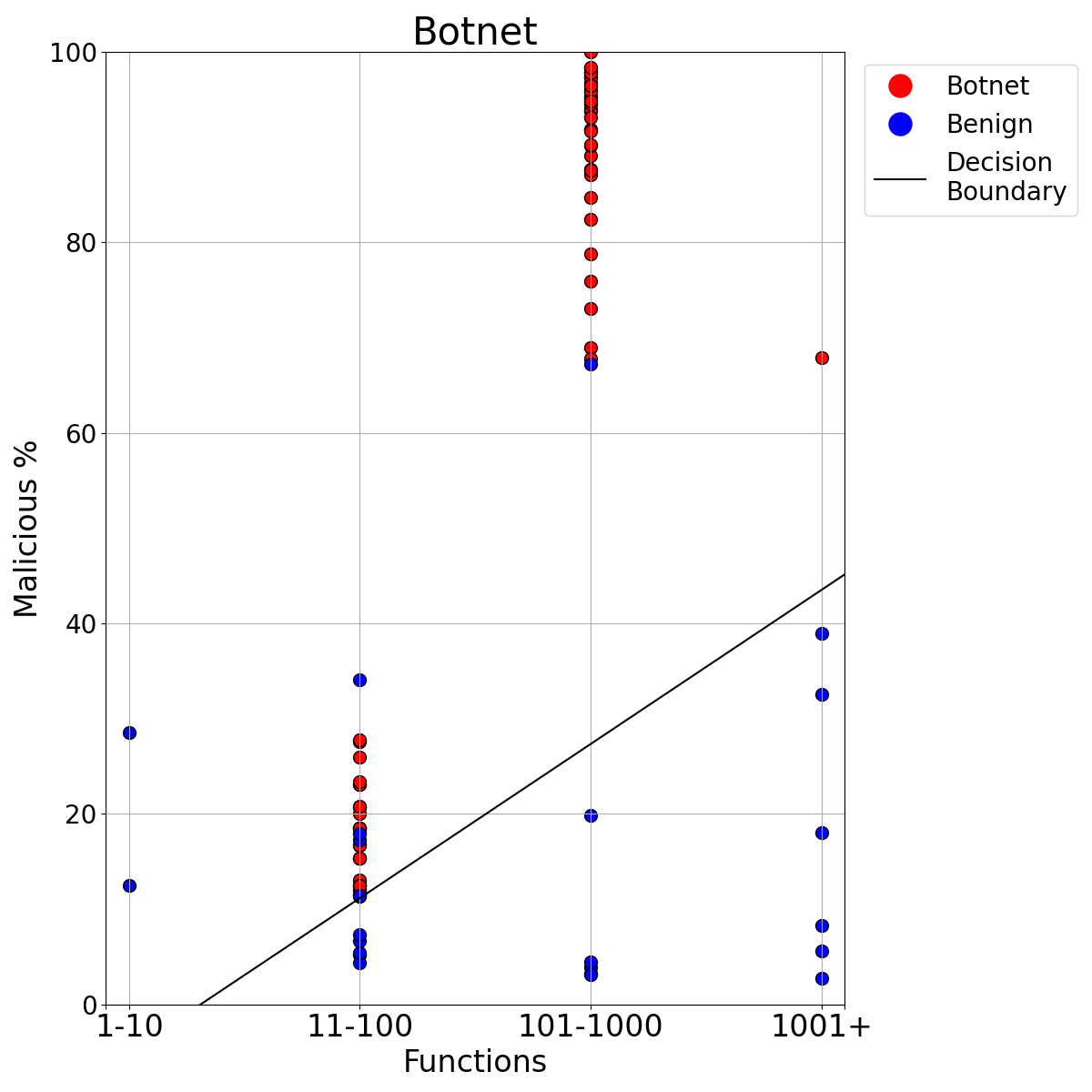}
        
    \end{minipage}

    \begin{minipage}{0.3\textwidth}
        \centering
        \includegraphics[width=\textwidth]{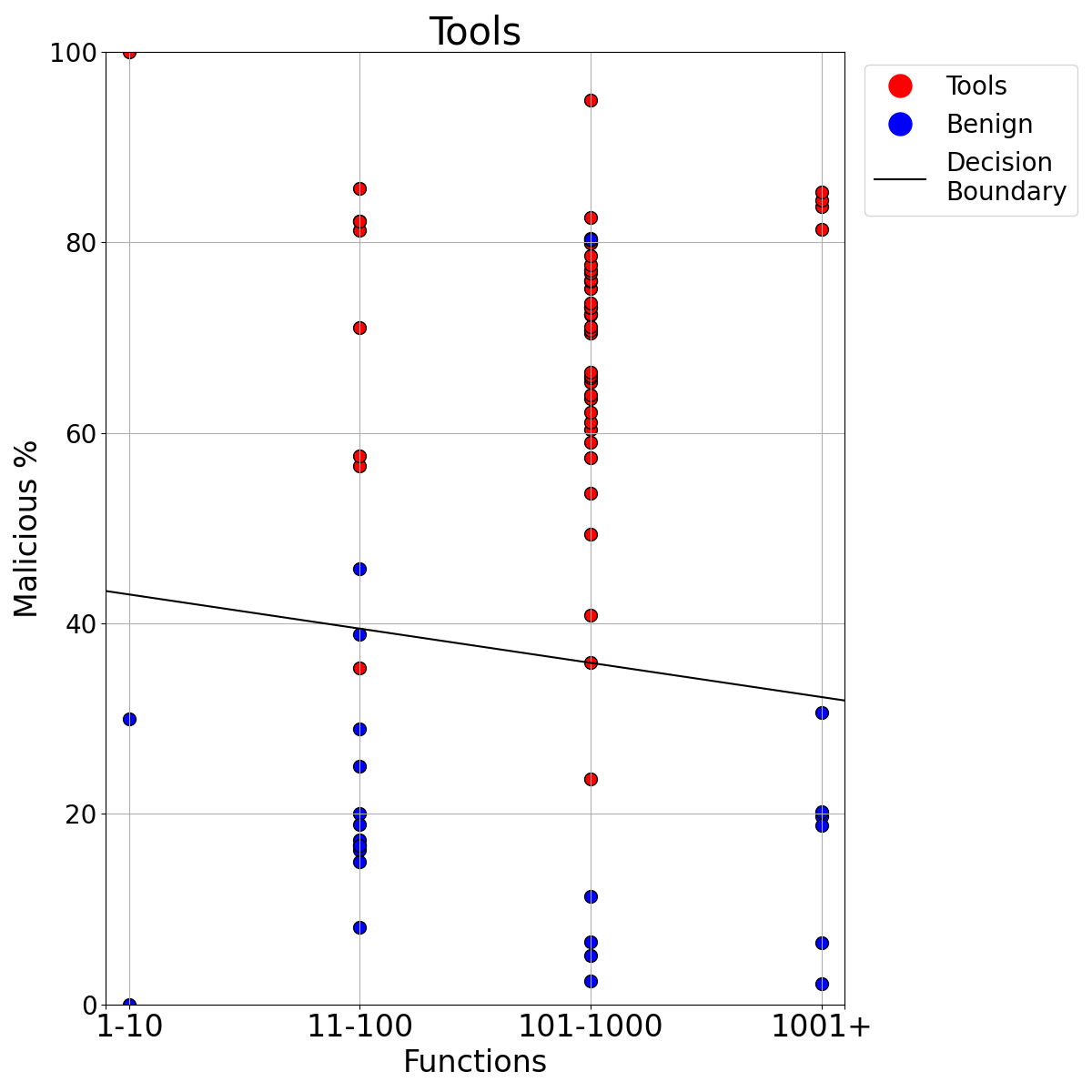}
        
    \end{minipage}
    \caption{Minute 3 data tested with trained models}
    \label{fig:experiment b minute 3 svm output}
\end{figure}

\begin{table}[hbp]
\centering
\footnotesize
\captionsetup{justification=centering}
\caption{Final Classification SVM performance metrics for minute 3}
\label{tab:exb final svm 3min}
\begin{tabular}{lcccccccc}
    \toprule
    \makecell{Type} & \makecell{TP} & \makecell{FN} & \makecell{FP} & \makecell{TN} & \makecell{Accuracy} & \makecell{Precision} & \makecell{Recall} & \makecell{F1 Score} \\
    \midrule
    \makecell{Ransomware} & \makecell{75} & \makecell{0} & \makecell{1} & \makecell{24} & \makecell{99.00} & \makecell{98.68} & \makecell{100} & \makecell{99.33} \\
    \makecell{Worm} & \makecell{21} & \makecell{0} & \makecell{0} & \makecell{25} & \makecell{100} & \makecell{100} & \makecell{100} & \makecell{100} \\
    \makecell{Trojan 35} & \makecell{80} & \makecell{4} & \makecell{14} & \makecell{11} & \makecell{83.49} & \makecell{85.11} & \makecell{95.24} & \makecell{89.97} \\
    \makecell{Spyware} & \makecell{132} & \makecell{1} & \makecell{10} & \makecell{15} & \makecell{93.04} & \makecell{92.96} & \makecell{99.25} & \makecell{95.92} \\
    \makecell{APT} & \makecell{14} & \makecell{0} & \makecell{1} & \makecell{24} & \makecell{97.44} & \makecell{93.33} & \makecell{100} & \makecell{96.55} \\
    \makecell{Botnet} & \makecell{61} & \makecell{0} & \makecell{8} & \makecell{17} & \makecell{90.07} & \makecell{88.41} & \makecell{100} & \makecell{93.74} \\
    \makecell{Tools} & \makecell{48} & \makecell{2} & \makecell{2} & \makecell{23} & \makecell{94.67} & \makecell{96.00} & \makecell{96.00} & \makecell{96.00} \\
    \bottomrule
\end{tabular}
\end{table}

In this experiment ASM instruction density and distribution were analyzed on a per minute basis, as shown in Figure \ref{fig:asm instruction density}. The methodology involved segmenting each test sample into discrete one minute intervals of ASM data, where each interval represented the unique functionalities and instructions extracted from that specific time frame. This segmentation approach allowed for a granular examination of whether a shorter time slice, as opposed to the complete sample, could still provide sufficient information for accurate classification. With minute 0 marking the start of execution, minutes 1, 2, and 3 exhibit the highest instruction density across most malware types and families, as shown in Figure \ref{fig:asm instruction density}. However, several families of Ransomware, Trojans, and the SnakeKeyLogger Spyware show elevated instruction densities at minute 4. Since a primary objective of this research was to evaluate whether DBI and Transformers could be applied effectively in a real-world application, specifically assessing their speed and efficiency, minutes 1 to 3 were selected for further analysis.

Maze ransomware, which exhibited the highest relative density of ASM instructions at minute 4, was selected to evaluate whether a 3-minute window from the start of execution, and specifically which minute within that window, would support accurate classification. As always, the data from these minute slices were filtered against the training dataset to retain only novel functions. The DistilBERT ransomware model, fine-tuned in Experiment A, that only used layer 2 and 3, as shown in Figure \ref{fig:Alpha Classifier}, was then used to test each minute slice. The results are shown in Figure \ref{fig:Maze ransomware per minute tests} and Table \ref{tab:maze 1 minute output}. The data from minutes 1 and 2 yielded poor predictions, barely exceeding random chance. In contrast, minute 3 alone and the full 3-minute window both achieved 100\% accuracy.

Filtered minute 3 slices for all test samples were then used as inputs for the corresponding DistilBERT model, which had been trained in Experiment A. The purpose was to determine whether the model could still perform accurate classification across all malware types when presented with only this 1 minute slice of data. This approach tested the robustness of the various model’s learning and their ability to generalize from limited information.

By examining the classification results for these time slices, this experiment evaluated the feasibility of using smaller data segments for malware and benign sample identification. The findings provided insights into whether minute 3, due to its higher instruction density, could serve as a reliable predictor for classification, thus potentially reducing the data processing overhead required for accurate zero day malware detection.

The number of test samples in this experiment differs from Experiment A, which used entire test samples. The one minute interval was filtered against the training data, to remove previously seen functions. As a result, some malware and benign test samples, were left with fewer than 3 functions and were excluded from the test.

Figure \ref{fig:experiment b minute 3 svm output} shows the final classification SVM results, indicating that the models achieved accurate predictions with just one minute of data, specifically from minute 2–3. It is also important to note that for the benign test samples, only data from minute 3 was utilized and as always it was filtered against the training dataset. Table \ref{tab:exb final svm 3min} shows the results for each type of malware. There was an overall decrease in accuracy compared to Experiment A, with Trojans dropping to 83.49\% with 4 FN, however the accuracy for the Worms increased to 100\%.

\subsection{Experiment C}
\begin{figure}[htbp]
    \centering
    \begin{minipage}{0.3\textwidth}
        \centering
        \includegraphics[width=\textwidth]{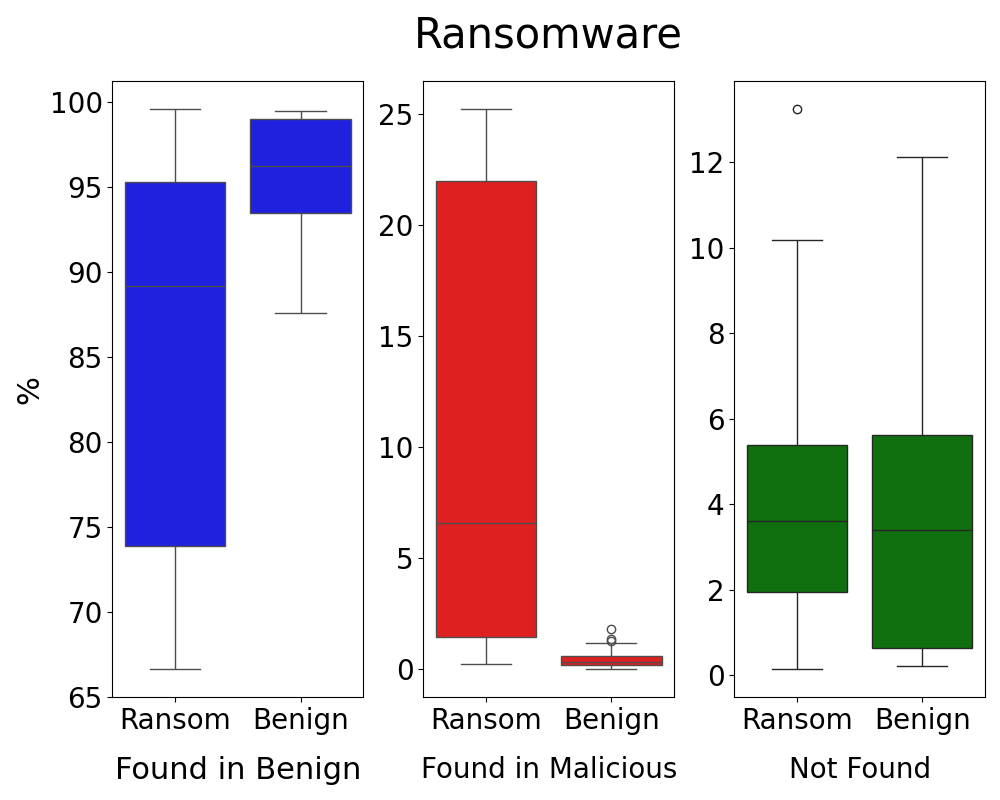}
    \end{minipage}
    \begin{minipage}{0.3\textwidth}
        \centering
        \includegraphics[width=\textwidth]{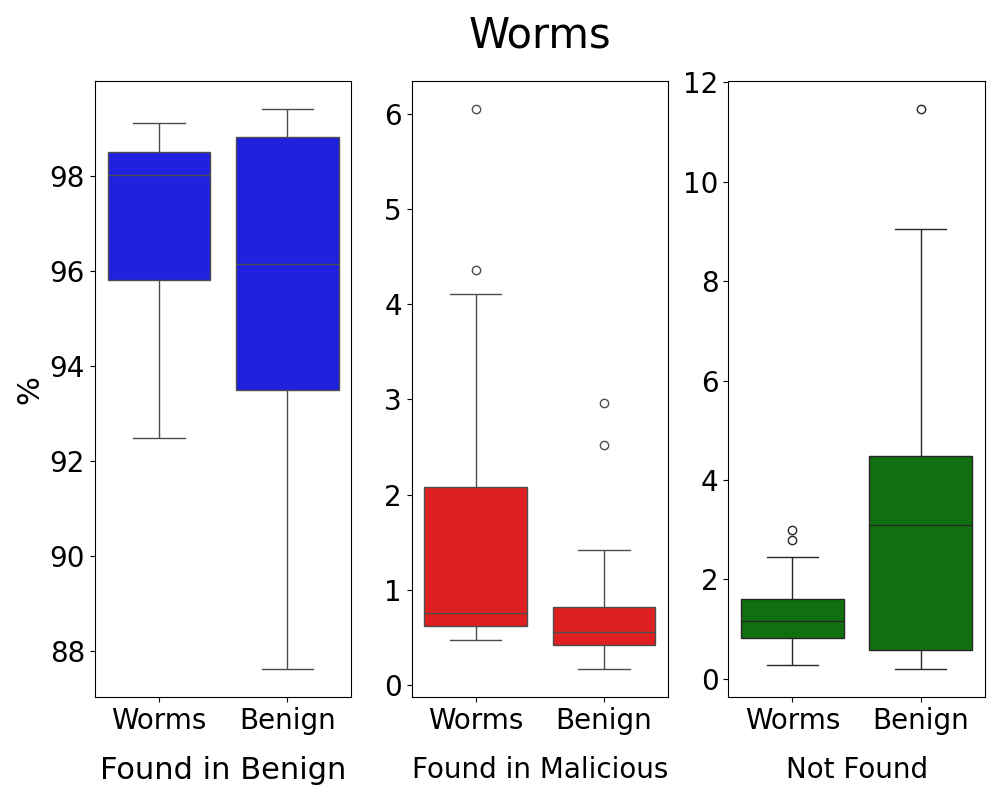}
    \end{minipage}
    
    \begin{minipage}{0.3\textwidth}
        \centering
        \includegraphics[width=\textwidth]{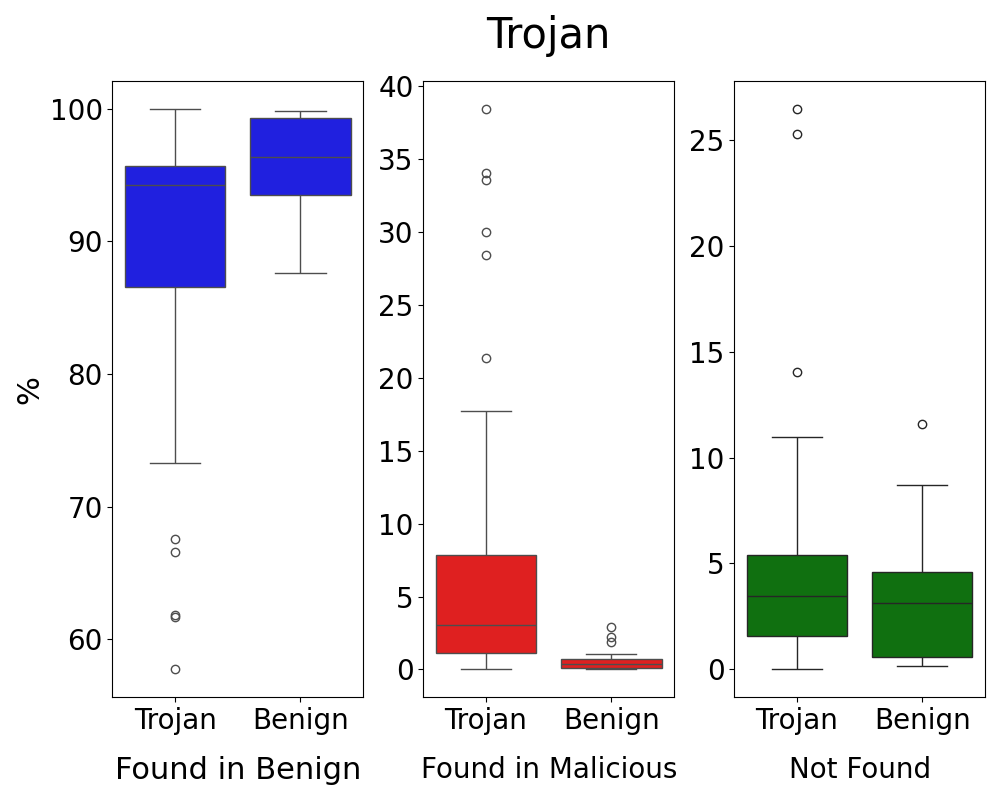}
    \end{minipage}
    \begin{minipage}{0.3\textwidth}
        \centering
        \includegraphics[width=\textwidth]{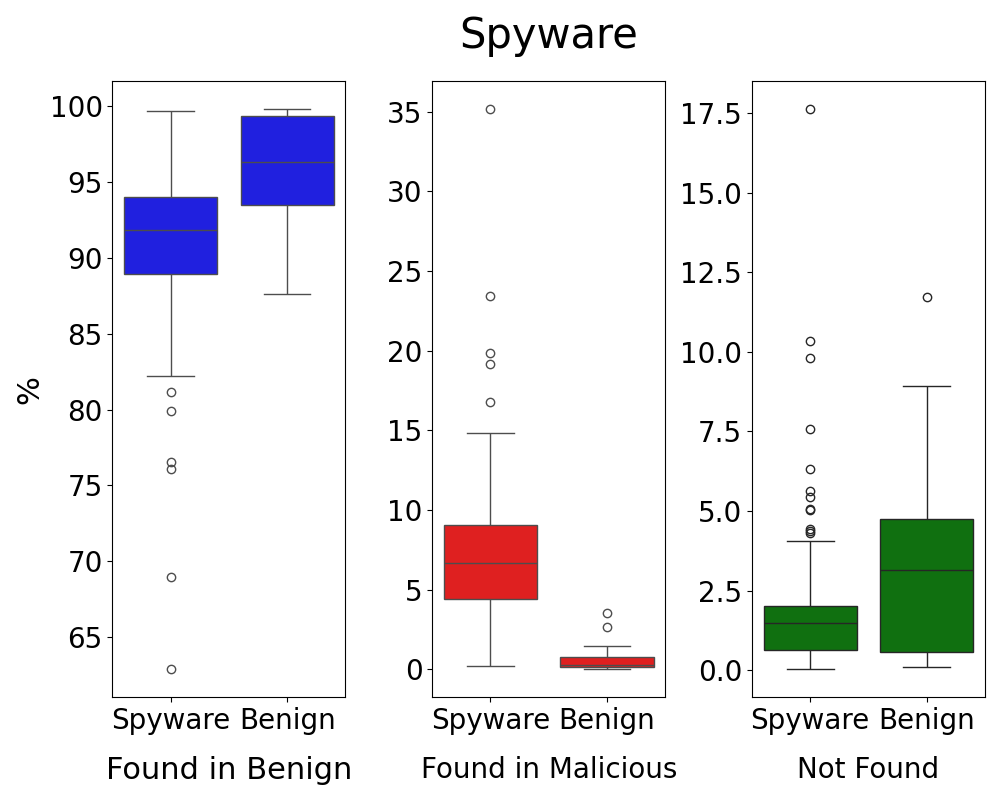}
    \end{minipage}
    \begin{minipage}{0.3\textwidth}
        \centering
        \includegraphics[width=\textwidth]{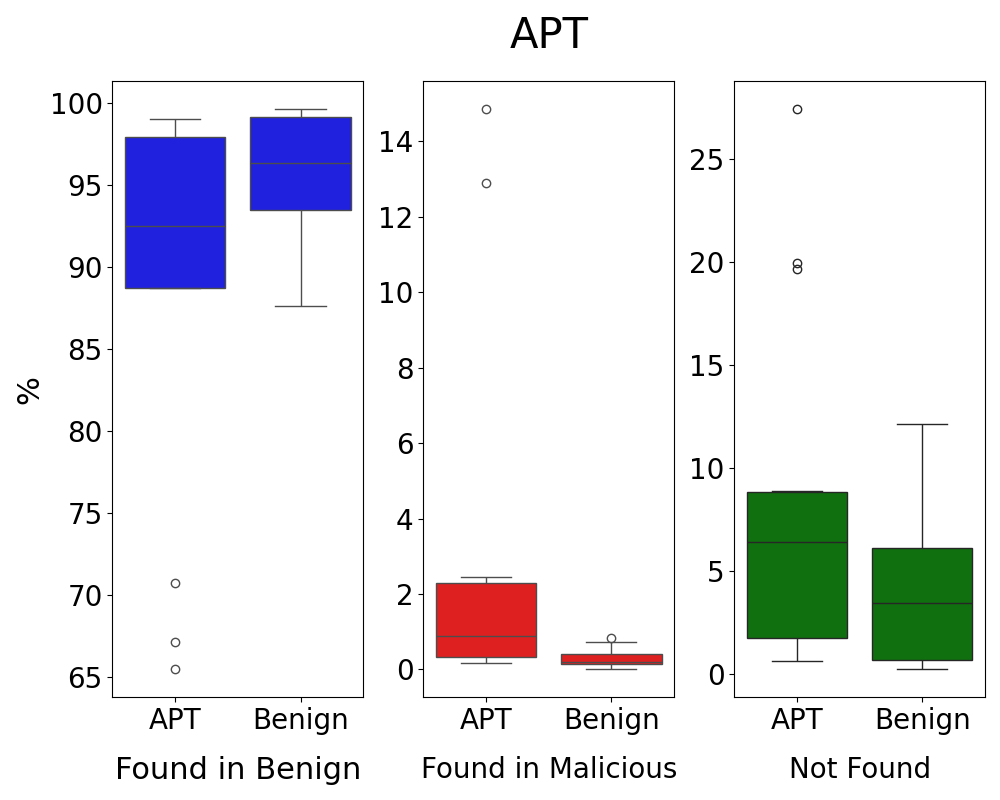}
    \end{minipage}
    
    \begin{minipage}{0.3\textwidth}
        \centering
        \includegraphics[width=\textwidth]{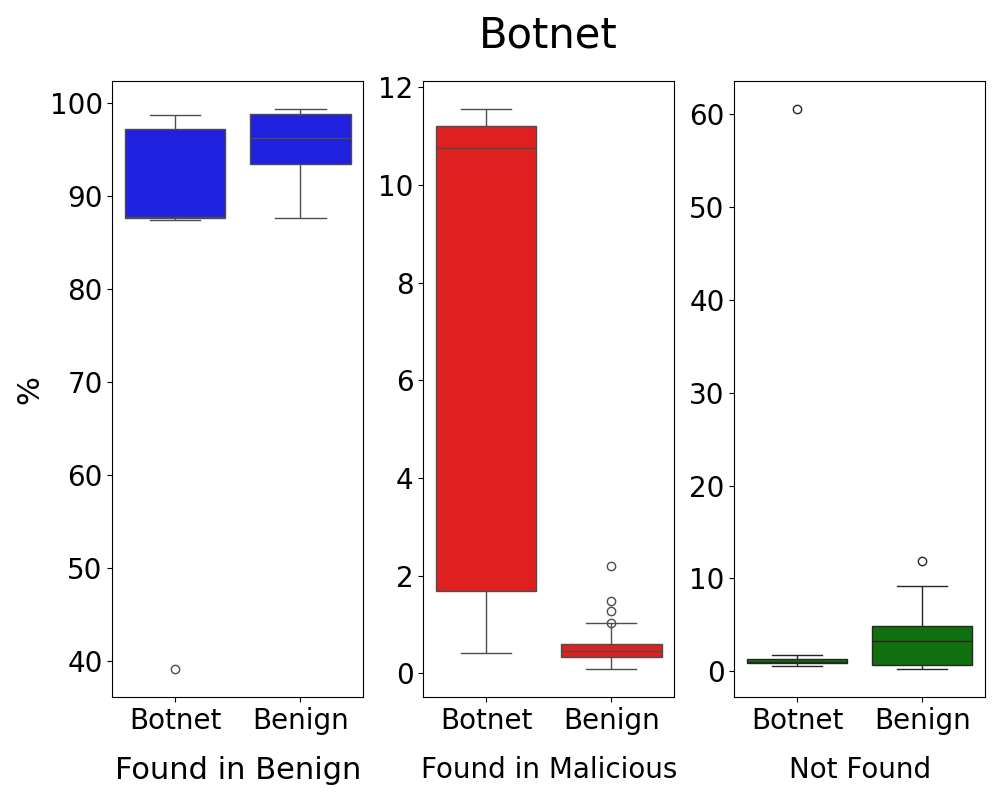}   
    \end{minipage}
    \begin{minipage}{0.3\textwidth}
        \centering
        \includegraphics[width=\textwidth]{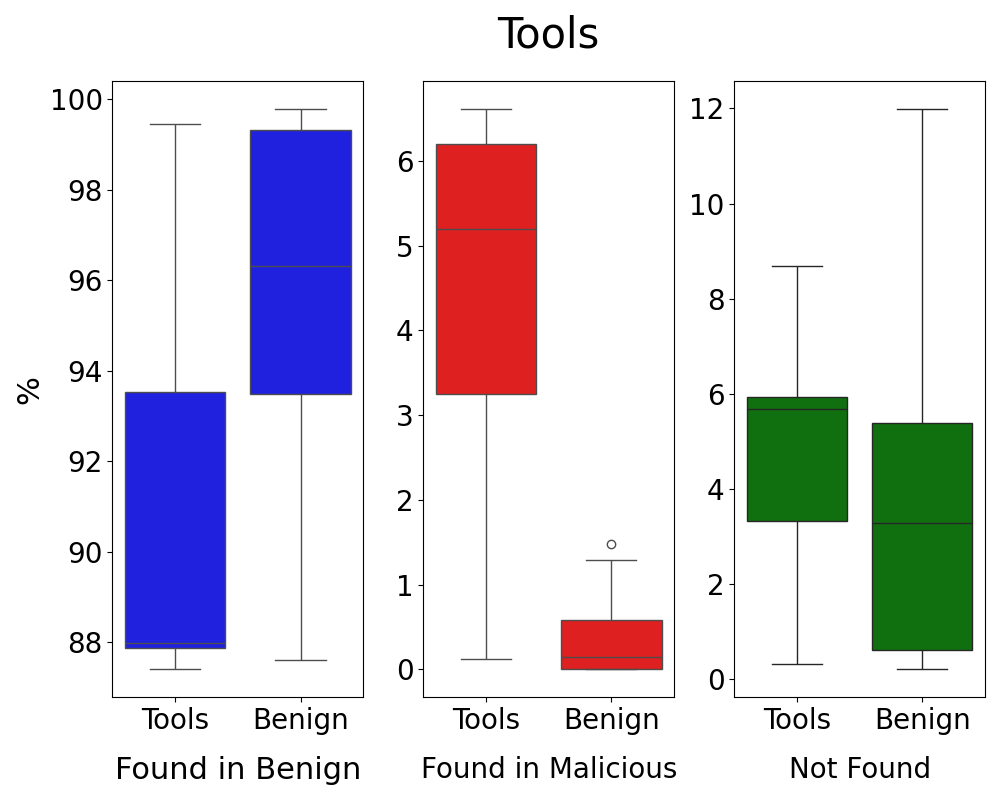}    
    \end{minipage}
    \caption{Malware and Benign test sample function loss}
    \label{fig:functionloss boxplots}
\end{figure}

When the minute 3 time slice of each test sample is filtered against the training data, a substantial amount of information is lost. Figure \ref{fig:functionloss boxplots} illustrates the percentage of functions removed from the test samples prior to classification with the DistilBERT model, as well as the reasons for their removal, that is whether the functions were labeled benign or malicious in the training data. The analysis shows that up to 25\% of the functions in malware test samples were present in the malicious training data, while up to 3\% of the functions in benign test samples were also found in the malicious training data. Further, up to 99\% of the functions in both malware and benign samples were found in the training data and labeled as benign. The \texttt{Not Found} category are the functions that remain in the test samples and are those that were used in Experiment B.

\begin{algorithm}[htbp]
\caption{Alpha}\label{alg:alpha-classification}
\raggedright
\textbf{Input:}
\begin{algorithmic}
\STATE Training dataset \( D \) of functions labeled as malicious or benign.
\STATE Fine-tuned DistilBERT model \( M \).
\STATE Test sample \( S \) to classify.\\
\end{algorithmic}
\textbf{Preprocessing:}
\begin{algorithmic}
\STATE Tokenize \( S \) into functions \( F = \{ f_1, f_2, \ldots, f_n \} \).\\
\end{algorithmic}
\textbf{Compare Functions to Training Dataset}
\begin{algorithmic}
\STATE Initialize \( \text{malicious\_count} = 0 \), \( \text{benign\_count} = 0 \).
\FOR{each function \( f_i \in F \)}
    \IF{\( f_i \) exists in \( D \)}
        \STATE \( l \leftarrow \text{label of } f_i; \text{Increment } \text{malicious\_count} \text{ if } l = \text{malicious, else } \text{benign\_count}. \)
    \ENDIF
\ENDFOR
\end{algorithmic}
\textbf{Layer 1 Function Loss Classification SVM}
\begin{algorithmic}
\FOR{each sample \( S \)}
\STATE Use the SVM model to classify \( S \) = SVM[\text{malicious\_count}, \text{benign\_count}]:\\
\STATE Compute \( \text{distance} \) of the feature vector to the SVM hyperplane.
\IF{\( (\text{classification} = \text{malicious} \text{ AND } \text{distance} \geq \text{upper\_threshold}) \)} 
    \STATE Classify \( S \) as \textbf{malicious}.
\ELSIF{\( (\text{classification} = \text{benign} \text{ AND } \text{distance} \leq \text{lower\_threshold}) \)}
    \STATE Classify \( S \) as \textbf{benign}.
\ELSE
    \STATE Proceed to \textbf{DistilBERT Classification}.
\ENDIF
\ENDFOR\\
\end{algorithmic}
\textbf{Layer 2 DistilBERT Function Classification}
\begin{algorithmic}
\FOR{each function \( f_i \in S \)}
    \STATE Use model \( M \) to classify \( f_i \).
\ENDFOR
\STATE Collect predictions \( P = \{ p_1, p_2, \ldots, p_n \} \) from model \( M \).\\
\end{algorithmic}
\textbf{Layer 3 Final Classification SVM}
\begin{algorithmic}
\STATE Use BERT predictions \( P \) for \( S \) to create feature set \( FS \), representing the malicious percentage.
\STATE Fit SVM hyperplane \( H \) to \( FS \) and predict the class label for \( S \).\\
\end{algorithmic}
\textbf{Output:}
\begin{algorithmic}
\STATE Class label for the test sample \( S \) (either malicious or benign).
\end{algorithmic}
\end{algorithm}

\begin{figure}[htbp]
    \centering
    \begin{minipage}{0.25\textwidth}
        \centering
        \includegraphics[width=\textwidth]{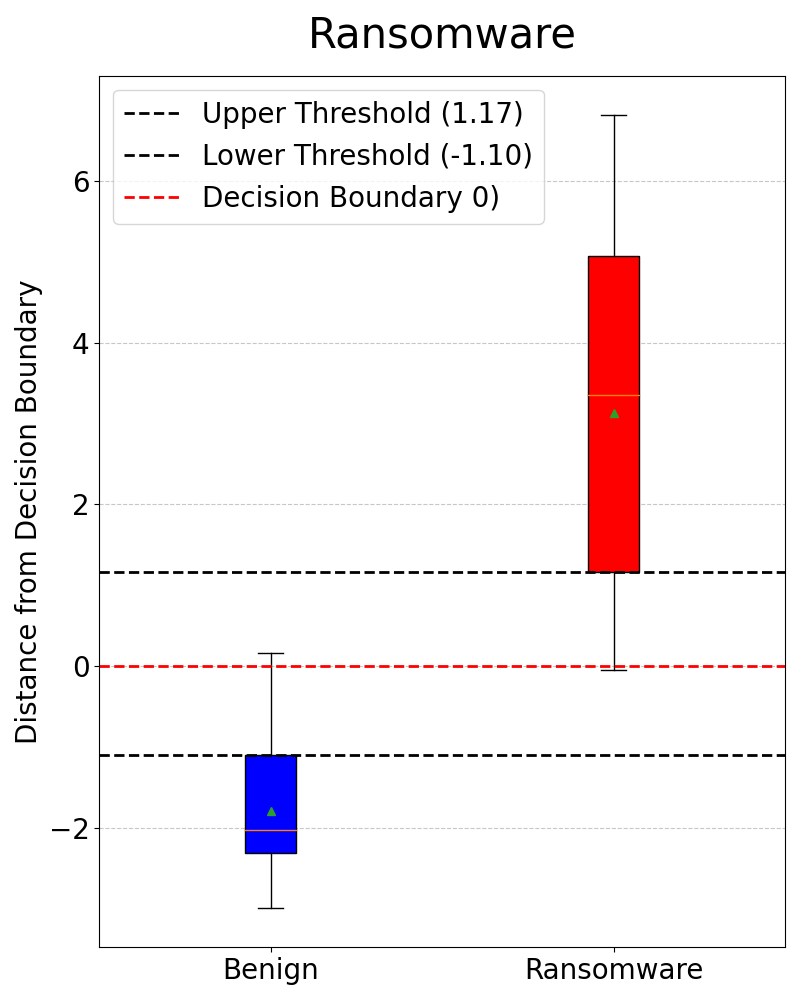}
    \end{minipage}
    \begin{minipage}{0.25\textwidth}
        \centering
        \includegraphics[width=\textwidth]{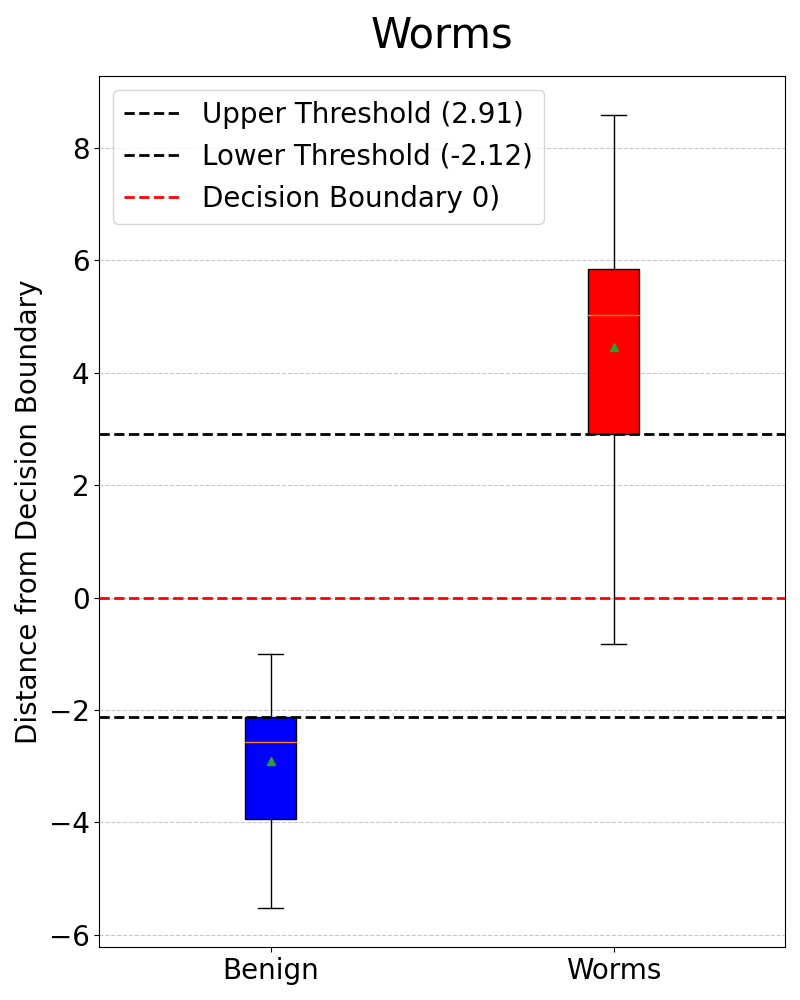}
    \end{minipage}
    
    \begin{minipage}{0.25\textwidth}
        \centering
        \includegraphics[width=\textwidth]{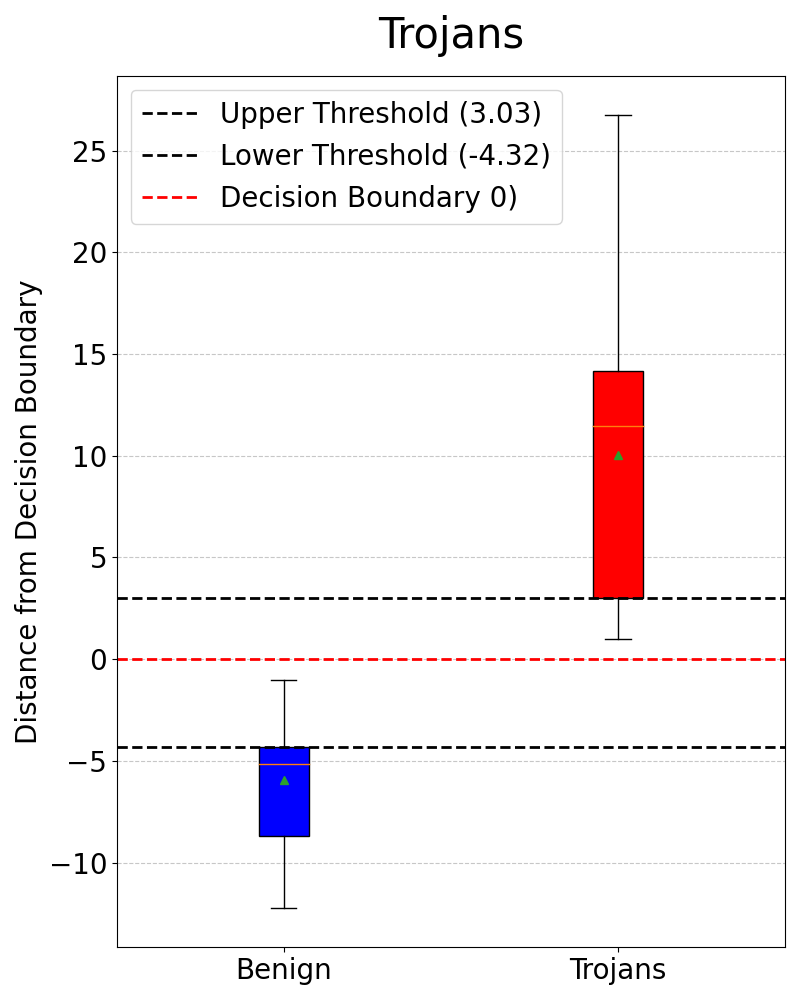}
    \end{minipage}
    \begin{minipage}{0.25\textwidth}
        \centering
        \includegraphics[width=\textwidth]{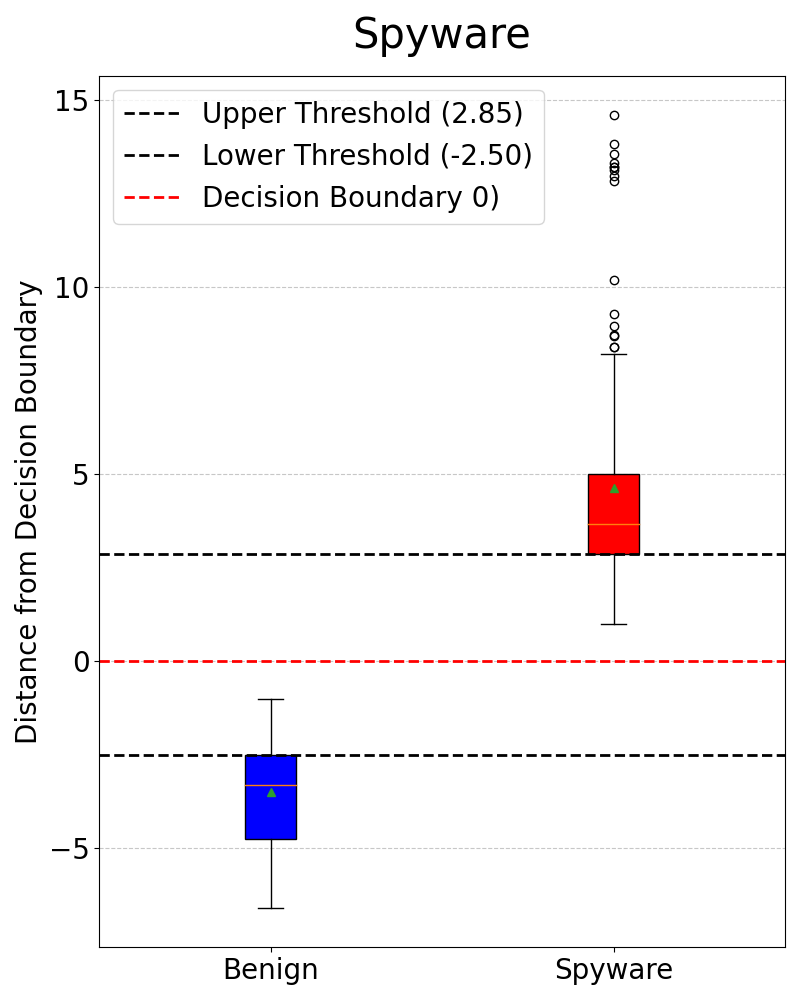} 
    \end{minipage}
    \begin{minipage}{0.25\textwidth}
        \centering
        \includegraphics[width=\textwidth]{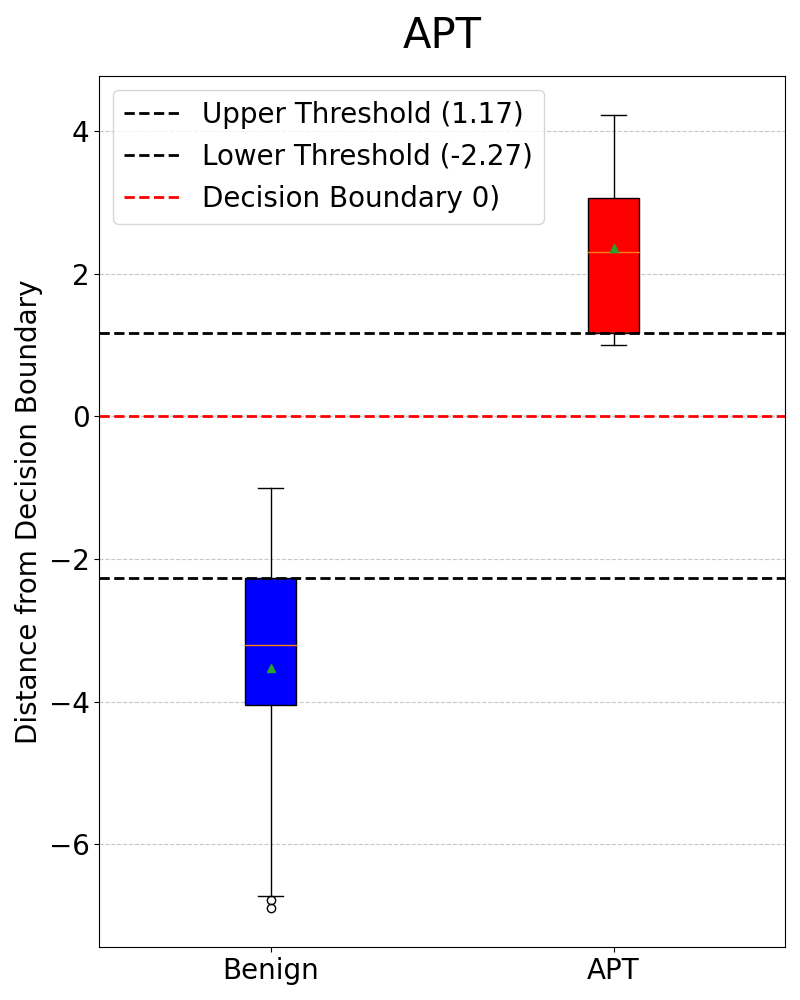}
    \end{minipage}
    
    \begin{minipage}{0.25\textwidth}
        \centering
        \includegraphics[width=\textwidth]{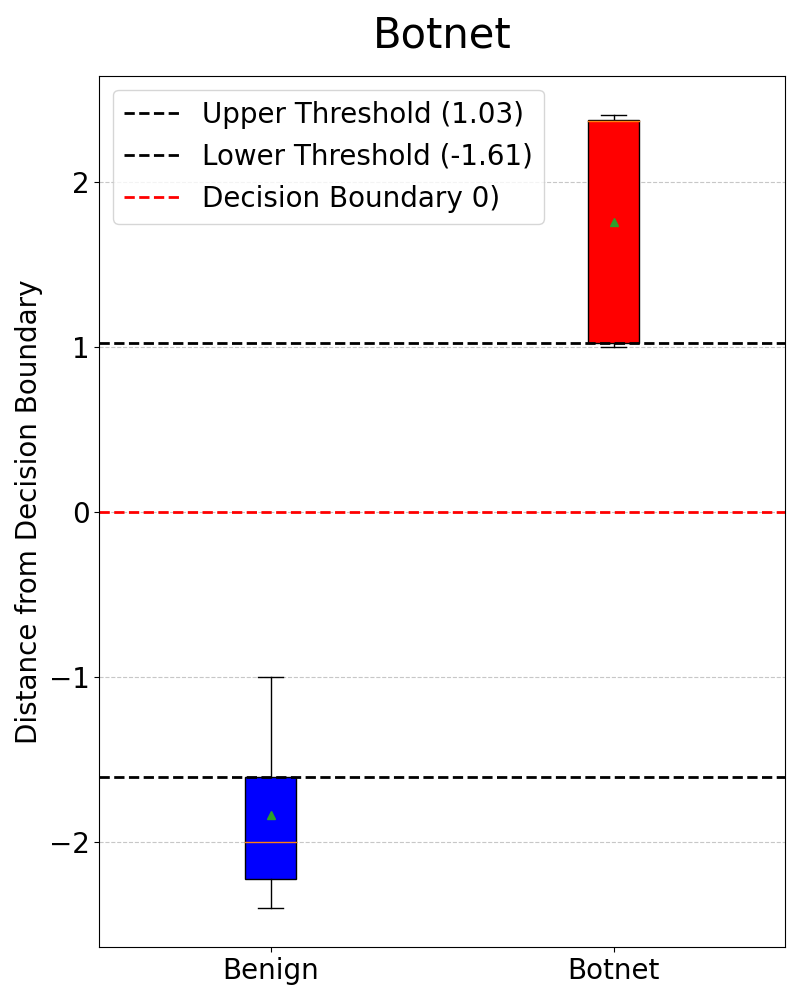}      
    \end{minipage}
    \begin{minipage}{0.25\textwidth}
        \centering
        \includegraphics[width=\textwidth]{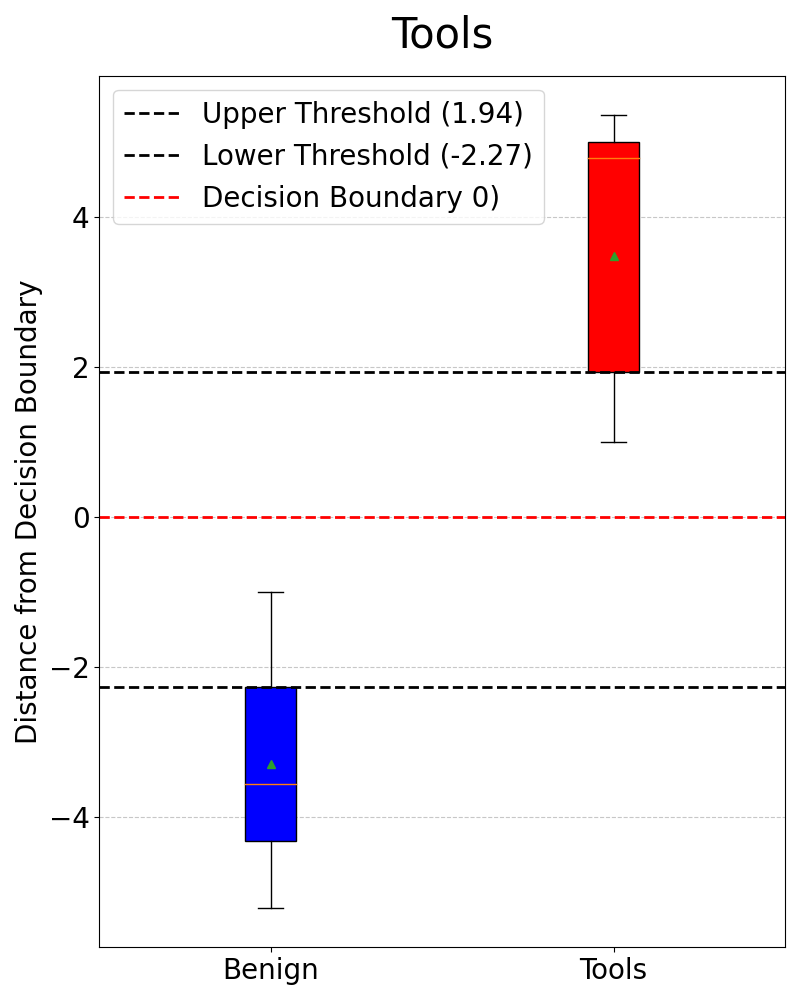}
    \end{minipage}
    \caption{Layer 1 Function Loss Classification SVM results   }
    \label{fig:Experiment C FuncLoss SVM}
\end{figure}

\begin{figure}[htbp]
    \centering
    \begin{minipage}{0.3\textwidth}
        \centering
        \includegraphics[width=\textwidth]{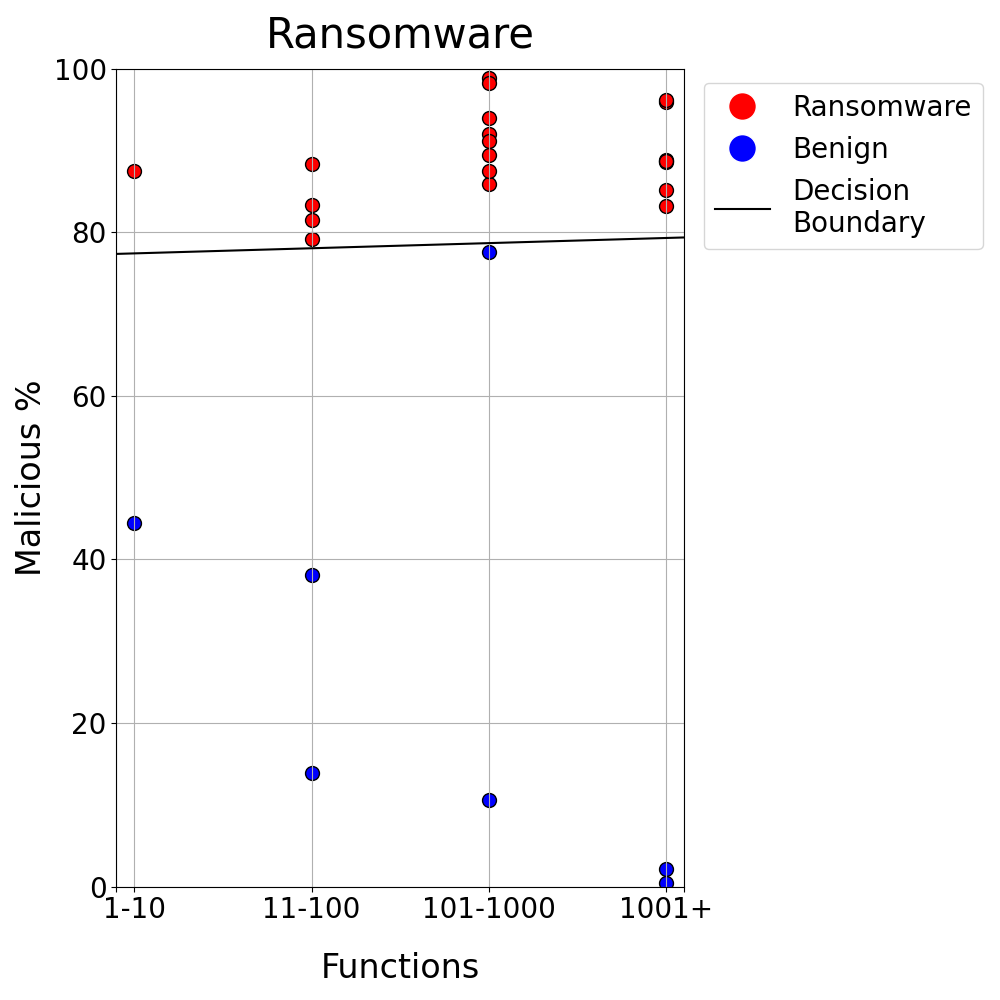}
    \end{minipage}
    \begin{minipage}{0.3\textwidth}
        \centering
        \includegraphics[width=\textwidth]{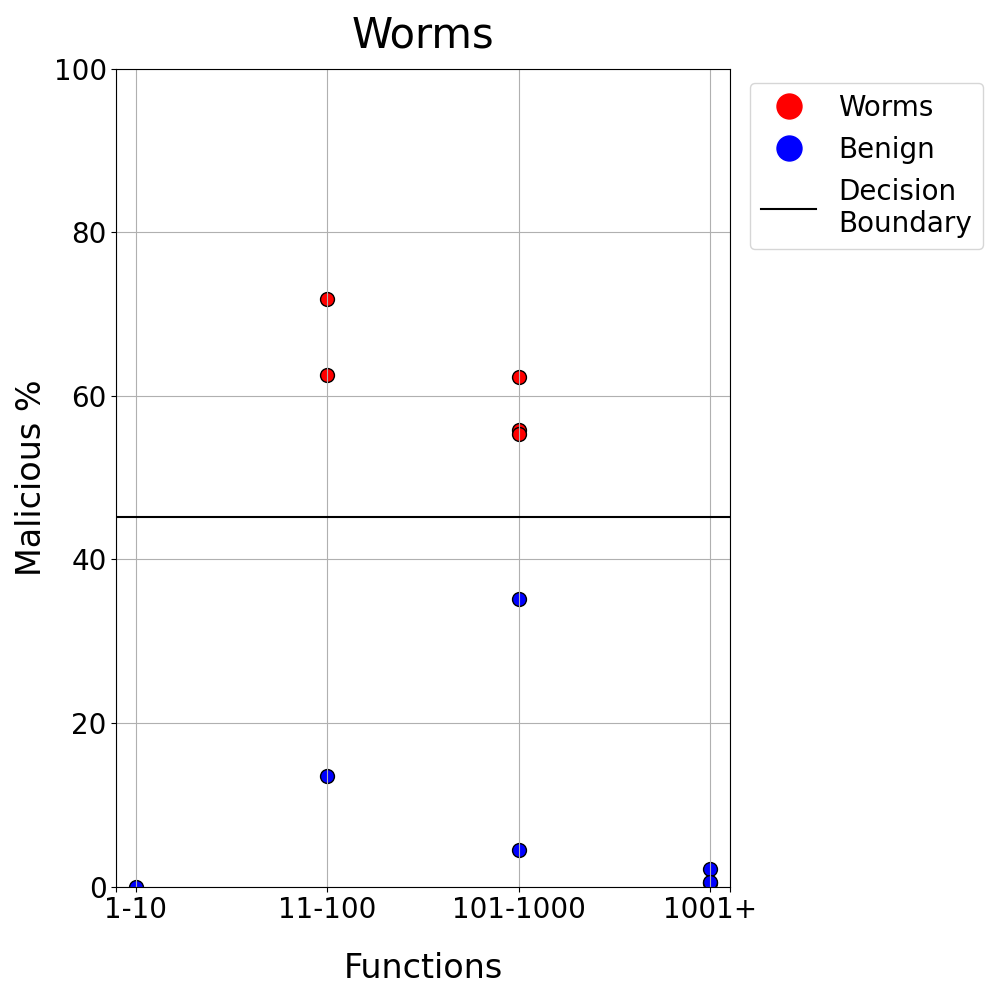}
    \end{minipage}
    
    \begin{minipage}{0.3\textwidth}
        \centering
        \includegraphics[width=\textwidth]{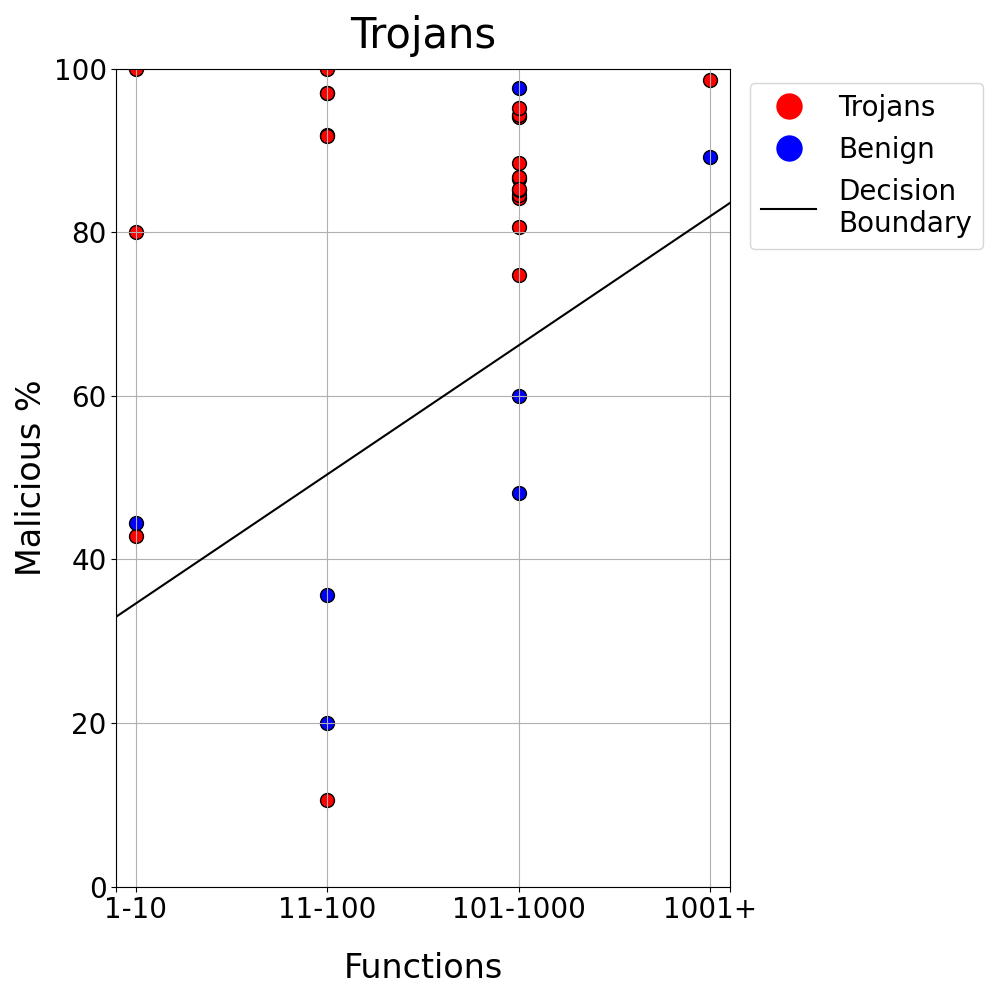}
    \end{minipage}
    \begin{minipage}{0.3\textwidth}
        \centering
        \includegraphics[width=\textwidth]{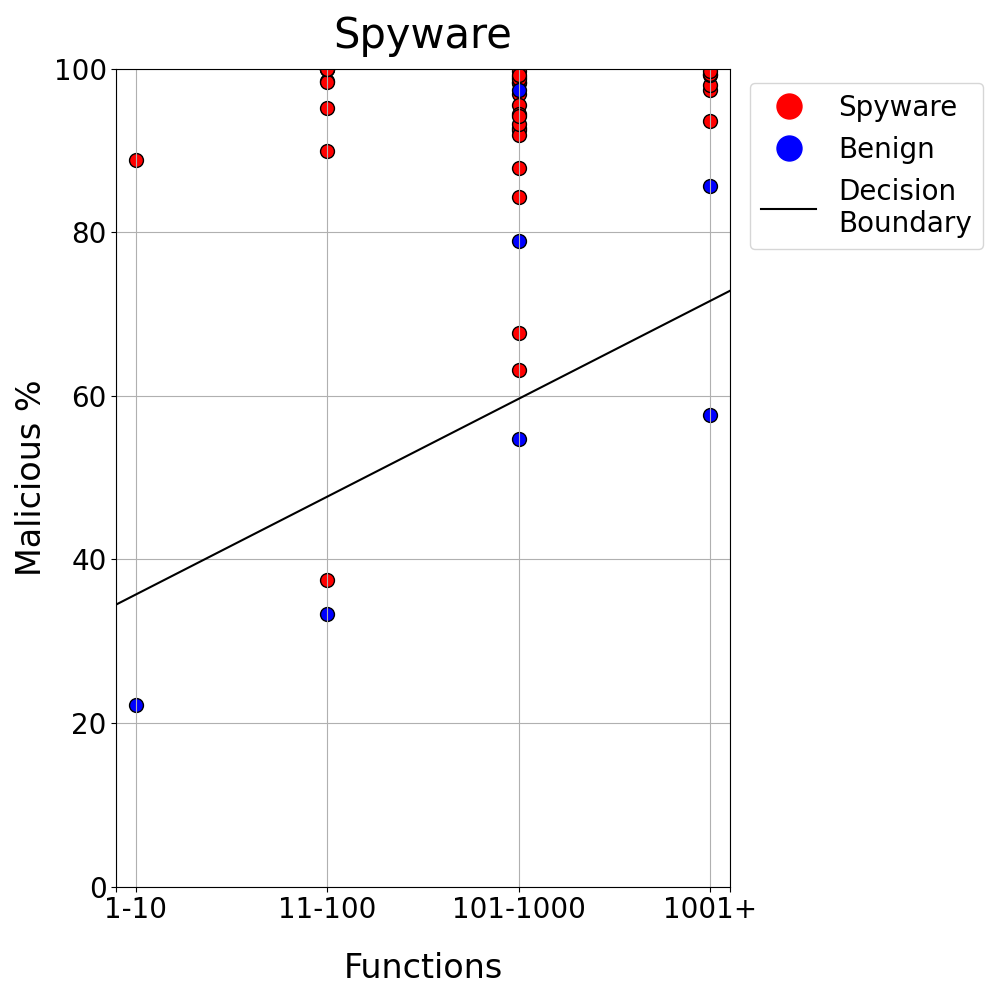}
    \end{minipage}
    \begin{minipage}{0.3\textwidth}
        \centering
        \includegraphics[width=\textwidth]{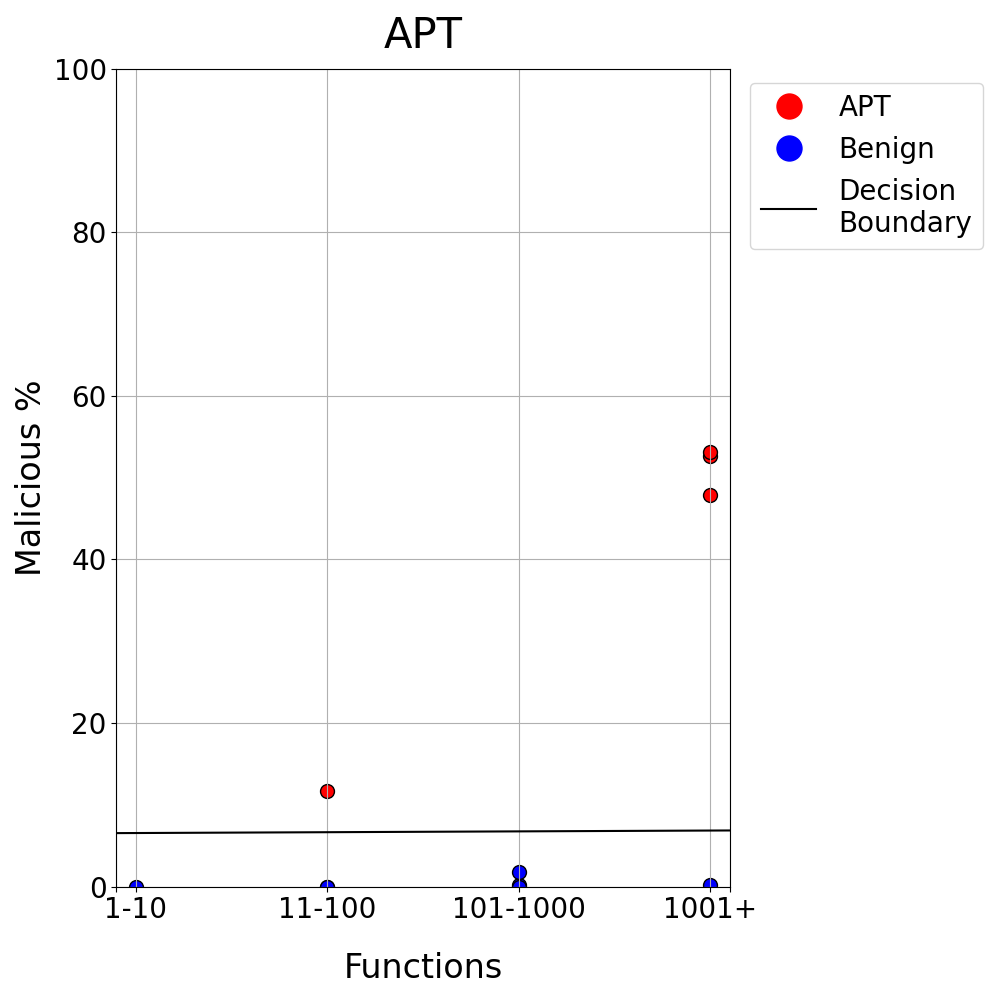}
    \end{minipage}
    
    \begin{minipage}{0.3\textwidth}
        \centering
        \includegraphics[width=\textwidth]{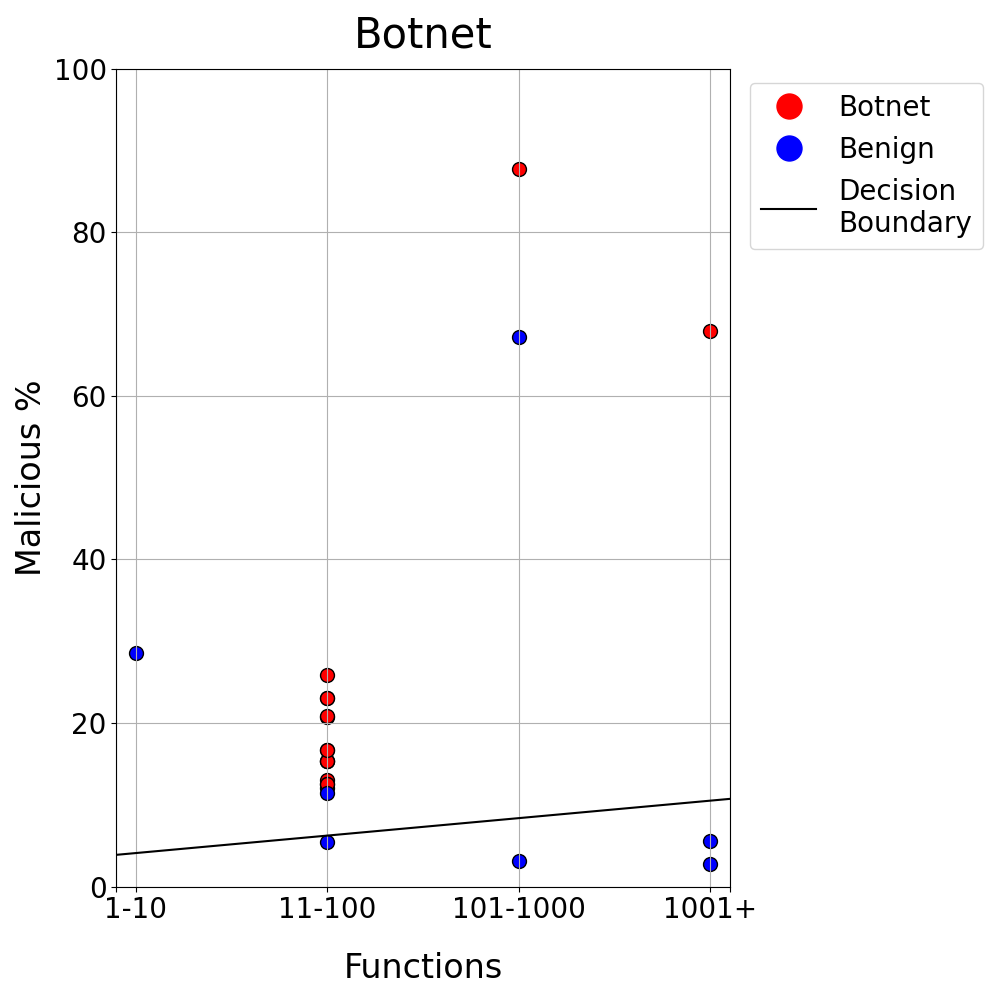}
    \end{minipage}
    \begin{minipage}{0.3\textwidth}
        \centering
        \includegraphics[width=\textwidth]{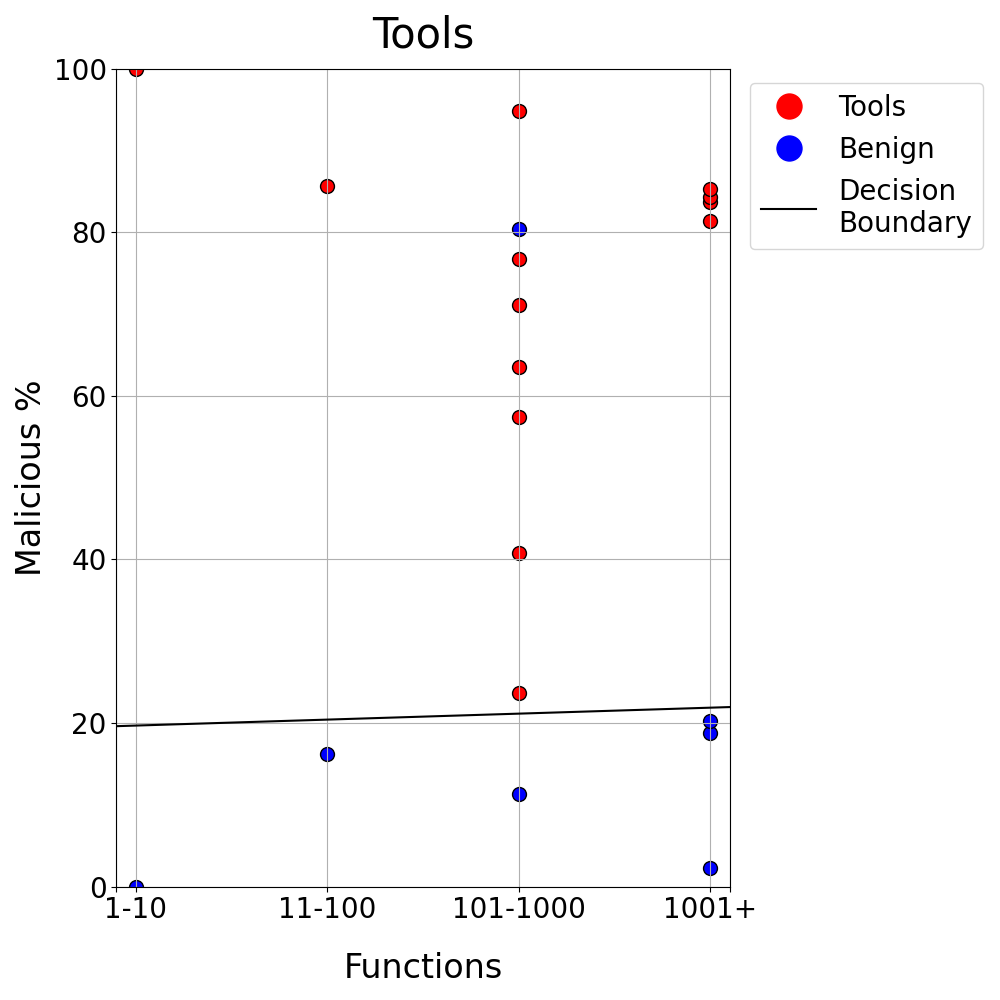}    
    \end{minipage}
    \caption{Final Alpha results for 1 minute of data}
    \label{fig:expc final svm}
\end{figure}

In this final experiment, a 1 minute data slice from minute 2 to 3 was used whenever available. For samples lacking this specific time slice, a 1 minute segment from the period just before their execution ended was used. As a result, the number of test samples in this experiment matches that of Experiment A. However, the number of test samples in Experiment B was smaller, as it exclusively relied on the time slice from minute 3, which was not always available.

Layer 1 Function Loss Classification SVM was incorporated into the model, as illustrated in \ref{fig:Alpha Classifier}, to address the significant information loss in some test samples. One of the main objectives of this experiment is to determine how fast accurate classification can be achieved, which raises the question: can a sample be classified based on its similarity to the training data? Specifically, when a test sample is filtered against the training data, how many functions are removed because they match benign functions in the training data, and how many are removed because they match malicious functions? This information loss is particularly significant for some test sample families that exhibit very similar functionality to other malware families of the same malware type present in the training data. The complete Alpha algorithm is presented in Algorithm \ref{alg:alpha-classification} and for interested readers the function loss details and numbers for Benign and Ransomware are presented in  the Appendix Tables \ref{tab:benign function loss} and \ref{tab:ransomware function loss}.

The results for Layer 1 Function Loss Classification SVM are shown in Figure \ref{fig:Experiment C FuncLoss SVM}. The upper and lower thresholds are determined using quartile-based metrics derived from the benign and malicious classifications. Specifically, if a sample's distance from the decision boundary exceeds the first quartile (Q1) for its predicted class, it is considered classified with high confidence. Conversely, if the sample's distance falls within the first quartile, it is flagged for further testing using the DistilBERT model. This cautious approach ensures robust classification.

The Function Loss Classification SVM demonstrated 100\% accuracy for the malware types Trojans, Spyware, APTs, Botnets, and Tools, with only 3 mis-classifications observed in the Ransomware and Worm categories. When the flagged Ransomware and Worm samples, identified based on Q1 thresholds, were subsequently tested using the corresponding DistilBERT  model from Experiment A, an overall accuracy of 100\% was achieved. However, for the other types of malware, although the SVM was 100\% accurate, the flagged samples passed to DistilBERT achieved slightly lower accuracies.

This layered classification strategy is justified, as it prioritizes minimizing both false positives and false negatives. In future research, the goal is to combine the models into a binary classifier that directly distinguishes between benign and malicious samples. Under this unified framework, any new sample tested will not have prior type information and could belong to any malware category. Establishing precise thresholds at this first layer is critical to ensure no errors propagate to subsequent stages, thereby maintaining the integrity of the classification process.

The results from Layer 3 Final Classification SVM for the flagged samples from Layer 1 are shown in Figure \ref{fig:expc final svm}. This shows how the flagged samples, those that were deemed uncertain by the initial classification thresholds in Layer 1 are handled by the DistilBERT models. The results demonstrate that, for the majority of malware types, the classification appears robust, with a clear and clean separation between the benign and malicious classes. This clean separation indicates that the DistilBERT model is effective in refining the classification decisions for these edge cases, improving the overall accuracy of the system. By addressing flagged samples in this way, the approach ensures minimal overlap between classes, reducing the likelihood of mis-classifications and enhancing the reliability of this final decision making process.

The final results and performance metrics are shown in Table \ref{tab:finalsvmexpc}. Alpha demonstrates perfect accuracy for certain types of malware. Ransomware, Worms and APT all achieved 100\% accuracy, precision, recall, and F1-score, indicating flawless classification for both malicious and benign samples.
These results highlight the model’s exceptional performance in separating these malware types from benign data. Alpha demonstrates high overall accuracy for Spyware, Trojans, Botnets and Tools. While not perfect, these categories achieved accuracy above 96\%, with F1-scores ranging from 97.64 to 99.01. These scores demonstrate that the model performs effectively even when dealing with more complex malware categories or those with functionality closely resembling legitimate updaters and utilities. Across all categories, precision and recall are consistently high, indicating minimal false positives (FP) and false negatives (FN), which are critical for reliable malware detection. The flagged samples from Layer 1 Function Loss SVM classification were effectively managed by Layer 2 DistilBERT and Layer 3 Final Classification SVM, leading to strong performance metrics.

\begin{table}[htbp]
\centering
\footnotesize
\captionsetup{justification=centering}
\caption{Final Alpha results and performance metrics for 1 minute of data}
\label{tab:finalsvmexpc}
\begin{tabular}{lccccccccccc}
    \toprule
    \makecell{Name} & \makecell{L1\\Malware} & \makecell{L1\\Benign} & \makecell{L1\\Flagged} & \makecell{L3\\TP} & \makecell{L3\\FN} & \makecell{L3\\FP} & \makecell{L3\\TN} & \makecell{Accuracy} & \makecell{Precision} & \makecell{Recall} & \makecell{F1}\\
    \midrule
    \makecell{Ransomware} & \makecell{68} & \makecell{18} & \makecell{27} & \makecell{20} & \makecell{0} & \makecell{0} & \makecell{7} & \makecell{100} & \makecell{100} & \makecell{100} & \makecell{100} \\
    \makecell{Worms} & \makecell{21} & \makecell{18} & \makecell{13} & \makecell{5} & \makecell{0} & \makecell{0} & \makecell{7} & \makecell{100} & \makecell{100} & \makecell{100} & \makecell{100} \\
    \makecell{Trojans} & \makecell{63} & \makecell{17} & \makecell{29} & \makecell{21} & \makecell{1} & \makecell{3} & \makecell{4} & \makecell{96.33} & \makecell{96.55} & \makecell{98.82} & \makecell{97.67} \\
    \makecell{Spyware} & \makecell{99} & \makecell{18} & \makecell{41} & \makecell{33} & \makecell{1} & \makecell{3} & \makecell{4} & \makecell{97.47} & \makecell{97.78} & \makecell{99.25} & \makecell{98.51} \\
    \makecell{APT} & \makecell{10} & \makecell{18} & \makecell{11} & \makecell{4} & \makecell{0} & \makecell{0} & \makecell{7} & \makecell{100} & \makecell{100} & \makecell{100} & \makecell{100} \\
     \makecell{Botnet} & \makecell{46} & \makecell{18} & \makecell{23} & \makecell{16} & \makecell{0} & \makecell{3} & \makecell{4} & \makecell{96.55} & \makecell{95.38} & \makecell{100} & \makecell{97.64} \\
     \makecell{Tools} & \makecell{37} & \makecell{18} & \makecell{20} & \makecell{13} & \makecell{0} & \makecell{1} & \makecell{6} & \makecell{98.67} & \makecell{98.04} & \makecell{100} & \makecell{99.01} \\
    \bottomrule
\end{tabular}
\end{table}
There are challenges with Trojans and Spyware, Trojans had 3 FP, reducing its precision to 96.55\%. Spyware also had 3 FP, with precision dropping to 97.78\%. Trojans and Spyware each had 1 FN, that is a malicious sample was incorrectly classified as benign. Further, Spyware and Trojan categories had the highest number of flagged samples, 41 for Spyware and 29 for Trojans, indicating that these types are more challenging to classify confidently in Layer 1. These issues indicate room for improvement in classifying malware categories that are more complex and functionally similar to benign software. While the recall for Botnets is 100\%, precision is slightly lower at 95.38\% due to 3 FP. This suggests that some benign samples are misclassified as Botnets, which could lead to unnecessary escalations. 

\section{Discussion}
\begin{table}[htbp]
\centering
\footnotesize
\captionsetup{justification=centering}
\caption{A comparison of the Alpha framework with state-of-the-art malware detection \\ \textbf{Note:} b=benign, m=malware, r=ransomware}
\label{tab:performance_comps}
\begin{tabular}{ccccc}
    \toprule
    \makecell{Paper} & \makecell{Dataset} & \makecell{Features} & \makecell{Model} & \makecell{Accuracy \%}\\
    \midrule
    \makecell{\cite{saracino2023}} & \makecell{568 B\\568 M} & \makecell{API Call\\Sequence} & \makecell{BERT\\Neural Net}& \makecell{97.62} \\\\
    \makecell{\cite{kimzeroday2021}} & \makecell{4,000 B\\4,000 M} & \makecell{Static} & \makecell{CNN}& \makecell{94.8} \\\\
    \makecell{\cite{kimzero2023}} & \makecell{8,000 B\\2,000 M} & \makecell{Static} & \makecell{AutoEncoder\\CNN}& \makecell{97.1} \\\\
    \makecell{\cite{deng2024}} & \makecell{27,118 B\\35,367 R} & \makecell{Static PE\\Header} & \makecell{Double Deep Q\\Learning Network}& \makecell{97.9} \\\\
    \makecell{\cite{zahoora2022}} & \makecell{942 B\\582 R} & \makecell{Sandbox} & \makecell{Auto Encoder\\Ensemble Classifier}& \makecell{92.8} \\\\
    \makecell{\cite{ayub2024}} & \makecell{101 B\\215 R} & \makecell{Sandbox} & \makecell{Random Forest}& \makecell{97.67} \\\\
    \makecell{\cite{zahoora2022pareto}} & \makecell{942 B\\582 R} & \makecell{Sandbox} & \makecell{Cost-Sensitive\\Pareto Ensemble}& \makecell{93.00} \\\\
    \makecell{\cite{demirkiran2022}} & \makecell{40,566 M} & \makecell{Cuckoo\\Sandbox} & \makecell{Transformers}& \makecell{61.49 F1} \\\\
    \makecell{\cite{rahali2021}} & \makecell{12,000 B\\10,000 M} & \makecell{Static\\Manifest.xml} & \makecell{BERT}& \makecell{97.61} \\\\
    \makecell{Alpha\\(our work)} & \makecell{ 308\\ 4,505 M} & \makecell{Peekaboo DBI} & \makecell{DistilBERT}& \makecell{98.43} \\
    \bottomrule
\end{tabular}
\end{table}

A comparison to other approaches is provided in Table \ref{tab:performance_comps}. The accuracy and effectiveness of an AI model depend on factors such as the dataset, the features used for training, and the model architecture \cite{gabercsur2024}. Comparing the performance of different models is inherently challenging, as each study employs distinct datasets, features, and AI techniques. Nevertheless, Alpha demonstrates superior performance in several key areas.

What sets Alpha apart is its ability to detect genuinely new and previously unseen samples while achieving the highest accuracy. Malware variants often display only minor differences, typically involving subtle modifications to key functionalities such as file enumeration, data exfiltration, or payload execution algorithms. As a result, models trained on one variant can easily classify similar test samples if overlapping features are present. In contrast, our approach removes any familiar functions from the test samples, compelling the model to identify malicious behavior based solely on context and novel ASM instruction patterns. This ensures a more robust detection of new threats.

Alpha demonstrates outstanding performance in classifying truly new and novel malware, achieving perfect scores for Ransomware, Worms, APT, and Tools. While Trojans, Spyware, and Botnets also perform exceptionally well, occasional false positives and false negatives in these categories highlight areas for further refinement. The high number of flagged samples for Spyware and Trojans indicates that these malware types are particularly challenging to classify accurately. This may stem from their functional similarities to benign software, making it harder for the model to distinguish between the two classes. Trojans, Spyware and Post Exploitation Tools exhibit functionality similar to updaters and utilities because they perform common actions such as downloading and executing files, modifying system settings, or operating in the background, tasks that are typical of legitimate software. For instance, these types of malware, like legitimate updaters, may download additional components; however, the components delivered  are malicious payloads. Similarly, they mimic utilities by performing tasks such as accessing system files and interacting with the registry, while covertly executing harmful activities like installing backdoors or stealing data. These overlapping behaviors at the functional level make it challenging to distinguish between benign software and malware. Addressing these issues would require a more balanced and diverse training dataset that includes a wider range of both benign and malicious samples. For benign samples, it’s important to incorporate examples that closely mimic the behavior or functionality of malware, such as legitimate updaters or utilities, to help the model learn subtle distinctions. Similarly, the inclusion of more diverse malicious samples, representing various behaviors and characteristics, would enhance the model’s ability to generalize and improve its performance in these borderline cases. This balance is crucial for reducing the number of flagged samples and improving the overall classification accuracy for these complex categories. Overall, Alpha effectively balances accuracy, precision, and recall, minimizing errors while ensuring robust zero day malware detection.

Alpha demonstrated resilience even against challenging samples, correctly classifying samples that only executed for a brief period. Notably, none of the ransomware samples performed encryption within the first three minutes of execution, showcasing the robustness of the classification process. Further, adversarial attacks against the DistilBERT model itself poses a significant challenge because the features used for classification are extracted dynamically under Peekaboo DBI. Any adversarial attempts to escape or compromise the DBI would likely stand out as suspicious behavior and act as additional indicators of malicious intent, rather than undermining the integrity of the extracted features. For example, long sleeps are a common evasion technique used by malware authors but such suspicious behavior would itself be a red flag. Extended idle periods or sleep calls are uncommon in legitimate processes and this functionality would make the sample stand out in analysis. By leveraging behavioral patterns under Peekaboo DBI at the ASM insturuction level, Alpha effectively detects and classifies zero day malware even when adversaries attempt to hide their malicious activity, ensuring a high degree of resilience against evasive tactics.

\section{Conclusion and Future Research}
This research leveraged DistilBERT Transformer models alongside innovative feature engineering techniques using Peekaboo DBI data. The test dataset comprised multiple malware types and families as well as benign samples. To emphasize the detection of truly novel malware, a key objective of this study, any function present in both the training and test data was excluded from the test samples. This forced the model to rely solely on identifying malicious patterns based on ASM instruction combinations and contextual information, rather than memorizing known functions.

The classification process involved a 3 layer approach. First, the function loss SVM used thresholds to make highly confident predictions based on the loss of functions when filtered against the training dataset. Secondly the DistilBERT model and function classification layer estimated the number of benign and malicious functions within each sample. Then, the final classification SVM used a hyperplane as the decision boundary to perform the final classification of samples as either benign or malicious.

Our experimental results demonstrate the exceptional effectiveness of Alpha in detecting novel malicious functions, showcasing its robustness and transferability to previously unseen threats. Alpha surpasses previous state-of-the-art methods for zero day malware detection by focusing exclusively on new ASM instruction patterns and behaviors, avoiding reliance on familiar functions. To the best of our knowledge, this study is the first to utilize ASM language with Transformer models for the detection of entirely new malicious functions and samples across diverse types of malware.

In future research, we aim to develop a binary classification model to distinguish between malware and benign samples, building upon the approach and algorithm presented in this study. Additionally, we plan to enhance Peekaboo's efficiency to capture the first 3 minutes of data more effectively. This will involve eliminating unnecessary logging of API and system calls, as well as avoiding instrumentation that does not directly impact Peekaboo.

\section{Data Availability Statement}
\label{sec:Data Availability Statement}
The fine tuned DistilBERT models and scripts are available in the Peekaboo Transformer Models repository \cite{gabertpro2024}.

\section{Author Contributions Statement}
\label{sec:Author Contributions Statement}
Matthew Gaber conceived of the presented idea, carried out the experiments, wrote the main manuscript text and prepared the figures. Mohiuddin Ahmed and Helge Janicke helped supervise the project. All authors discussed the results and contributed to the manuscript.

\bibliographystyle{elsarticle-num} 
\bibliography{sample-base-num}

\section{Appendix}
\begin{table}[htbp]
\centering
\footnotesize
\captionsetup{justification=centering}
\caption{Experiment C benign function loss, found in benign or malicious training data}
\label{tab:benign function loss}
\begin{tabular}{lccccc}
    \toprule
    Filename & Initial Length & \makecell{After\\Deduplication\\Length} & \makecell{Found\\in Benign} & \makecell{Found\\in Malicious} & \makecell{Final\\ Instructions\\Left} \\
    \midrule
    b092b0 & 17784 & 12613 & 11906 & 148 & 559 \\ 
    eefac8 & 22875 & 15871 & 14759 & 220 & 892 \\ 
    c906bb & 18520 & 13289 & 12751 & 87 & 451 \\ 
    aa10bf & 15300 & 10756 & 10668 & 39 & 49 \\ 
    6caffa & 15013 & 10649 & 10579 & 34 & 36 \\ 
    5e87e2 & 13939 & 10119 & 10016 & 39 & 64 \\
    551a62 & 13569 & 9945 & 9892 & 32 & 21 \\ 
    1217e3 & 11609 & 8566 & 8521 & 25 & 20 \\ 
    b4e744 & 11557 & 8530 & 8487 & 25 & 18 \\ 
    86ee28 & 10966 & 8041 & 7781 & 148 & 112 \\ 
    7297a4 & 531 & 486 & 467 & 0 & 19 \\ 
    0e3de1 & 605 & 576 & 557 & 0 & 19 \\ 
    29ae90 & 708 & 619 & 604 & 0 & 15 \\ 
    147192 & 780 & 693 & 638 & 3 & 52 \\ 
    a1c81c & 839 & 702 & 648 & 1 & 53 \\ 
    cb7b6a & 898 & 759 & 665 & 2 & 92 \\ 
    b4853f & 944 & 785 & 730 & 8 & 47 \\ 
    54e561 & 964 & 798 & 746 & 1 & 51 \\ 
    63bf3a & 984 & 828 & 747 & 5 & 76 \\ 
    e58771 & 1235 & 1039 & 1001 & 4 & 34 \\ 
    9698fe & 1391 & 1135 & 1079 & 1 & 55 \\ 
    8cf5d3 & 1351 & 1114 & 1072 & 3 & 39 \\ 
    262aa8 & 1444 & 1169 & 1107 & 15 & 47 \\ 
    0fbe33 & 1725 & 1316 & 1283 & 6 & 27 \\ 
    4c33d7 & 1879 & 1578 & 1566 & 3 & 9 \\ 
    \bottomrule
\end{tabular}
\end{table}

\begin{table}[htbp]
\centering
\footnotesize
\captionsetup{justification=centering}
\caption{Experiment C Ransomware function loss, found in benign or malicious training data}
\label{tab:ransomware function loss}
\begin{tabular}{lccccc}
    \toprule
    Filename & Initial Length & \makecell{After\\Deduplication\\Length} & \makecell{Found\\in Benign} & \makecell{Found\\in Malicious} & \makecell{Final\\ Instructions\\Left} \\
    \midrule
111093 & 2979 & 2242 & 2202 & 27 & 13 \\ 
8d62ed & 18371 & 12019 & 9577 & 1675 & 767 \\ 
1520e4 & 10851 & 7643 & 6927 & 558 & 158 \\ 
a88e9c & 3647 & 2668 & 2628 & 28 & 12 \\ 
1866b2 & 2847 & 2147 & 2121 & 15 & 11 \\ 
802206 & 10108 & 7205 & 6480 & 528 & 197 \\ 
cea3e8 & 11892 & 7988 & 7318 & 526 & 144 \\ 
8b1380 & 10531 & 7385 & 6782 & 480 & 123 \\ 
1c2a51 & 10563 & 7389 & 6781 & 477 & 131 \\ 
f3e891 & 14611 & 9914 & 8596 & 591 & 727 \\ 
cc58dc & 2835 & 2132 & 2108 & 11 & 13 \\ 
b9d3d5 & 14652 & 9930 & 8572 & 591 & 767 \\ 
ea6153 & 1808 & 1508 & 1501 & 4 & 3 \\ 
0b02c5 & 1789 & 1492 & 1486 & 4 & 2 \\ 
a5e6df & 10530 & 7376 & 6770 & 475 & 131 \\ 
acfe3f & 2857 & 2161 & 2139 & 17 & 5 \\ 
74437a & 14635 & 9972 & 8601 & 598 & 773 \\ 
8d864c & 2848 & 2142 & 2119 & 12 & 11 \\ 
175944 & 10535 & 7378 & 6776 & 473 & 129 \\ 
8958d7 & 18383 & 12062 & 9606 & 1696 & 760 \\ 
fef1f9 & 14556 & 9935 & 8580 & 590 & 765 \\ 
0dd36a & 10838 & 7635 & 6922 & 561 & 152 \\ 
bcbdc1 & 10539 & 7379 & 6773 & 482 & 124 \\ 
989007 & 3571 & 2661 & 2602 & 23 & 36 \\ 
a95000 & 18345 & 12016 & 9568 & 1662 & 786 \\ 
f7c432 & 10949 & 7763 & 6812 & 538 & 413 \\ 
f03172 & 11042 & 7910 & 6918 & 554 & 438 \\ 
e8a091 & 10983 & 7893 & 6975 & 578 & 340 \\ 
4263ea & 13806 & 9747 & 8291 & 871 & 585 \\ 
ac457f & 12793 & 9107 & 7636 & 805 & 666 \\ 
6a2222 & 10599 & 7585 & 6895 & 498 & 192 \\ 
388558 & 10554 & 7596 & 6877 & 508 & 211 \\ 
d65fa9 & 11253 & 8038 & 7046 & 558 & 434 \\ 
b34569 & 14632 & 10258 & 8731 & 905 & 622 \\ 
4e2554 & 10996 & 7915 & 6922 & 567 & 426 \\ 
b5ef16 & 5677 & 4171 & 3875 & 54 & 242 \\ 
a1f30c & 5686 & 4161 & 3869 & 55 & 237 \\ 
36ae29 & 26834 & 18379 & 17461 & 311 & 607 \\ 
e079fa & 3159 & 2408 & 2378 & 26 & 4 \\ 
389a7d & 5692 & 4152 & 3867 & 50 & 235 \\ 
dff27b & 7571 & 5117 & 4917 & 75 & 125 \\ 
9fb6be & 13879 & 9130 & 6440 & 2220 & 470 \\ 
09ba59 & 13830 & 9263 & 6262 & 2141 & 860 \\ 
d6c5d9 & 12752 & 8408 & 6177 & 1855 & 376 \\ 
76bac3 & 7432 & 5398 & 4277 & 406 & 715 \\ 
011c24 & 12322 & 8386 & 5936 & 2104 & 346 \\ 
e8952e & 13737 & 9067 & 6393 & 2260 & 414 \\ 
c94e64 & 12295 & 8362 & 5973 & 2095 & 294 \\ 
c53eb1 & 13534 & 8897 & 6415 & 2160 & 322 \\ 
f4b761 & 19500 & 13048 & 12137 & 801 & 110 \\ 
be2264 & 4860 & 3766 & 3214 & 169 & 383 \\ 
fc8497 & 7573 & 5127 & 4913 & 78 & 136 \\ 
7ec640 & 7692 & 5183 & 4937 & 85 & 161 \\ 
ec700d & 5435 & 3872 & 3693 & 32 & 147 \\ 
38de37 & 12266 & 8339 & 5962 & 2099 & 278 \\ 
b04012 & 7560 & 5154 & 4935 & 78 & 141 \\ 
8768b3 & 13883 & 9105 & 6418 & 2231 & 456 \\ 
8341e1 & 13851 & 9120 & 6413 & 2227 & 480 \\ 
ce2194 & 13957 & 9169 & 6449 & 2255 & 465 \\ 
9ba201 & 12844 & 8412 & 6239 & 1841 & 332 \\ 
994b41 & 13847 & 9122 & 6418 & 2214 & 490 \\ 
cb11ca & 13799 & 9059 & 6428 & 2286 & 345 \\ 
adaaf5 & 7527 & 5116 & 4918 & 78 & 120 \\ 
    \bottomrule
\end{tabular}
\end{table}

\end{document}